\documentclass[12pt]{article}
\usepackage{setspace, amsmath, graphicx, multirow, bigdelim, float, amsfonts, caption, color, colortbl, lscape, subcaption,bm, array,natbib,url}

\usepackage{amssymb, amsthm}
\usepackage{booktabs}
\usepackage{geometry}
\usepackage{multirow}
\usepackage{comment}
\usepackage{ltablex}
\usepackage{ragged2e}
\def \IR{\hbox{{\rm I}\kern-.2em\hbox{{\rm R}}}}

\newcommand{\N}{\mbox{$N$}}

\newcommand{\bmx}{\mbox{\boldmath $x$}}

\newcommand{\btheta}{\mbox{\boldmath $\theta$}}

\usepackage{tensor}

\newcommand{\image}[2]{\includegraphics[#1]{#2}}
\usepackage{placeins}
\usepackage{wrapfig}
\usepackage{lscape}
\usepackage{rotating}

\defcitealias{KDHS2014}{KDHS, 2014}
\defcitealias{KenyaCensus:2009}{Kenya National Bureau Of Statistics, 2014}

\begin{document}
\title{Design- and Model-Based Approaches to Small-Area Estimation in a
Low and Middle Income Country Context: Comparisons and Recommendations}

\author{John Paige, Geir-Arne Fuglstad, Andrea Riebler, Jon Wakefield\thanks{John Paige was supported by The National Science Foundation Graduate Research Fellowship Program under award DGE-1256082, and Jon Wakefield was supported by the National Institutes of Health under award R01CAO95994.}}
\date{}
\maketitle

\begin{abstract}

\noindent
The need for rigorous and timely health and demographic summaries has provided the impetus for an explosion in geographic studies, with a common approach being the production of pixel-level maps, particularly in low and middle income countries. 
In this context, household surveys are a major source of data, usually with a two-stage cluster design with stratification by region and urbanicity. Accurate estimates are of crucial interest for precision public health policy interventions, but many current studies take a cavalier approach to acknowledging the sampling design, while presenting results at a fine geographic scale. 
In this paper we investigate the extent to which accounting for sample design can affect predictions at the aggregate level, which is usually the target of inference. We describe a simulation study in which realistic sampling frames are created for Kenya, based on population and demographic information,  with a survey design that mimics a Demographic Health Survey (DHS). We compare the predictive performance of various commonly-used models. We also describe a cluster level model with a discrete spatial smoothing prior that has not been previously used, but provides reliable inference. We find that including stratification and cluster level random effects can improve predictive performance. Spatially smoothed direct (weighted) estimates were robust to priors and survey design. Continuous spatial models performed well in the presence of fine scale variation; however, these models require the most ``hand holding". Subsequently, we examine how the models perform on real data; specifically we model the prevalence of secondary education for women aged 20--29 using data from the 2014 Kenya DHS. 

\end{abstract}


\section{Introduction}
\renewcommand{\arraystretch}{.6} % not sure why this command is necessary to shrink tables

Complex, multi-stage household surveys play an important role in producing a variety of estimates of health and demographic quantities of interest, especially in  low and middle income countries (LMICs). Examples of such surveys include Demographic Health Surveys (DHS) \citep{dhs}, Multiple Indicator Cluster Survey (MICS) \citep{MICS}, AIDS Indicator Surveys (AIS) \citep{AIS}, and Living Standard Measurement Surveys (LSMSs) \citep{LSMS}. The lack of high quality vital registration (VR) data often necessitates the use of these household surveys in LMICs \citep{li:etal:19,wagner:etal:18}. For instance, it has been estimated that only 4\% of neonatal deaths (deaths in the first 28 days of life) are recorded via high quality VR data \citep{lawn:etal:14}, while in 2012, 35\% of births remained unregistered within a year, and 89\% of these occurred in South Asia and Sub-Saharan Africa \citep{lawn:etal:14}. In general, VR data is more sparse and of lower quality in LMICs than in high income countries, making household surveys especially useful in these contexts.

The Sustainable Development Goals (SDGs) specify targets for a variety of health outcomes \citep{sdgsWeb}. Household surveys are used to estimate these indices and attainment of the SDGs can then be assessed. In particular, SDG 3 calls for an end to preventable deaths of newborns and children under 5 years of age and states that all countries should aim to reduce neonatal mortality to below 12 deaths per 1,000 live births. Additionally, SDG 4 calls for improved education for all, and for all people to complete their secondary education with, in particular, the elimination of inequalities in education due to gender or location. As important as it is to estimate relevant indicators at the country level, the SDGs specifically call for estimates at finer spatial scales. Hence, developing statistical models that can accurately account for the sampling design, while producing estimates at subnational scales, is of great importance.

Estimates of demographic indicators can also be used to highlight areas in need of intervention and  to examine associations between relevant covariates and health outcomes. The Equitable Impact Sensitive Tool (EQUIST) \citep{equist:19}, for instance, is designed to inform decision-makers using a variety of data and model output visualizations. Policymakers can use such tools, as well as other forms of model output and analysis to, for example, create vaccination initiatives targeting areas with a high disease burden, identify possible factors influencing disease prevalence and mortality risk, and erect community-based care programs in order to improve quality of care while increasing coverage to those that need it most while reducing cost. Some of these applications, including community-based care programs, are discussed in detail in Chapters 14 and 15 of \cite{black:etal:16}.

In spite of the ubiquity of multi-stage household surveys and their importance in estimating health indices and planning interventions, classical techniques for analyzing survey data that can account for the survey design often have difficulty producing estimates at the required spatial resolutions (interventions are often made at the  Admin2 level). For example, weighted (direct) estimates, often have large associated variances at the Admin2 level, due to data sparsity. To improve estimation in such situations, a number of small area estimation (SAE) methods have been proposed \citep{rao:molina:15} including those that extend upon the seminal Fay-Herriot model \citep{fay:herriot:79}. These methods  not only acknowledge the survey design, but also ``borrow strength'' from data in nearby areas to produce reliable estimates with smaller uncertainty intervals \citep{marhuenda:etal:13,chen:etal:14,mercer:etal:15,congdon:lloyd:10,you:zhou:11,porter:etal:14,vandendijck:etal:16,watjou:etal:17}. These approaches are all based on discrete spatial models, which are based on arbitrary neighborhood structures, which may be more or less realistic, depending on the context and geography.

A number of papers have used continuous spatial models to analyze health and demographic outcomes using survey data \citep{wardrop:etal:18,gething:etal:16,golding:etal:17,utazi:etal:18,gething:etal:15,osgood:etal:18,graetz:etal:18,diggle:giorgi:16,giorgi:etal:18,diggle2019model}. Included in this list are publications from WorldPop and the Institute for Health Metrics and Evaluation (IHME), both of which are large-scale producers of health and demographic maps. In these references, all of the models ignore the stratification; in addition WorldPop  routinely ignore the clustering also. In general, ignoring the design results in biased estimates and inaccurate uncertainty intervals. 
%This is similarly true for discrete spatial statistical models that do not directly acknowledge the design such as the Besag-York-Molli\'{e} (BYM) model introduced in \citet{besag:york:mollie:91} and its variations, some of which are reviewed in \citet{riebler:etal:16}. 
%It is not currently known how to account for design weights using these models except through the methods referenced previously based on the Fey-Harriot model  The 2012 DHS Sampling and Household Listing Manual states that failing to use the correct sampling weights can lead to invalid statistical inference, biased estimates, and more generally a set of estimates that are not representative of the population in question. 
No study has been conducted to explore the effects of ignoring design stratification and cluster level overdispersion in the LMIC context, and here we aim to fill this gap in the literature, by comparing a variety of spatial modeling approaches, under different levels of stratification and clustering. 

%We will discuss how to aggregate cluster level predictions to get area level estimates, this is not always straightforward since information on unsampled clusters within the areas is not available. This will be discussed in detail in Section \ref{sec:models}.

%Regardless of the class of model used, accounting for within-cluster dependence is especially important in the context of multi-stage household surveys with cluster sampling \citep{fitzmaurice1997detecting,zhang:etal:10}.
% If the overdispersion occurs at the cluster level, then classical survey sampling techniques can account for this trivially. For spatial statistical models that do not directly account for the design it is usually simple to add in cluster level random effects to induce overdispersion due to correlation within clusters, but less clear is how to aggregate cluster level predictions to get area level estimates, since information on unsampled clusters within the areas is not available. This will be discussed in detail in Section \ref{sec:models}.
%

In this paper we explore the performance of different design- and model-based  methods, when applied to simulated data. We will also apply these methods to analyze two outcomes recorded in the 2014 Kenya DHS, the proportion of women between the ages of 20 and 29 who have secondary education, and the neonatal mortality rate (NMR).
 Section \ref{sec:data} describes the data we will use in this analysis and Section \ref{sec:models} introduces the models that we compare and apply. Section \ref{sec:simulation} describes the simulation study and in Section \ref{sec:application} we apply the models to the secondary education  outcome, reporting predictions, uncertainties, and using cross validation to assess the out of sample performance of each of the models. We discuss the results as well as our conclusions in Section \ref{sec:discussion}. Appendices A and B give details on the modeling and the simulation study respectively, and Appendix C gives additional results related to the secondary education example.%Section \ref{sec:application}.
  Lastly, Appendix D describes an application of the models to NMRs in Kenya from 2010--2014.

\section{Data} \label{sec:data}

The DHS Program uses a set of consistent sampling approaches from country to country, with methods described in the 2012 DHS Sampling and Household Listing Manual \citep[Sec. 5.2, p.~80--85]{samplingManualDHS}. This standard design is a stratified two-stage cluster sampling scheme with stratification by county crossed with urban/rural. The first sampling stage involves selecting enumeration areas (EAs) using probability proportional to size (PPS) sampling, where the probability of sampling each EA is proportional to the listed number of households in that EA, and the second stage involves simple random sampling of (typically) 25 households within each EA. Mothers within the household are then asked a number of questions about their children, and, if any died, the mothers are asked about those children's deaths. The 2014 Kenya DHS \citepalias{KDHS2014} follows the typical DHS scheme,  though is powered to the county level %(there are 47 counties in Kenya) 
so that 1,612 clusters are sampled out of the 96,251 total EAs that were in the 2009 Kenya Population and Housing Census \citepalias{KenyaCensus:2009}. Of these clusters, 995 are urban while 617 are rural, with urban areas oversampled in the majority of the 47 counties. Mombasa and Nairobi are entirely urban and the remaining 45 counties have both urban and rural areas, so that there are 92 strata in total.

%
%
%In order to maintain confidentiality, the geographical locations of the clusters are displaced: urban clusters by up to 2km, rural clusters by up to 5km, with a further 1\% of these clusters by up to 10km. Although we do not jitter the spatial coordinates of the clusters in the simulation study, we do not believe that the effect of jittering on the predictions would be large, since the true spatial range was fixed at 150km. Additionally, in the two applications considered based on true DHS data, the spatial range parameter estimates were both over 200km, suggesting that the relatively small amounts of jittering in the DHS data would have little effect on the results, especially considering the fact that the stratum of the clusters is known rather than being estimated from each cluster's spatial location.

%
%We use the 2009 Kenya Population and Housing Census \citepalias{KenyaCensus:2009} as well as the 2014 Kenya DHS to estimate empirical distributions of the number of households per EA, the number of mothers per household, and the number of children per mother. These distributions were used to simulate a realistic true population for the simulation study describe in Section \ref{sec:simulation}.

\begin{figure}
\centering
\image{width=3.16in}{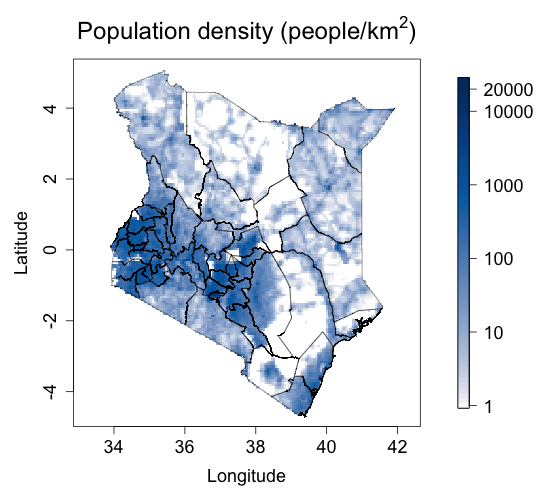} \image{width=2.89in}{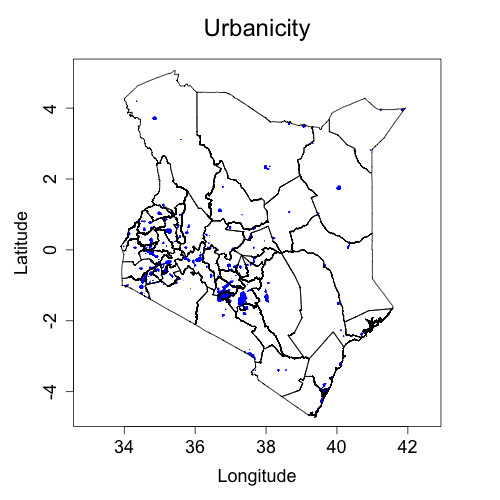}
\caption{Left: WorldPop based population density estimates. Right: urban areas in Kenya used in this analysis are depicted in blue. Areas are determined to be urban versus rural based on thresholding population density.}
\label{fig:urbanMap}
\end{figure}

In order to be able to generate spatial maps of urbanicity, we use 1km $ \times $ 1km gridded population density surfaces from WorldPop \citep{stevens2015disaggregating,tatem2017worldpop} as plotted in the left panel of Figure \ref{fig:urbanMap}. The 2010 and 2015 population density maps are interpolated assuming a constant rate of population growth to produce the 2014 population density map used throughout this paper. 
The 2009 Kenya Population and Housing Census provides information on the proportion of the population within each county that is urban and rural, and we generate urbanicity maps by thresholding the population density maps within each county at the level required to achieve this proportion. This results in the urbanicity map given in the right panel of Figure \ref{fig:urbanMap}.

\section{Methods} \label{sec:models}

\subsection{Models}
We first describe notation in the scenario in which we wish to estimate NMRs, so that the denominators are the number of children that were born in the relevant period, and the response is whether a death occurred in the first month after birth. Let $Y_{ck}=0/1$ represent the binary response for child $k$ in cluster $c$ with the total number of deaths in each cluster being $Y_c= \sum_{k \in \mathcal{B}_c} Y_{ck}$ where $\mathcal{B}_c$ is the set of indices of the children in cluster $c$ that are sampled.  
%Also let $Y_{i} = \sum_{c \in \mathcal{C}_i} Y_c$ be the total number of deaths in county $i$, where $\mathcal{C}_i$ is the set of clusters that are selected. 
We let $\bmx_c$ be the spatial location of cluster $c$. 
Associated with location $\bmx_c$ is a county, which will be denoted $i[c]$, and the set of spatial locations that are urban is denoted $U$. 
We now describe the different models considered; in general, we focus on inference at the county level, since this is often the target of inference.

%\begin{comment}
%We will denote the 2014 Kenya DHS weights $w_{ck}$, which are the product of the inverse sampling probability and an adjustment for non-response. Specifically,
%\begin{eqnarray*}
%w_{ck} = w_{ck}^{\mbox{\tiny{DES}}} \times w_{ck}^{\mbox{\tiny{NON}}}  =  \pi_{ck}^{-1} \times r_{ck}^{-1},
%\end{eqnarray*}
%where $w_{ck}^{\mbox{\tiny{DES}}} $ is the design weight equal to the inverse of the probability the child was included in the sample, $\pi_{ck}$, and $w_{ck}^{\mbox{\tiny{NON}}} $ is the nonresponse weight equal to the inverse of the response rate $r_{ck}$.
%\end{comment}

\medskip
\noindent
{\bf Naive:} Ignoring the sampling design, we fit a binomial model to the county-level data, without accounting for the sampling design. 
%See \citet[p. 49--53]{wakefield:13:BOOK} for a description of quasi-likelihood, including the quasi-binomial  family. 
In this case, we assume the probability of mortality for child $k$ in cluster $c$ is the same for all children in county $i$,
$$\log \left( \frac{p_{ck}}{1-p_{ck}} \right) = \beta_{i[c]}, $$
where the models are fit independently to the data from each county. The targets
of inference are the county-level probabilities $\text{expit}(\beta_i)$, $i=1,\dots,47$.

\medskip
\noindent
{\bf Direct:} County-level direct estimates,  $\widehat p^{~\mbox{\tiny{DIR}}} _i$, are calculated using a weighted estimator that account for the survey design. The weights are proportional to the inverse of the probability of sampling each child.  This estimator is reliable for large samples, but for small samples, will have high variance \citep{rao:molina:15}. Weighted estimators can yield estimates that lie on the boundary    and variance instability  in small samples. These problems mean that yearly estimates at the Admin2 level are typically not reliable when based on a DHS with around 400 clusters (a typical design).

\medskip
\noindent
{\bf Smoothed Direct:} Following the approach of \cite{mercer:etal:15} we calculate $Z_i~=~\mbox{logit}(\widehat p^{~\mbox{\tiny{DIR}}} _i )$ along with its associated (design-based) variance estimate $\widehat{V}_i$. We assume $Z_i | \eta_i~\sim_{ind}~\N(\eta_i, \widehat{V}_i)$ with linear predictor, 
\begin{align*}
\eta_i &=  \beta_0 + \frac{1}{\sqrt{ \tau }}(\sqrt{ \phi }S_i + \sqrt{1 -  \phi }\delta_i) ,%&& i=1,\dots,47,
\end{align*}
where $S_1, \dots, S_{47}$ and $\delta_1, \dots,  \delta_{47}$ are respectively mean zero county level intrinsic conditional autoregressive (ICAR) terms and independent and identically distributed (iid) Gaussian random effects. The ICAR model, described in \citet{besag:etal:91}, is a discrete spatial model and assumes the effect in each area arises from a normal whose mean is the  average of the effects in neighboring areas. We apply a sum-to-zero constraint $\sum_{i=1}^{47} S_i =0$ to the ICAR terms to make the intercept identifiable.
%, and scale the ICAR effects so that the geometric mean of the marginal variances of its components is equal to 1. This allows $ \tau $ to be interpreted as the marginal variance of the sum of the pair of random effects in  \citep{riebler:etal:16}. 
The parameterization adopted is a variation of the model introduced in \citet{simpson:etal:17} and named the BYM2 model in \citet{riebler:etal:16}, since it is a reparameterization of the model originally introduced by \cite{besag:etal:91}.  The total precision of the county level components of the model is $ \tau$ and $ \phi $ represents the proportion of the total variance that is spatial. %Since this is a two stage (Fay-Herriot type) model that involves smoothing the direct estimates, we call this the smoothed direct model.
 Under this approach the posterior distribution is obtained for the county level probabilities: 
$ %\begin{align*}
p^{~\mbox{\tiny{SM-DIR}}} _i =\mbox{expit} (  \eta_i)
%{1+\exp ( \eta_i )}
,$, $i=1,\dots,47.$
%\end{align*}
This model produces a design consistent estimate of the NMR, since if the entire population is sampled, the direct estimate has an associated variance estimate  $\widehat{V}_i=0$ for $i=1,\dots,47$, and so is immovable with respect to the prior. The space-time version of this model has been used in an extensive study of under 5 mortality in 35 countries in Africa over the period 1980--2015 \citep{li:etal:19}. This model can alleviate some of the high variance problems of the direct estimates, but still struggles with boundary estimates and undefined variances.

\medskip
\noindent
{\bf Model-based approaches:}
For the model-based spatial approaches, we assume that
$Y_c | p(\bm{x}_c) \sim \mbox{Bin}(n_c,p(\bm{x})),$
where $n_{c}$ is the total number of children sampled  in cluster $c$. 
The underlying risk at location $\bm{x}_c$ for cluster $c$ is modeled as
% while $n_{i} = \sum_{c \in \mathcal{C}_i} n_c$ is the total number of children surveyed in county $i$. 
\begin{equation}
	\log \left( \frac{p(\bm{x}_c)}{1-p(\bm{x}_c)} \right) = \beta_0 + u(\boldsymbol{x}_c) + \beta^{\mbox{\tiny{URB}}} I(\bm{x}_c \in U) + \epsilon_c,
	\label{eq:modelB:main}
\end{equation}
where $\beta_0$ is the intercept, $u(\boldsymbol{x}_c)$ is a spatial random effect, $\beta^{\mbox{\tiny{URB}}}$ is the association with the cluster being urban (as compared to rural), and $\epsilon_c$ is an iid Gaussian cluster random effect with variance $ \sigma_\epsilon^2$. This term is sometimes described  as the ``nugget" and is often taken to reflect the combination of unmodeled sampling variability and small-scale variation. 

The first model-based approach is termed BYM2 and uses the spatial random effect
\(
	u(\bm{x}) = \frac{1}{\sqrt{ \tau }}(\sqrt{ \phi }S_{i[\bm{x}]} + \sqrt{1 -  \phi }\delta_{i[\bm{x}]}),
\)
where we use $i[\bm{x}]$ to denote the county which which $\bm{x}$ belongs, and the structure of the
model follows the description for the smoothed direct model. This binomial model naturally deals with 0 or $n_c$ outcomes.

The second model-based approach is termed SPDE and uses a Gaussian process (GP) for
the spatial random effect, 
$u( \cdot ) \sim \mbox{GP}(0,\btheta)$ with
 $\btheta =[\sigma_s^2, \rho ]$. The marginal variance is  $\sigma_s^2$ and the spatial range 
 at which the correlation is approximately 0.1 is $ \rho $. Note that the GP we use is the 
 solution to a stochastic partial differential equation (SPDE) which is approximated by 
 a particular Gaussian Markov Random Field (GMRF) defined on a fine triangular mesh \citep{lindgren:etal:11}.

For both BYM and SPDE, we consider four variations of  \eqref{eq:modelB:main}: including or not including the association with urban, and including or not including the cluster (nugget) effect. 
Models without and with urban effects are labeled as `u' and `U', respectively. Similarly, models without and with cluster effects are labeled as `c' and `C', respectively. Table
\ref{tab:BYM2variations} summarizes the eight alternatives for BYM2 and
SPDE.

\begin{table}[ht]
\centering
\begin{tabular}{l | l}
Models & Linear predictor extra effects  \\ 
\\[-.3em]
  \hline \\[-.4em]
  BYM2\textsubscript{uc}/SPDE\textsubscript{uc} &-- \\\\[-.4em]
  BYM2\textsubscript{uC}/SPDE\textsubscript{uC} &  $\epsilon_c$   \\ \\[-.4em]
  BYM2\textsubscript{Uc}/SPDE\textsubscript{Uc} & $ \beta^{\mbox{\tiny{URB}}} I(\bmx_c \in U) $   \\ \\[-.4em]
  BYM2\textsubscript{UC}/SPDE\textsubscript{UC} &  $ \beta^{\mbox{\tiny{URB}}} I(\bmx_c \in U) + \epsilon_c$ 
  \end{tabular}
  \caption{Variations of model-based approaches used in the analysis. The subscript symbols `U' and `u' indicate whether or not an urban effect is present in the linear predictor, while `C' and `c' indicate whether or not an iid cluster effect is included. All models include intercept and the spatial effect.}
  \label{tab:BYM2variations}
\end{table}

%For the discrete (BYM2) model aggregation to the same level of the spatial model is straightforward, since the surface is constant albeit with a shift if there is an urban/rural association included. In the later case, the aggregation is the weighted combination of the  respective urban and rural prevalences, where the weights are the proportion urban/rural.
For the continuous (SPDE) model, if we knew the complete list of EA locations in the sampling frame, we could perform predictions
for the county level  by using the posterior distribution of a weighted sum
over the EA locations. In the absence of such a list, we can describe the probability
surface at unobserved locations by $p(\bm{x})$ via  \eqref{eq:modelB:main}, and 
%there is no obvious way to incorporate the cluster effects. We therefore leave out cluster effects in the aggregated county level predictive distributions. 
aggregate by continuously integrating the spatial probability surface with respect to population density. Let $p_i$ denote the county level estimates for county $i$, then
\begin{equation}
p_i = \int_{A_i} p(\bmx) \times q(\bmx) ~ d\bmx \approx \sum_{j=1}^{m_i} p(\bmx_j) \times q(\bmx_j),
\label{eq:popDensityIntegral}
\end{equation}
where $A_i$ is the geographical extent of area $i$, $q(\bmx)$ is the population density at location $\bmx$, and $m_i$ is the number of grid cells with centroids in area $i$ that is used to approximate the continuous integral. 

For the BYM2 model, a continuous population density surface is not needed since the probabilities are modeled as constant within each area, and we can use
\begin{eqnarray*}
 p_i &=& \mbox{expit}\left( \beta_0+\frac{1}{\sqrt{ \tau }}(\sqrt{ \phi }S_{i} + \sqrt{1 -  \phi }\delta_{i}\right)  q_i \\&+ &
  \mbox{expit}\left(\beta_0+\beta^{\mbox{\tiny{URB}}} +\frac{1}{\sqrt{ \tau }}(\sqrt{ \phi }S_{i} + \sqrt{1 -  \phi }\delta_{i}\right) (1-q_i),
 \end{eqnarray*}
where $q_i$ is the proportion of the target population in county $i$ that is rural. Further details of accounting for the cluster effects and performing the spatial aggregation may be found  in Appendix A.

Both Worldpop and IHME use a continuous GP model in the context of the analysis of DHS (and other) household survey data, without adjustment for the stratification. IHME include a cluster effect in their model, and for aggregation add a nugget contribution at the pixel-level, while Worldpop do not include a nugget.

\subsection{Inference}

Penalized complexity (PC) priors were introduced in \citet{simpson:etal:17}, and penalize a model's ``distance", on an appropriate scale, from a simple ``base" model. For example, for iid random effects arising from a zero mean Gaussian distribution with variance $\sigma^2$, the base model corresponds to $\sigma=0$. Following \cite{fuglstad:etal:19a}, we set a joint PC prior on the continuous spatial standard deviation and effective range parameters $ \sigma_s$ and $  \rho  $,  respectively.  We use the joint PC prior described by \citet{riebler:etal:16} on the BYM2 standard deviation $1/\sqrt{\tau}$ and the proportion of variation that is spatial, $\phi$.  We also set a marginal PC prior  on the cluster effect standard deviation parameter $ \sigma_\epsilon$.  The priors in the simulation study and in the application are set so that the median of the prior for $\rho$ is at roughly one fifth the diameter of the spatial domain, and so that $P( \sigma_s>1)=P(1/\sqrt{ \tau }>1)=P( \sigma_\epsilon>1)=0.01$. This implies that the continuous spatial effects, BYM2 effects, and cluster effects for the spatial smoothing models each have a  95\% prior chance of lying between  0.5 and 2 on an odds scale. The  PC prior for the spatial proportion in the BYM2 model, $\phi$, is given a $2 / 3$ prior probability of being less than $1/2$, implying that we slightly favor the iid county level effects when apportioning residual variation. We choose this prior on $ \phi $ in order to promote less complex models with a smaller  spatial contribution.

 All design-based estimates were obtained using the {\tt svyglm} function within the {\tt survey} package \citep{lumley:04,Lumley:2018aa} in the {\tt R} programming language.
Each of the spatial models can be fitted using the integrated nested Laplace approximation (INLA) approach introduced in \cite{rue:etal:09}, a method for fitting Bayesian models without the computational difficulties of Markov Chain Monte Carlo (MCMC) and implemented in the {\tt INLA} package in {\tt R}.  The direct, smoothed direct and binomial BYM2 models are available in the SUMMER package \citep{martin:etal:18}. Code to reproduce the results can be found at \url{https://github.com/paigejo/NMRmanuscript}, and the 2014 Kenya DHS data can be requested from \url{https://dhsprogram.com/}.

%This approach to computation uses a clever combination of numerical integration and Laplace approximations, exploiting the sparse matrices that encode the conditional independencies in various precision matrices \citep{rue:knorrheld:05}.

\section{Simulation Study} \label{sec:simulation}

\subsection{Comparison Measures}

In this section, we describe an extensive simulation study in which we compare various models, in particular with respect to the inclusion of strata and cluster effects. We do this for multiple simulated populations and survey designs in order to test the models under a variety of circumstances. As in Section \ref{sec:models}, the nominal response is a binary indicator of whether or not death occurred within the first month of life.

We evaluate the model predictions at the county level using bias, mean squared error (MSE), the continuous rank probability score (CRPS), coverage of 80\%  intervals, and width of 80\%  intervals. All scoring rules are calculated on the probability scale. Note that CRPS is a strictly proper scoring rule \citep{gneiting:raftery:07} that accounts for both predictive accuracy as well as accuracy of the uncertainty of the predictive distribution. Given the cumulative distribution function of the predictive distribution of a proportion in the finite population, $F$, and an empirical proportion response $y / n$, the CRPS is defined as:
\begin{align*}
\mbox{CRPS}(F, y) 
%&\equiv \int_{-\infty }^\infty  (F(\tilde{y} / n) - 1\{\tilde{y}\geq y\})^2 \ d\tilde{y}  \\
&= \sum_{\tilde{y} = 0}^n (F(\tilde{y} / n) - 1\{\tilde{y}\geq y\})^2. 
\end{align*}
%For each predictive distribution $F$ that is  binomial (conditional on the fixed probability $p$), we obtain an estimate, $\hat{F}$, of $F$ by numerically integrating $F$ over the posterior distribution of $p$.
Small values of CRPS are desirable. 

%We estimate each scoring rule $S(F, y)$ with $S(\hat{F}, y)$. 
The reported scoring rules are calculated using predictive distributions that have been calculated at the county level. The reported scores are the averages over counties and repeated surveys, and the full set of calculated scoring rules are given in Appendix B.3.
%
%{\color{red} Johnny: I dropped some details which I think only adds confusion. You should look
%through and see if you agree.}
\subsection{Simulation Setup}

In order to generate a true, underlying population from which we can draw surveys, we first spatially partition Kenya into urban and rural zones by thresholding population density so that the proportion of population in each county that is urban matches the 2009 Kenya Population and Housing Census. We then simulate all 96,251 census EA locations such that the number in each of the 92 strata matches the true number, as given in the 2009 census. The EA locations in our simulated population are drawn proportional to population density within each strata. This information is all available in the Kenya DHS final report \citepalias{KDHS2014}.

The number of households in each EA, as well as the number of mothers per household and children born per mother per year, are simulated based on the corresponding empirical distributions in the true population stratified by urban/rural. In order to estimate the empirical distribution for the number of households per EA, we take the maximum household ID sampled per cluster in the 2014 Kenya DHS as an estimate for the number of households in each EA. For empirical distributions of the number of children born per mother per year we only include children living in the same house as their mother, since children living in a different house than their mother only account for about 5\% of all children according to available census data.

\begin{figure}
\centering
\includegraphics[width=6.5in]{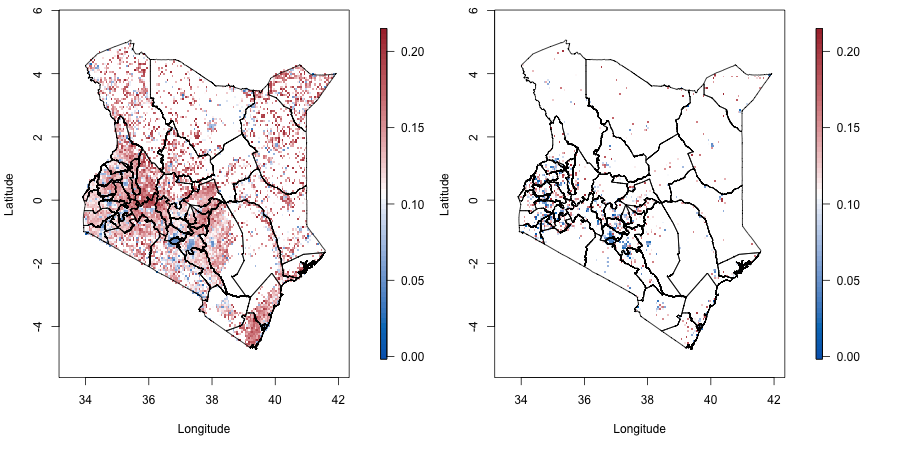}
\caption{Simulated population of Kenya and associated NMRs (left), and an example simulated dataset based on the ``Stratified" design (right).}
\label{simulation}
\end{figure}

We compared the models in eight distinct scenarios. We simulate four different complete populations from the models (with associated names): constant risk (Pop\textsubscript{suc}), spatially-varying risk (Pop\textsubscript{Suc}), spatially-varying risk with an urban association (Pop\textsubscript{SUc}), and spatially-varying risk with an urban association and a cluster effect (Pop\textsubscript{SUC}).  Note that in the subscript labels for the populations, we again use U/u and C/c to respectively indicate the presence of urban and cluster effects, and we additionally use S/s to indicate presence of a continuous spatial effect. In the case where we include spatial, urban, and cluster effects, we simulate NMRs at all 96,251 spatial locations using the SPDE model described above with an effective correlation range of 150km, and with parameters $\beta_0=-1.75$, $\sigma_s^2 = 0.15^2$, $\sigma^2_\epsilon= 0.1^2$, and $\beta^{\mbox{\tiny{URB}}}=-1$. For ``typical" rural/urban  areas, with random effects of zero, the prevalences are 17\%/6\%. Within each of these four scenarios we carry out  ``Unstratified" and ``Stratified" sampling, always taking 1,612 clusters to match the 2014 Kenya DHS. In the Unstratified design, we fix the total number of clusters in each county to be the same as in the Kenya DHS, and choose the proportion of urban and rural cluster within each county to match the proportion of the urban and rural population in that county. In the Stratified design, we sample urban and rural clusters at different rates for each county so as to match the proportion of urban clusters in each county of the 2014 Kenya DHS, obstaining 995 and 617 urban and rural clusters respectively. Conditional on the total number of urban and rural clusters for each of the 92 strata, we use PPS sampling to determine which clusters are included in the surveys, sampling clusters with probability proportional to the number of households in each strata. Within each EA, 25 households are chosen at random to be included in the cluster sample. The simulated population and a single simulated survey based on the Stratified design are shown in Figure \ref{simulation}.

% design in which the number of clusters sampled in each county matches the numbers 
%
%There are four different fixed populations for which we simulate NMRs. For each of the eight scenarios, 100 surveys were simulated.
%
%Given the fixed population, we simulate 100 surveys mimicking the design of the 2014 Kenya DHS and 100 ``representative'' surveys, where urban and rural areas are sampled proportionately. 

%In the representative case, instead of fixing the number of urban and rural clusters to be the same as in the 2014 Kenya DHS,. We call this the representative case even though it is technically cluster sampling as a simple way to emphasize that, unlike for the DHS design, the urban and rural areas are sampled on average in a representative fashion.

\subsection{Simulation Results}

The scoring rules summarizing the main results for the stratified design are plotted in Figure \ref{fig:scoringRules}, and parameter summary statistics are given in Appendix B.3. Scoring rules for additional model variations and designs are compared in Appendix B.2.
%Figures \ref{fig:scoringRules1Full} and \ref{fig:scoringRules2Full}. 
When interpreting these scoring rules, it is important to keep in mind that SDG 3 calls for a reduction in NMRs to 12 deaths per 1,000 children, which corresponds to 120 $\times 10^{-4}$ children. When absolute bias is large relative to this number, it is an indicator of poor model performance. Since we are especially interested in the performance of the models in a not atypical scenario in which spatial, urban and cluster effects must be accounted for, we will be discussing Pop\textsubscript{SUC} under a Stratified design unless we state otherwise.

Of the direct, smoothed direct, BYM2\textsubscript{UC}, and SPDE\textsubscript{UC} models, the BYM2\textsubscript{UC} model performed the best in terms of CRPS, MSE, coverage, and CI width.  Although the BYM2\textsubscript{UC} model was slightly positively biased, the precision of its central predictions and the well-calibrated predictive distribution and uncertainty intervals led to accurate coverages and good predictive performance. 

Interestingly, although the SPDE\textsubscript{UC} model matched the model used to simulate the data, it did not perform well in terms of MSE. Its MSE was $1.24\times10^{-4}$ compared to $0.41\times10^{-4}$ for the BYM2\textsubscript{UC} model, $0.63\times10^{-4}$ for the smoothed direct, and $0.72\times10^{-4}$ for the direct model. Although SPDE\textsubscript{UC} model estimates were somewhat positively biased, the high level of MSE was mainly due to 
%between county variability in the residuals, as evidenced by the 
high variances (see results in Appendix B.3.2).
%, as seen in Table \ref{tab:scoresStratifiedSUC}.
 In spite of this, the SPDE\textsubscript{UC} model had a CRPS of $4.7\times 10^{-3}$, which was comparable to the value of $4.6 \times 10^{-3}$ for the smoothed direct model, and was better than the direct model value of $4.9 \times 10^{-3}$. Additionally, the coverage of the SPDE\textsubscript{UC} model was 82\%, second in accuracy only to the BYM2\textsubscript{UC} model. Hence, although the central predictions of the SPDE model were somewhat variable, the uncertainty of the predictive distribution was well-calibrated.

The smoothed direct and BYM2\textsubscript{Uc} models had the smallest magnitude of bias. An advantage of the smoothed direct model is that, regardless of the population and survey scheme considered, the model performed well from the standpoints of MSE, CRPS, bias, and coverage. Although the coverage for the Pop\textsubscript{suc} (constant risk) population was over 90\%, so were the coverages of the BYM2 models. Additionally, the constant risk setting is not realistic, and therefore should not be focused on too much.  Overall, the smoothed direct model was robust in terms of its predictive accuracy and uncertainty.

The tables in Appendix B.3 show that the spatial smoothing models were better at estimating the urban effect than the intercept, although inclusion of an urban effect improved intercept estimation. For instance, for the SUC population under the Stratified design, the BYM2\textsubscript{UC} model average intercept estimate was -1.67, whereas the average BYM2\textsubscript{uC} estimate was -1.99. The equivalent values for the SPDE\textsubscript{UC} and SPDE\textsubscript{uC} models were respectively -1.69 and -1.86. Although the models including urban effects were generally closer to the truth than models without urban effects, there still was some bias in the parameter estimates. The urban effect tended to be better estimated, however. The average estimate for the BYM2 and SPDE `UC' models were both -1.0.

Models that did not account for urbanicity either indirectly via sampling weights or directly as a covariate had relatively poor performance from MSE, bias, CRPS, and coverage standpoints. Even for populations without urban associations or under the unstratified design, there was little downside to including urban effects so long as the proportion of children in urban and rural areas was not poorly estimated (so that the area averaging was poorly performed). Including urban effects led to MSE, bias, CRPS, credible interval width, and coverage that was on average either better or nearly equal to the corresponding models without urban effects throughout all simulated populations and designs. The benefit of including urbanicity as a covariate was increased under the Stratified design relative to the Unstratified design since urban and rural areas were not sampled proportionately in that case.

In general, the inclusion of a cluster effect led to better or equally good predictions, compared to when cluster effects were not included, in terms of MSE and CRPS, as shown in Appendix B.2.
%Figures \ref{fig:scoringRules1Full} and \ref{fig:scoringRules2Full}.  
This is especially true for the SPDE model, whose predictions were more influenced by the inclusion of cluster effects.
% due to the model's flexibility.  
Although the MSE and bias of the BYM2\textsubscript{uc} and BYM2\textsubscript{uC} models were essentially the same, the inclusion of the cluster effect led to a dramatic improvement in coverage from 55\% to 68\%, indicating that cluster effects can lead to more accurate measures of uncertainty.
% in the presence of confounders. 
Although the BYM2\textsubscript{Uc} model arguably performs slightly better than the BYM2\textsubscript{UC} model in terms of its MSE and CRPS, the coverage of the BYM2\textsubscript{UC} model is better, and the uncertainty intervals are more conservative. To summarize, these simulations suggest that, amongst the BYM2 models, 
% Considering the benefit of including cluster effects,
% in the presence confounders, 
 the BYM2\textsubscript{UC} model  is a robust choice for the analysis of 
% most suited of the models tested for handling 
DHS household survey data.

\begin{figure}
\centering
\image{width=3in}{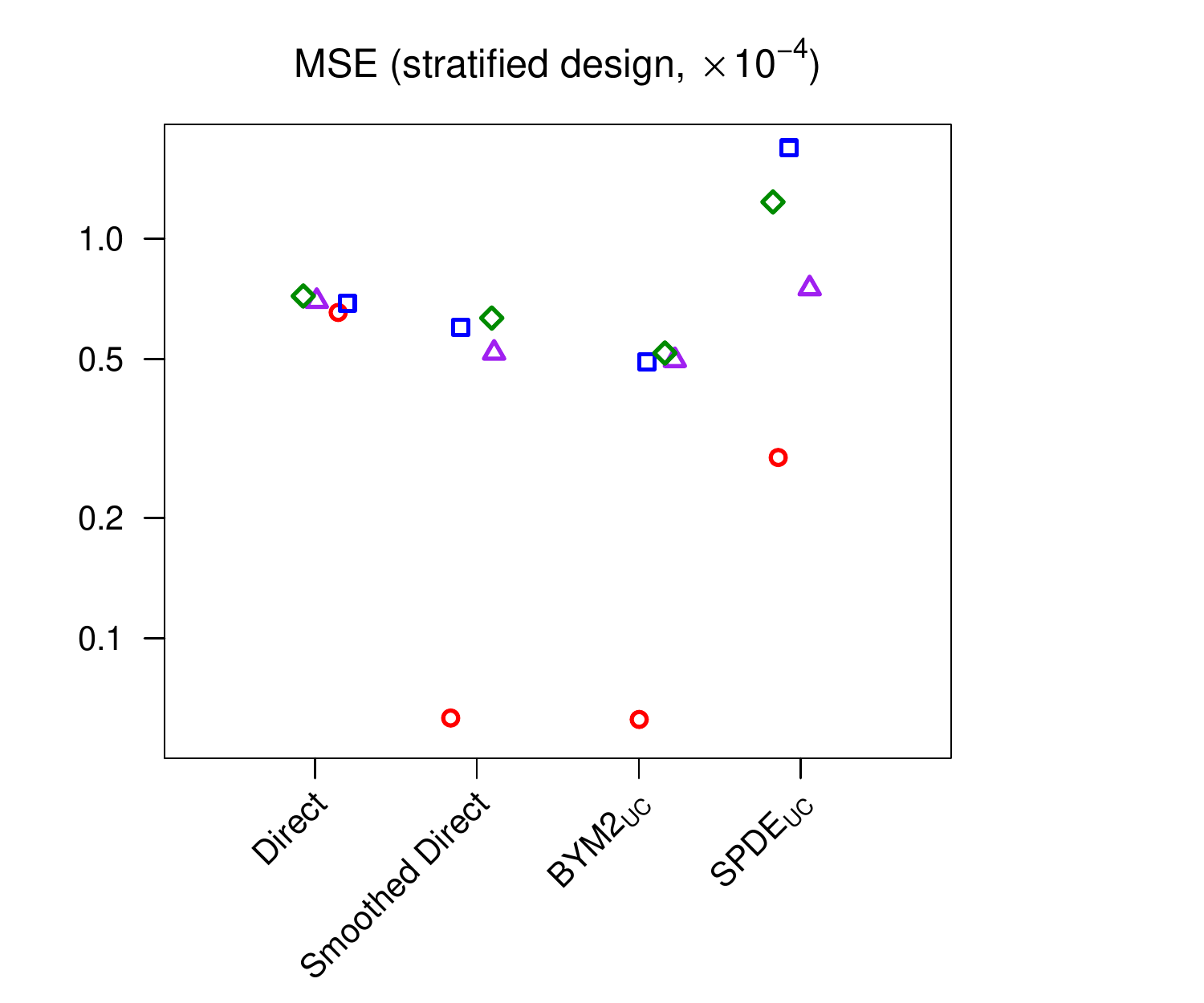} \image{width=3in}{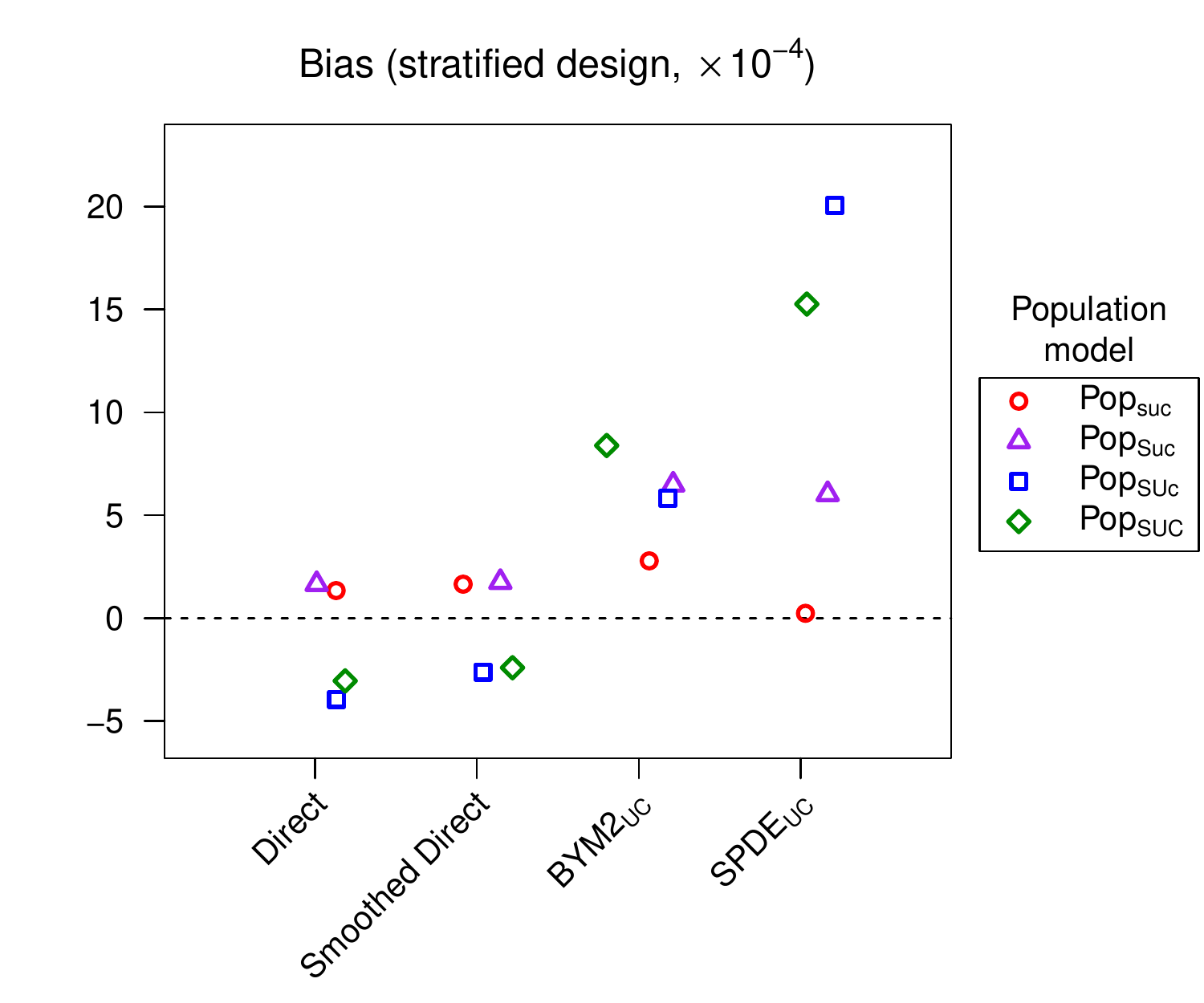} \\
\image{width=3in}{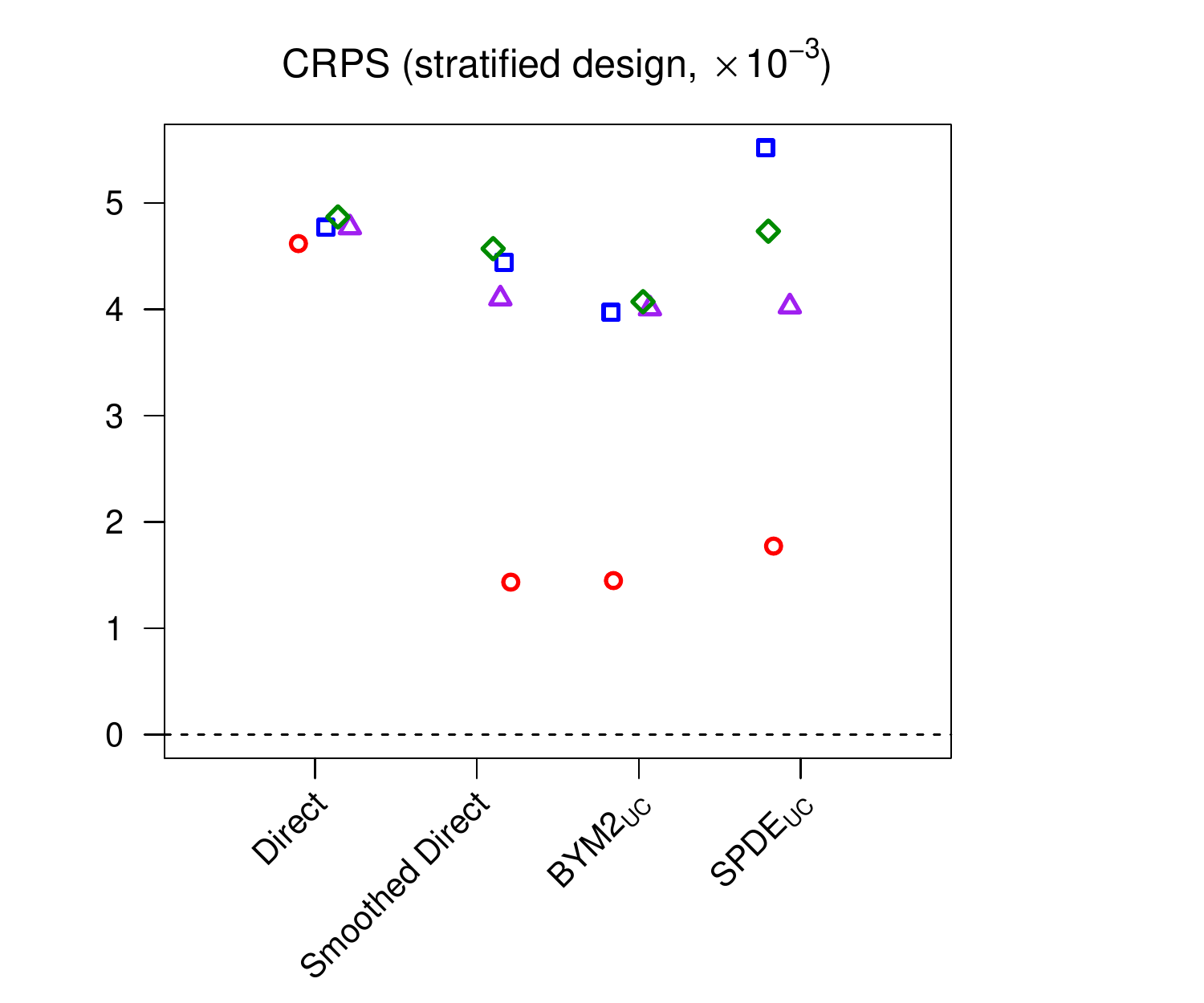} \image{width=3in}{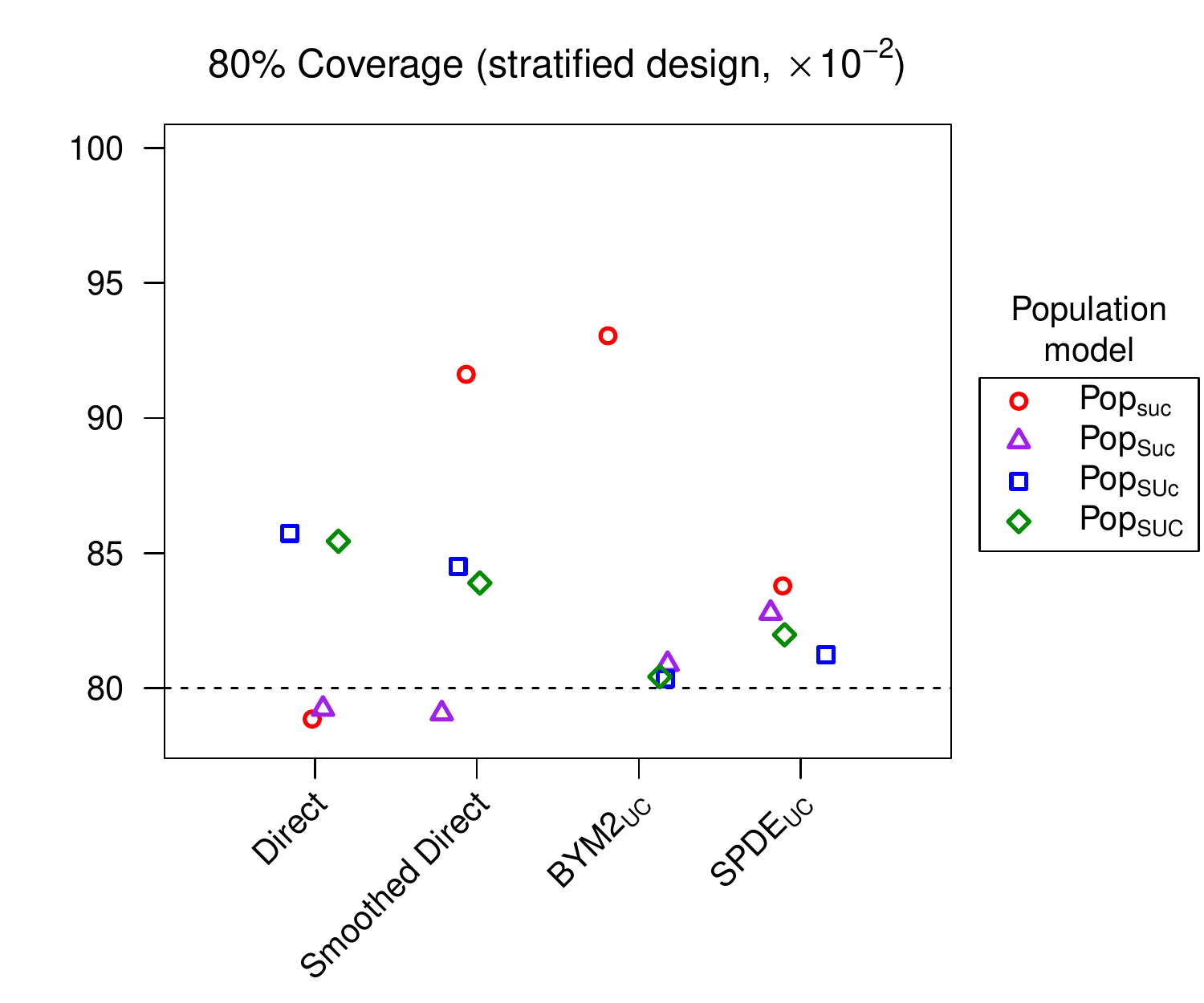} \\
\image{width=3in}{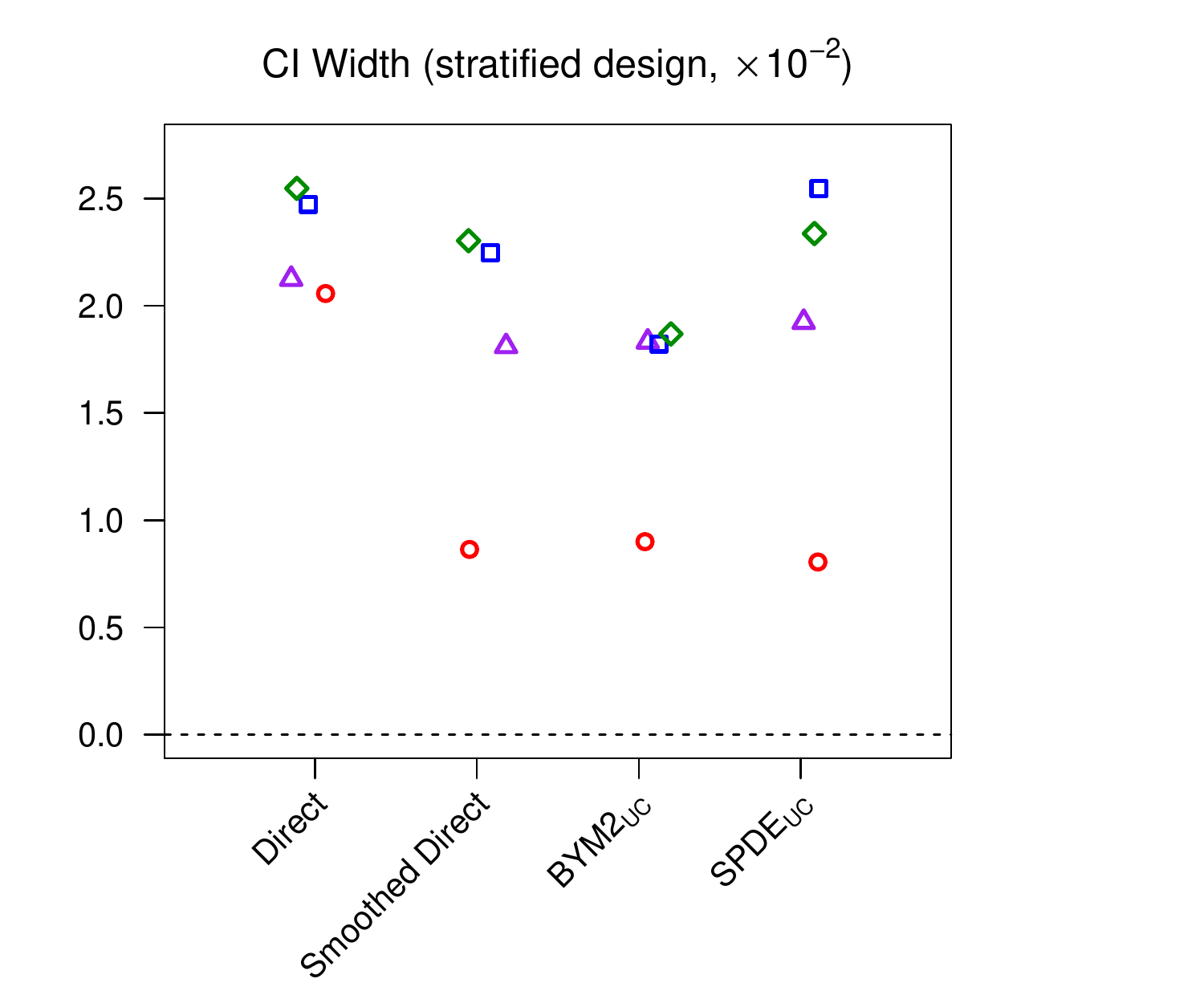} \\
%\image{width=3in}{figures/varPlotSRS.pdf} \image{width=3in}{figures/varPlotDHS.pdf} \\
\caption{County level scoring rules plotted for each of the simulated populations and the main models considered for the Stratified design. The labels s/S, u/U, and c/C denote whether or not spatial, urban, and cluster effects are included in the models respectively. The ``Population model" denotes the method by which the data were generated.}
\label{fig:scoringRules}
\end{figure}

\section{Prevalence of Secondary Education for Women in Kenya} \label{sec:application}

Although some prior results suggest that under-5 mortality rates and other health outcomes are influenced by urbanicity in Sub-Saharan Africa \citep{wakefield:etal:19,ntenda2014factors,antai2010urban,balk:2004,root:1997,defo1994determinants,pezzulo:etal:16}, we did not find strong evidence of a marginal association between NMRs and urbanicity  in Kenya. An analysis of NMRs from the 2014 Kenya DHS is presented in Appendix D. In this section, we focus on secondary education completion rates  for women aged 20--29 in 2014 using the Kenya DHS. This outcome displays a strong association with urbanicity.
%We choose this specific application to skew the impact of the design in a situation with large urban/rural effect, and due to the importance of educational attainment among young women for SDGs. 
We focus on the age range because most women that will complete their secondary education have already done so by that age, and also because we found evidence of generational differences in secondary education levels among women.
 Weighted estimates of the secondary education levels at the county level are plotted in the top left panel of Figure \ref{fig:Kenya-comparison-admin1-ed}, and we see large variability in the estimates, though there is a large amount of uncertainty in many of these estimates (bottom left panel).

%\begin{figure}
%\centering
%\image{width=3in}{figures/Ed/empiricalEducation.png} \image{width=3in}{figures/Ed/empiricalEducationDiscrete.png}
%\caption{Empirical average of secondary education prevalences among Kenyan women aged 20--29, based on the 2014 Kenya DHS. Values are shown at both the cluster (left) and county levels (right).}
%\label{fig:SEPs}
%\end{figure}

\subsection{Prevalence Mapping}

\begin{sidewaysfigure}
\centering
\includegraphics[width=1.05\textwidth]{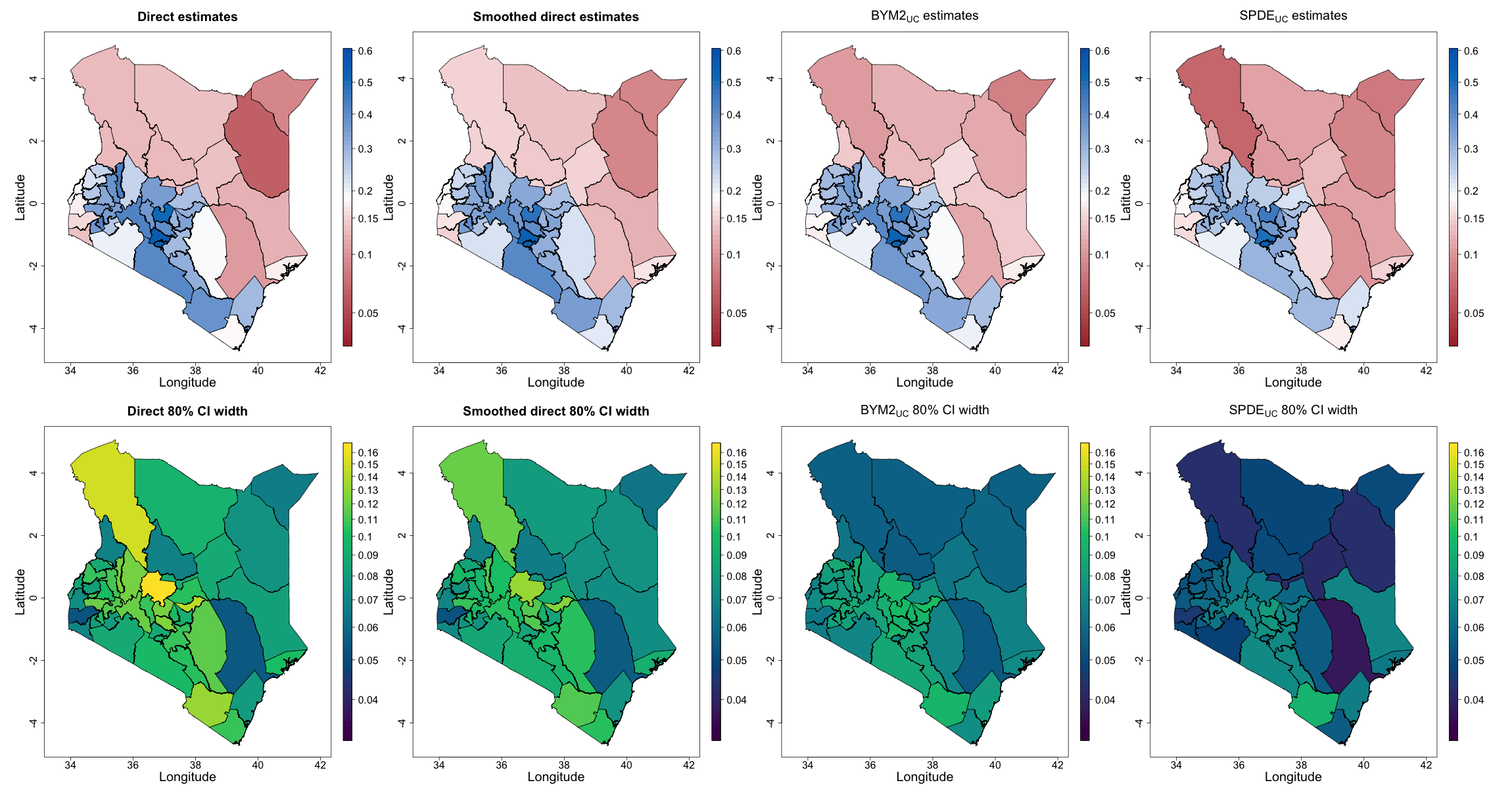}
\caption{Kenya county level 2014 secondary education predictive estimates (top) and 80\% uncertainty interval width (bottom) for women aged 20--29. The BYM2 and SPDE model predictions include both urban and cluster effects.}\label{fig:Kenya-comparison-admin1-ed}
\end{sidewaysfigure}

Here we provide only a small number of summaries, Appendix C gives more detailed results.
Central predictions as well as interval widths for the direct, naive, smoothed direct, and the full (`UC') spatial smoothing models are shown in Figure \ref{fig:Kenya-comparison-admin1-ed}. The top row (point) estimates are quite similar, since there are a large number of clusters, but close examination shows there are differences.
% while 10th and 90th percentiles are given in Figure \ref{fig:Kenya-comparison-quantiles-admin1-ed} in Appendix C. County level central predictions as well as 80\% credible intervals for the smoothed direct, BYM2\textsubscript{UC}, and SPDE\textsubscript{UC} models are given in Appendix C in Table \ref{tab:countyPredictionsSEP}, and parameter summary statistics for the BYM2 and SPDE models including both urban and cluster effects are given in Table \ref{tab:parameterPredictionsSEP} in Appendix C.
Prevalence tended to be higher in the central, southern, and western counties, and tended to be lower and with greater uncertainty in the more rural counties to the north and east.  Appendix C gives full numerical results and here we summarize.
%Table \ref{tab:parameterPredictionsSEP} 
 The odds (with associated 80\% CIs)  of young women in urban clusters completing their secondary education are  larger, relative to rural clusters, by 210\% (185\%, 236\%) or 170\% (148\%, 193\%) as respectively estimated by the BYM2\textsubscript{UC} and SPDE\textsubscript{UC} models.

%Table \ref{tab:countyPredictionsSEP} 
Results in Appendix C shows that the smoothed direct, SPDE\textsubscript{UC}, and BYM2\textsubscript{UC} models all estimate that the secondary education levels for young women in Kenya was highest in Nairobi, with point estimates (80\% CIs) of 0.54 (0.49, 0.58), 0.55 (0.51, 0.58), and 0.53 (0.50, 0.57) respectively. On the other hand, Mandera was estimated to have the lowest secondary education levels for all models except for the SPDE\textsubscript{UC} model (for which Turkana was estimated to have the lowest secondary education levels) with point estimates (80\% CIs) of 0.088 (0.061, 0.13), 0.081 (0.058, 0.11), and 0.085 (0.060, 0.12) respectively. While Nairobi is designated as completely urban, approximately 18\% of the population of Mandera is urban, which is very close to the median for counties in Kenya. This suggests there are other factors in Mandera that are reducing the secondary education prevalence for the women living there.

The credible interval widths were largest for the direct model and smallest for the SPDE\textsubscript{UC} model. Of the displayed spatial smoothing models, the smoothed direct model had the largest predictive variances. Both the smoothed direct and the BYM2 models had relatively high uncertainties for counties with fewer neighbors, whereas the SPDE model variances tended to be high near the edges and where the number of clusters were spatially distant from each other. In central and west Kenya, where the sampled clusters tended to be denser, the SPDE model had lower predictive variances relative to the discrete spatial models.

\begin{figure}
\centering
\includegraphics[width=6in]{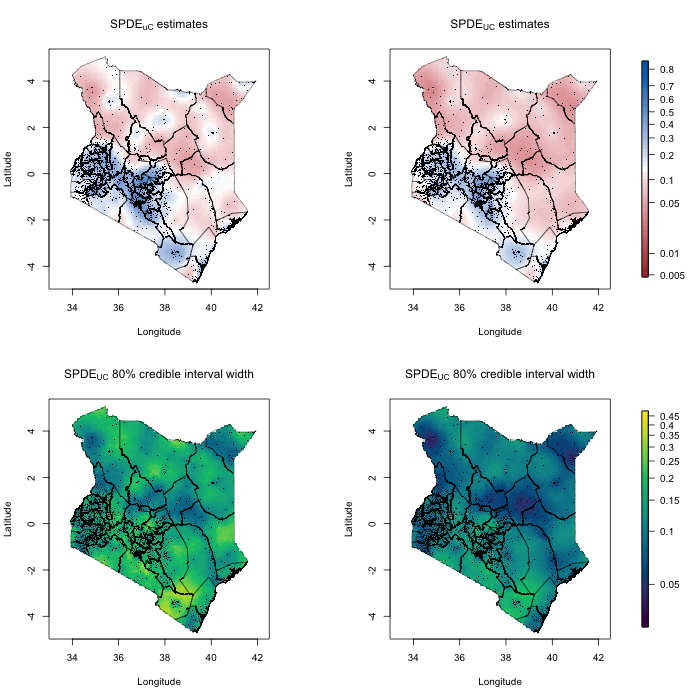}
\caption{Kenya 5km $\times$ 5km pixel level 2014 secondary education predictive mean (top) and 80\% uncertainty interval width (bottom) for women aged 20--29. Results are shown for the SPDE\textsubscript{uC} (left) and SPDE\textsubscript{UC} (right) models.}\label{fig:Kenya-comparison-pixel-ed}
\end{figure}

Figure \ref{fig:Kenya-comparison-pixel-ed} shows the continuous, 5km$\times$5km pixel level predictions and credible interval widths for the SPDE\textsubscript{uC} and SPDE\textsubscript{UC} models. The urban effect is especially visible in the predictions of the SPDE\textsubscript{uC}, since it oversmooths the effect of the urban areas into nearby rural regions.  Interestingly, secondary education prevalences appear to be high not only in urban areas, but also in rural areas bordering urban centers. 
This  indicates that  some of the urban/rural association (which was smaller for the SPDE model) is being absorbed into the spatial field.
 In general, we are nervous about presenting pixel-level maps, and far more comfortable with area-level summaries.  Figure \ref{fig:Kenya-comparison-pixel-ed} clearly shows that care that must be taken with stratification. In order to maintain confidentiality, the geographical locations of the DHS clusters are displaced (jittered): urban clusters by up to 2km, rural clusters by up to 5km, with a further 1\% by up to 10km, which is another reason that pixel-level maps should not be over-interpreted.

%\begin{comment}
% \begin{figure}
% \centering
%\includegraphics[width=0.82\textwidth]{figures/Ed/fullDirectNaiveEducation.png}
%    \caption{Kenya county level 2014 secondary education predictive summary statistics for women aged 20--29. Shown are central estimates (top), standard deviation of the logit predictions (second row), 10th percentiles (third row), and 90th percentiles (bottom). Summary statistics are given for the naive (left) and direct (right) models.}\label{fig:Kenya-comparison-admin1-directNaive-ed}
%%    \vspace{-0.5cm}
%\end{figure}
%
% \begin{figure}
% \centering
%\includegraphics[width=\textwidth]{figures/Ed/fullSmoothedEducation.png}
%    \caption{Kenya county level 2014 secondary education predictive summary statistics for women aged 20--29. Shown are central estimates (top), standard deviation of the logit predictions (second row), 10th percentiles (third row), and 90th percentiles (bottom). Summary statistics are given for the smoothed direct (left), full BYM2 (middle column), and full SPDE (right) models.}\label{fig:Kenya-comparison-admin1-smooth-ed}
%%    \vspace{-0.5cm}
%\end{figure}
%\end{comment}

\subsection{Validation}

We calculate a number of scoring rules at the cluster level to evaluate the spatial smoothing models. We compute the scoring rules by leaving out data from one county at a time and averaging the scoring rules over all 47 such experiments. We carry out the validation out at the county level, since this is generally the target of inference. In addition to calculating MSE (broken down into variance and bias and in urban as well as rural areas) we also compute CRPS, the deviance information criterion (DIC), and the conditional predictive ordinate (CPO). The naive, direct, and smoothed direct models are fit at the county level, so we did not include their validation results, since they are not comparable with the cluster level data models.

The SPDE\textsubscript{Uc} and SPDE\textsubscript{UC} models had the two best average MSEs, and the SPDE\textsubscript{Uc} model has the best CPO and CRPS. The SPDE\textsubscript{UC} had better MSE than the Uc model, but it had worse CPO and CRPS. In terms of MSE and CRPS, the BYM2\textsubscript{Uc} model also performed well, and had the smallest magnitude of bias.
The good performance of the SPDE models may be due to their ability to model continuous changes in secondary education near the borders of each county, whereas the BYM2 models are unable to distinguish between clusters at the border of a county versus clusters in the center. In Appendix C we see that the spatial standard deviation (SD) of the SPDE\textsubscript{UC} model is estimated to be 0.91, whereas the cluster effect is estimated to have a SD of 0.65. This is a higher proportion of variability going into the spatial term than in the BYM2\textsubscript{UC} model, which has a total variance of county level random effects estimated to be 0.57 and a cluster variance estimated to be 0.71. This suggests the ability of the SPDE model to predict continuously through space gives it an advantage when making predictions at the cluster level.

\begin{table}[H]
\centering
\begin{tabular}{lrrrrrrrr}
\toprule
\multicolumn{1}{c}{\em{\textbf{ }}} & \multicolumn{4}{c}{\em{\textbf{BYM2}}} & \multicolumn{4}{c}{\em{\textbf{SPDE}}} \\
\cmidrule(l{3pt}r{3pt}){2-5} \cmidrule(l{3pt}r{3pt}){6-9}
  & uc & uC & Uc & UC & uc & uC & Uc & UC\\
\midrule
\addlinespace[0.3em]
\multicolumn{9}{l}{\textbf{MSE ($\times 10^{-2}$)}}\\
\hspace{1em}Average & 5.9 & \em{6.0} & 5.1 & 5.2 & 5.6 & 5.4 & 5.0 & \textbf{4.9}\\
\hspace{1em}Urban & \em{7.2} & \em{7.2} & 6.2 & 6.2 & 7.1 & 6.8 & 6.2 & \textbf{6.0}\\
\hspace{1em}Rural & 5.0 & \em{5.2} & 4.5 & 4.5 & 4.6 & 4.5 & \textbf{4.2} & \textbf{4.2}\\
\addlinespace[0.3em]
\multicolumn{9}{l}{\textbf{Var ($\times 10^{-3}$)}}\\
\hspace{1em}Average & 59 & \em{59} & 51 & 52 & 56 & 53.8 & 50 & \textbf{49}\\
\hspace{1em}Urban & 61 & 62 & 62 & 62 & \em{63} & \textbf{60} & 62 & 60\\
\hspace{1em}Rural & 45 & 45 & 45 & 45 & \em{45} & 42 & 42 & \textbf{42}\\
\addlinespace[0.3em]
\multicolumn{9}{l}{\textbf{Bias ($\times 10^{-3}$)}}\\
\hspace{1em}Average & 6.0 & 10.5 & \textbf{1.0} & 5.6 & \em{-15.0} & -6.2 & -3.6 & -3.2\\
\hspace{1em}Urban & \em{-105} & -103 & 11.1 & 8.3 & -93 & -91 & \textbf{1.0} & 3.1\\
\hspace{1em}Rural & 77 & \em{83} & -5.4 & \textbf{4.0} & 35 & 48 & -6.5 & -7.2\\
\addlinespace[0.3em]
\multicolumn{9}{l}{\textbf{CPO}}\\
\hspace{1em}Average & 0.22 & \em{0.21} & 0.25 & 0.24 & 0.26 & 0.25 & \textbf{0.27} & 0.26\\
\addlinespace[0.3em]
\multicolumn{9}{l}{\textbf{CRPS}}\\
\hspace{1em}Average & 0.17 & \em{0.21} & 0.16 & 0.18 & 0.17 & 0.18 & \textbf{0.15} & 0.17\\
\bottomrule
\end{tabular}
\caption{Validation results calculated at the cluster level when leaving out one county at a time for the 2014 Kenya DHS secondary education data for women aged 20--29. The worst entries in each row are in {\it italics}, while the best entries in each row are in {\bf bold}. In the table, the figures are rounded, but minimum and maximum were evaluated with more significant figures.}
\label{tab:leaveOutCounty}
\end{table}

\section{Discussion and Conclusions} \label{sec:discussion}

Direct estimators remain the gold standard, provided there are sufficient data for an associated variance that is of acceptable size. The smoothed direct estimator can reduce the variance using the totality of data, albeit with an introduction of small bias, due to the smoothing. When the direct estimates are unreliable, one is lead to modeling at the cluster-level, and one must use a model that is consistent with the design. In this paper, we introduced a binomial sampling model with discrete spatial random effects,  and it performed well in the simulations and in the real application. We have also been experimenting with a beta-binomial that allows for overdispersion (within-cluster variation).
Such approaches with discrete spatial models do not deal well with combining data at different geographical resolutions, and this is where the continuous spatial model is appealing. Unfortunately, aggregation with this model is the most difficult, since a population surface is required, and this needs, in general, to be stratified by urban/rural. 

%This work has provided a framework with which to assess statistical model fit with multi-stage survey data. A large scale simulation study involving simulating the population of Kenya at over 96,000 enumeration areas and simulating hundreds of surveys from two different designs provided a basis for comparing model predictions when aggregated to the county level.  Left out data was used to determine model performance at the cluster level.

In the simulations and application, we found that accounting for the design nearly always improved predictions. This was true whether the design was accounted for using sampling weights or by including stratification indicators as covariates. Although not included in our results. we have found that when the proportions urban required for aggregating strata predictions to the county level were estimated poorly, including stratification effects in the BYM2 model sometimes made the predictions worse. This implies that not only must design stratification be accounted for, but in the case where it is included as a covariate, it is important to make an effort to obtain good estimates of  the proportions of the studied population in each strata.

Although we expected the SPDE models without urban effects to have better predictions, because urbanicity is a spatial variable, we instead found that the spatial component of the SPDE models without urban effects had difficulty handling the sharp changes of urbanicity in space as well as the fact that the urban areas were so localized. As we mentioned in the introduction, WorldPop and IHME do not adjust for the urban/rural stratification, but they do include extensive covariates, including population. density, which will, to some extent at least adjust, for urban/rural.

%This also suggests that modeling design induced bias continuously and smoothly through space as in \citet{giorgi:etal:15} could lead to problems in these settings. Although county boundaries are discrete, we believe it is unlikely that county boundaries consistently lead to as large of an impact as urbanicity and smooth variation in space. Additionally, counties are often larger than individual urban areas, which could allow for continuous spatial effects to more accurately vary at their boundaries. In the cases that county effects appear large, however, it may be important to consider including county effects as well in the SPDE models.

%We found that discrete model predictions in general improved or performed nearly the same when they accounted for cluster effects.  A similar result held true for continuous models in the case where the cluster variance made up a relatively small proportion of the total variance. Care must be taken for continuous spatial models when choosing whether or not and how to include cluster effects as the proportion of total variance taken by the cluster effect increases. In this case, important considerations include prior selection for the cluster effect as well as how cluster effects are accounted for (and if they should be included at all) when aggregating predictions to areal units. A pragmatic approach to allowing for extra variation is to include an area-level iid random effect in the model, at the level at which aggregations are primarily required, for example Admin2.
%

A remaining open avenue of research is to determine how best to include cluster effects in area-level aggregated predictions from spatial models. 
%Answering this question would enable continuous spatial models to better account for overdispersion induced by cluster effects. 
Since the SPDE model predictions are aggregated to the county level by numerically integrating predictions on a spatial grid, whereas cluster effects are modeled discretely at cluster and EA point locations, it is unclear how to accurately proceed when the EA locations are unknown. Simply leaving out the cluster effects when aggregating predictions spatially leads to undercoverage and also bias, whereas using Monte Carlo methods to sample possible EA locations can be computationally expensive. 
The simulation study and prevalence application indicated that the smoothed direct model was the most reliable, performing well in nearly all circumstances, whereas the SPDE and BYM2 models that included stratification and cluster effects performed especially well when there was a stratification effect in the population. This was especially the case if the proportion of the population of interest (i.e.,~children, or women aged 20--29) that is urban in each county is accurately known. The BYM2 model including urban and cluster effects performed the best in in the simulation study, in terms of MSE, CRPS, and credible interval width, in many scenarios. The SPDE model including urban and cluster effects performed better in the cluster level validation, but care must be taken in selecting a prior due to its flexibility, and in generating spatially aggregated predictions when the estimated cluster effect accounts for a large proportion of the total variation.

In the simulation study, the DHS we emulated was powered to the Admin2 level, which coincided the level of inference. More commonly, DHSs are powered to the Admin1 level and it is an open question as to what the recommendations are in this case if inference is still required at Admin2.  In other work \citep{li:etal:19} we could only perform Admin1 level inference for countries in Africa using the majority of the DHSs, because there were insufficient samples to applied the  direct and smoothed direct methods.

%Although it is not considered in this analysis, depending on whether the goal of a study is to estimate the true population census value or the unknown population risk, it may also be important to account for binomial variation. Unfortunately, the binomial denominator (i.e., the number of children or young women) is often not known at the EA level even if the exact EA locations are known, which they often are not. This makes accounting for binomial variation very difficult in household surveys. Better ways of taking binomial variation into account could reduce undercoverage for the SPDE and BYM2 models.

\bibliographystyle{chicago}
\bibliography{spatepi.bib}

\begin{thebibliography}{}

\bibitem[\protect\citeauthoryear{Antai and Moradi}{Antai and
  Moradi}{2010}]{antai2010urban}
Antai, D. and T.~Moradi (2010).
\newblock Urban area disadvantage and under-5 mortality in {N}igeria: the
  effect of rapid urbanization.
\newblock {\em Environmental Health Perspectives\/}~{\em 118}, 877--883.

\bibitem[\protect\citeauthoryear{Balk, Pullum, Storeygard, Greenwell, and
  Neuman}{Balk et~al.}{2004}]{balk:2004}
Balk, D., T.~Pullum, A.~Storeygard, F.~Greenwell, and M.~Neuman (2004).
\newblock A spatial analysis of childhood mortality in {West Africa}.
\newblock {\em Population, Space and Place\/}~{\em 10}, 175--216.

\bibitem[\protect\citeauthoryear{Besag, York, and Molli\'e}{Besag
  et~al.}{1991}]{besag:etal:91}
Besag, J., J.~York, and A.~Molli\'e (1991).
\newblock Bayesian image restoration with two applications in spatial
  statistics.
\newblock {\em Annals of the Institute of Statistics and Mathematics\/}~{\em
  43}, 1--59.

\bibitem[\protect\citeauthoryear{Black, Laximinarayan, Temmerman, and
  Walker}{Black et~al.}{2016}]{black:etal:16}
Black, R.~E., R.~Laximinarayan, M.~Temmerman, and N.~Walker (2016).
\newblock {\em Reproductive, Maternal, Newborn, and Child Health, Third
  Edition}.
\newblock World Bank Group.

\bibitem[\protect\citeauthoryear{Chen, Wakefield, and Lumley}{Chen
  et~al.}{2014}]{chen:etal:14}
Chen, C., J.~Wakefield, and T.~Lumley (2014).
\newblock The use of sample weights in {B}ayesian hierarchical models for small
  area estimation.
\newblock {\em Spatial and Spatio-Temporal Epidemiology\/}~{\em 11}, 33--43.

\bibitem[\protect\citeauthoryear{Congdon and Lloyd}{Congdon and
  Lloyd}{2010}]{congdon:lloyd:10}
Congdon, P. and P.~Lloyd (2010).
\newblock Estimating small area diabetes prevalence in the {US} using the
  behavioral risk factor surveillance system.
\newblock {\em Journal of Data Science\/}~{\em 8}, 235--252.

\bibitem[\protect\citeauthoryear{Defo}{Defo}{1994}]{defo1994determinants}
Defo, B.~K. (1994).
\newblock Determinants of infant and early childhood mortality in {C}ameroon:
  the role of socioeconomic factors, housing characteristics, and immunization
  status.
\newblock {\em Social Biology\/}~{\em 41}, 181--211.

\bibitem[\protect\citeauthoryear{{DHS Program}}{{DHS Program}}{2019}]{AIS}
{DHS Program} (2019).
\newblock The {DHS} program--{AIDS} indicator surveys ({AIS}).
\newblock {https://dhsprogram.com/What-We-Do/Survey-Types/AIS.cfm}.

\bibitem[\protect\citeauthoryear{Diggle and Giorgi}{Diggle and
  Giorgi}{2016}]{diggle:giorgi:16}
Diggle, P. and E.~Giorgi (2016).
\newblock Model-based geostatistics for prevalence mapping in low-resource
  settings.
\newblock {\em Journal of the American Statistical Association\/}~{\em 111},
  1096--1120.

\bibitem[\protect\citeauthoryear{Diggle and Giorgi}{Diggle and
  Giorgi}{2019}]{diggle2019model}
Diggle, P.~J. and E.~Giorgi (2019).
\newblock {\em Model-based Geostatistics for Global Public Health: Methods and
  Applications}.
\newblock Boca-Raton: Chapman and Hall/{CRC}.

\bibitem[\protect\citeauthoryear{Fay and Herriot}{Fay and
  Herriot}{1979}]{fay:herriot:79}
Fay, R. and R.~Herriot (1979).
\newblock Estimates of income for small places: an application of
  {James--Stein} procedure to census data.
\newblock {\em Journal of the American Statistical Association\/}~{\em 74},
  269--277.

\bibitem[\protect\citeauthoryear{Fuglstad, Simpson, Lindgren, and Rue}{Fuglstad
  et~al.}{2019}]{fuglstad:etal:19a}
Fuglstad, G.-A., D.~Simpson, F.~Lindgren, and H.~Rue (2019).
\newblock Constructing priors that penalize the complexity of {G}aussian random
  fields.
\newblock {\em Journal of the American Statistical Association\/}~{\em 114},
  445--452.

\bibitem[\protect\citeauthoryear{Gething, Tatem, Bird, and
  Burgert-Brucker}{Gething et~al.}{2015}]{gething:etal:15}
Gething, P., A.~Tatem, T.~Bird, and C.~Burgert-Brucker (2015).
\newblock Creating spatial interpolation surfaces with {DHS} data.
\newblock Technical report, ICF International.
\newblock DHS Spatial Analysis Reports No. 11.

\bibitem[\protect\citeauthoryear{Gething, Casey, Weiss, Bisanzio, Bhatt,
  Cameron, Battle, Dalrymple, Rozier, Rao, Kutz, Barber, Huynh, Shackleford,
  Coates, Nguyen, Fraser, Kulikoff, Wang, Naghavi, Smith, Murray, Hay, and
  Lim}{Gething et~al.}{2016}]{gething:etal:16}
Gething, P.~W., D.~C. Casey, D.~J. Weiss, D.~Bisanzio, S.~Bhatt, E.~Cameron,
  K.~E. Battle, U.~Dalrymple, J.~Rozier, P.~C. Rao, M.~Kutz, R.~Barber,
  C.~Huynh, K.~Shackleford, M.~Coates, G.~Nguyen, M.~Fraser, R.~Kulikoff,
  H.~Wang, M.~Naghavi, D.~Smith, C.~Murray, S.~Hay, and S.~Lim (2016).
\newblock Mapping plasmodium falciparum mortality in {A}frica between 1990 and
  2015.
\newblock {\em New England Journal of Medicine\/}~{\em 375}, 2435--2445.

\bibitem[\protect\citeauthoryear{Giorgi, Diggle, Snow, and Noor}{Giorgi
  et~al.}{2018}]{giorgi:etal:18}
Giorgi, E., P.~J. Diggle, R.~W. Snow, and A.~M. Noor (2018).
\newblock Geostatistical methods for disease mapping and visualization using
  data from spatio-temporally referenced prevalence surveys.
\newblock {\em International Statistical Review\/}~{\em 86}, 571--597.

\bibitem[\protect\citeauthoryear{Gneiting and Raftery}{Gneiting and
  Raftery}{2007}]{gneiting:raftery:07}
Gneiting, T. and A.~E. Raftery (2007).
\newblock Strictly proper scoring rules, prediction, and estimation.
\newblock {\em Journal of the American Statistical Association\/}~{\em 102},
  359--378.

\bibitem[\protect\citeauthoryear{Golding, Burstein, Longbottom, Browne,
  Fullman, Osgood-Zimmerman, Earl, Bhatt, Cameron, Casey, Dwyer-Lindgren,
  Farag, Flaxman, Fraser, Gething, Gibson, Graetz, Krause, Kulikoff, Lim,
  Mappin, Morozoff, Reiner, Sligar, Smith, Wang, Weiss, Murray, Moyes, and
  Hay}{Golding et~al.}{2017}]{golding:etal:17}
Golding, N., R.~Burstein, J.~Longbottom, A.~Browne, N.~Fullman,
  A.~Osgood-Zimmerman, L.~Earl, S.~Bhatt, E.~Cameron, D.~Casey,
  L.~Dwyer-Lindgren, T.~Farag, A.~Flaxman, M.~Fraser, P.~Gething, H.~Gibson,
  N.~Graetz, L.~Krause, X.~Kulikoff, S.~Lim, B.~Mappin, C.~Morozoff, R.~Reiner,
  A.~Sligar, D.~Smith, H.~Wang, D.~Weiss, C.~Murray, C.~Moyes, and S.~Hay
  (2017).
\newblock Mapping under-5 and neontal mortality in {A}frica, 2000--15: a
  baseline analysis for the {S}ustainable {D}evelopment {G}oals.
\newblock {\em The Lancet\/}~{\em 390}, 2171--2182.

\bibitem[\protect\citeauthoryear{Graetz, Friedman, Osgood-Zimmerman, Burstein,
  Biehl, Shields, Mosser, Casey, Deshpande, Earl, Reiner, Ray, Fullman, Levine,
  Stubbs, Mayala, Longbottom, Browne, Bhatt, Weiss, Gething, Mokdad, Lim,
  Murray, Gakidou, and Hay}{Graetz et~al.}{2018}]{graetz:etal:18}
Graetz, N., J.~Friedman, A.~Osgood-Zimmerman, R.~Burstein, M.~H. Biehl,
  C.~Shields, J.~F. Mosser, D.~C. Casey, A.~Deshpande, L.~Earl, R.~Reiner,
  S.~Ray, N.~Fullman, A.~Levine, R.~Stubbs, B.~Mayala, J.~Longbottom,
  A.~Browne, S.~Bhatt, D.~Weiss, P.~Gething, A.~Mokdad, S.~Lim, C.~Murray,
  E.~Gakidou, and S.~Hay (2018).
\newblock Mapping local variation in educational attainment across {A}frica.
\newblock {\em Nature\/}~{\em 555}, 48.

\bibitem[\protect\citeauthoryear{{ICF International}}{{ICF
  International}}{2012}]{samplingManualDHS}
{ICF International} (2012).
\newblock {\em Demographic and Health Survey Sampling and Household Listing
  Manual}.
\newblock Calverton, Maryland, USA: ICF International.

\bibitem[\protect\citeauthoryear{{Kenya National Bureau of Statistics, Ministry
  of Health/Kenya, National AIDS Control Council/Kenya, Kenya Medical Research
  Institute, and National Council For Population And Development/Kenya}}{{Kenya
  National Bureau of Statistics, Ministry of Health/Kenya, National AIDS
  Control Council/Kenya, Kenya Medical Research Institute, and National Council
  For Population And Development/Kenya}}{2009b}]{KenyaCensus:2009}
{Kenya National Bureau of Statistics, Ministry of Health/Kenya, National AIDS
  Control Council/Kenya, Kenya Medical Research Institute, and National Council
  For Population And Development/Kenya} (2009b).
\newblock {\em The 2009 Kenya Population and Housing Census Volume {IC}:
  Population Distribution by Age, Sex, and Administrative Units}.
\newblock Nairobi: Kenya National Bureau of Statistics.

\bibitem[\protect\citeauthoryear{{Kenya National Bureau of Statistics, Ministry
  of Health/Kenya, National AIDS Control Council/Kenya, Kenya Medical Research
  Institute, and National Council For Population And Development/Kenya}}{{Kenya
  National Bureau of Statistics, Ministry of Health/Kenya, National AIDS
  Control Council/Kenya, Kenya Medical Research Institute, and National Council
  For Population And Development/Kenya}}{2015a}]{KDHS2014}
{Kenya National Bureau of Statistics, Ministry of Health/Kenya, National AIDS
  Control Council/Kenya, Kenya Medical Research Institute, and National Council
  For Population And Development/Kenya} (2015a).
\newblock {\em Kenya Demographic and Health Survey 2014}.
\newblock Rockville, Maryland, USA.

\bibitem[\protect\citeauthoryear{Lawn, Blencowe, Oza, You, Lee, Waiswa, Lalli,
  Bhutta, Barros, Christian, et~al.}{Lawn et~al.}{2014}]{lawn:etal:14}
Lawn, J.~E., H.~Blencowe, S.~Oza, D.~You, A.~C. Lee, P.~Waiswa, M.~Lalli,
  Z.~Bhutta, A.~J. Barros, P.~Christian, et~al. (2014).
\newblock Every newborn: progress, priorities, and potential beyond survival.
\newblock {\em The Lancet\/}~{\em 384}, 189--205.

\bibitem[\protect\citeauthoryear{Li, Hsiao, Godwin, Martin, Wakefield, and
  Clark}{Li et~al.}{2019}]{li:etal:19}
Li, Z.~R., Y.~Hsiao, J.~Godwin, B.~D. Martin, J.~Wakefield, and S.~J. Clark
  (2019).
\newblock Changes in the spatial distribution of the under five mortality rate:
  small-area analysis of 122 {DHS} surveys in 262 subregions of 35 countries in
  {A}frica.
\newblock {\em {PLoS One}\/}.
\newblock Published January 22, 2019.

\bibitem[\protect\citeauthoryear{Lindgren, Rue, and Lindstr\"{o}m}{Lindgren
  et~al.}{2011}]{lindgren:etal:11}
Lindgren, F., H.~Rue, and J.~Lindstr\"{o}m (2011).
\newblock An explicit link between {G}aussian fields and {G}aussian {M}arkov
  random fields: the stochastic differential equation approach (with
  discussion).
\newblock {\em Journal of the Royal Statistical Society, Series B\/}~{\em 73},
  423--498.

\bibitem[\protect\citeauthoryear{Lumley}{Lumley}{2004}]{lumley:04}
Lumley, T. (2004).
\newblock Analysis of complex survey samples.
\newblock {\em Journal of Statistical Software\/}~{\em 9}, 1--19.

\bibitem[\protect\citeauthoryear{Lumley}{Lumley}{2018}]{Lumley:2018aa}
Lumley, T. (2018).
\newblock survey: analysis of complex survey samples.
\newblock R package version 3.35.

\bibitem[\protect\citeauthoryear{Marhuenda, Molina, and Morales}{Marhuenda
  et~al.}{2013}]{marhuenda:etal:13}
Marhuenda, Y., I.~Molina, and D.~Morales (2013).
\newblock Small area estimation with spatio-temporal {F}ay--{H}erriot models.
\newblock {\em Computational Statistics and Data Analysis\/}~{\em 58},
  308--325.

\bibitem[\protect\citeauthoryear{Martin, Li, Hsiao, Godwin, Wakefield, and
  Clark}{Martin et~al.}{2018}]{martin:etal:18}
Martin, B.~D., Z.~R. Li, Y.~Hsiao, J.~Godwin, J.~Wakefield, and S.~J. Clark
  (2018).
\newblock {\em SUMMER: Spatio-Temporal Under-Five Mortality Methods for
  Estimation}.
\newblock R package version 0.2.1.

\bibitem[\protect\citeauthoryear{Mercer, Wakefield, Pantazis, Lutambi, Mosanja,
  and Clark}{Mercer et~al.}{2015}]{mercer:etal:15}
Mercer, L., J.~Wakefield, A.~Pantazis, A.~Lutambi, H.~Mosanja, and S.~Clark
  (2015).
\newblock Small area estimation of childhood mortality in the absence of vital
  registration.
\newblock {\em Annals of Applied Statistics\/}~{\em 9}, 1889--1905.

\bibitem[\protect\citeauthoryear{Ntenda, Chuang, Tiruneh, and Chuang}{Ntenda
  et~al.}{2014}]{ntenda2014factors}
Ntenda, P. A.~M., K.-Y. Chuang, F.~N. Tiruneh, and Y.-C. Chuang (2014).
\newblock Factors associated with infant mortality in {M}alawi.
\newblock {\em Journal of Experimental \& Clinical Medicine\/}~{\em 6},
  125--131.

\bibitem[\protect\citeauthoryear{Osgood-Zimmerman, Millear, Stubbs, Shields,
  Pickering, Earl, Graetz, Kinyoki, Ray, Bhatt, Browne, Burstein, Cameron,
  Casey, Deshpande, Fullman, Gething, Gibson, Henry, Herrero, Krause,
  Letourneau, Levine, Liu, Longbottom, Mayala, Mosser, Noor, Pigott, Piwoz,
  Rao, Rawat, Reiner, Smith, Weiss, Wiens, Mokdad, S.S., Murray, Kassebaum, and
  Hay}{Osgood-Zimmerman et~al.}{2018}]{osgood:etal:18}
Osgood-Zimmerman, A., A.~I. Millear, R.~W. Stubbs, C.~Shields, B.~V. Pickering,
  L.~Earl, N.~Graetz, D.~K. Kinyoki, S.~E. Ray, S.~Bhatt, A.~Browne,
  R.~Burstein, E.~Cameron, D.~Casey, A.~Deshpande, N.~Fullman, P.~Gething,
  H.~Gibson, N.~Henry, M.~Herrero, L.~Krause, I.~Letourneau, A.~Levine, P.~Liu,
  J.~Longbottom, B.~Mayala, J.~Mosser, A.~Noor, D.~Pigott, E.~Piwoz, P.~Rao,
  R.~Rawat, R.~Reiner, D.~Smith, D.~Weiss, K.~Wiens, A.~Mokdad, L.~S.S.,
  C.~Murray, N.~Kassebaum, and S.~Hay (2018).
\newblock Mapping child growth failure in {A}frica between 2000 and 2015.
\newblock {\em Nature\/}~{\em 555}, 41.

\bibitem[\protect\citeauthoryear{Pezzulo, T.Bird, Edson, Utazi, Sorichetta,
  Tatem, Yourkavitch, and Burgert-Brucker}{Pezzulo
  et~al.}{2016}]{pezzulo:etal:16}
Pezzulo, C., T.Bird, C.~Edson, C.~Utazi, A.~Sorichetta, A.~Tatem,
  J.~Yourkavitch, and C.~Burgert-Brucker (2016).
\newblock Geospatial modeling of child mortality across 27 countries in
  sub-{S}aharan {A}frica.
\newblock Technical report, {ICF} International.
\newblock DHS Spatial Analysis Reports No. 13.

\bibitem[\protect\citeauthoryear{Porter, Holan, Wikle, and Cressie}{Porter
  et~al.}{2014}]{porter:etal:14}
Porter, A.~T., S.~H. Holan, C.~K. Wikle, and N.~Cressie (2014).
\newblock Spatial {F}ay--{H}erriot models for small area estimation with
  functional covariates.
\newblock {\em Spatial Statistics\/}~{\em 10}, 27--42.

\bibitem[\protect\citeauthoryear{Rao and Molina}{Rao and
  Molina}{2015}]{rao:molina:15}
Rao, J. and I.~Molina (2015).
\newblock {\em Small Area Estimation, Second Edition}.
\newblock New York: John Wiley.

\bibitem[\protect\citeauthoryear{Riebler, S{\o}rbye, Simpson, and Rue}{Riebler
  et~al.}{2016}]{riebler:etal:16}
Riebler, A., S.~S{\o}rbye, D.~Simpson, and H.~Rue (2016).
\newblock An intuitive {B}ayesian spatial model for disease mapping that
  accounts for scaling.
\newblock {\em Statistical Methods in Medical Research\/}~{\em 25}, 1145--1165.

\bibitem[\protect\citeauthoryear{Root}{Root}{1997}]{root:1997}
Root, G. (1997).
\newblock Population density and spatial differentials in child mortality in
  {Z}imbabwe.
\newblock {\em Social Science and Medicine\/}~{\em 44}, 413--421.

\bibitem[\protect\citeauthoryear{Rue, Martino, and Chopin}{Rue
  et~al.}{2009}]{rue:etal:09}
Rue, H., S.~Martino, and N.~Chopin (2009).
\newblock Approximate {B}ayesian inference for latent {G}aussian models using
  integrated nested {L}aplace approximations (with discussion).
\newblock {\em Journal of the Royal Statistical Society, Series B\/}~{\em 71},
  319--392.

\bibitem[\protect\citeauthoryear{Simpson, Rue, Riebler, Martins, and
  S{\o}rbye}{Simpson et~al.}{2017}]{simpson:etal:17}
Simpson, D., H.~Rue, A.~Riebler, T.~Martins, and S.~S{\o}rbye (2017).
\newblock Penalising model component complexity: A principled, practical
  approach to constructing priors (with discussion).
\newblock {\em Statistical Science\/}~{\em 32}, 1--28.

\bibitem[\protect\citeauthoryear{Stevens, Gaughan, Linard, and Tatem}{Stevens
  et~al.}{2015}]{stevens2015disaggregating}
Stevens, F.~R., A.~E. Gaughan, C.~Linard, and A.~J. Tatem (2015).
\newblock Disaggregating census data for population mapping using random
  forests with remotely-sensed and ancillary data.
\newblock {\em PloS One\/}~{\em 10}, e0107042.

\bibitem[\protect\citeauthoryear{Tatem}{Tatem}{2017}]{tatem2017worldpop}
Tatem, A.~J. (2017).
\newblock World{P}op, open data for spatial demography.
\newblock {\em Scientific data\/}~{\em 4}.

\bibitem[\protect\citeauthoryear{{UNICEF}}{{UNICEF}}{2019}]{equist:19}
{UNICEF} (2019).
\newblock {\em {EQUIST} Analyst User Guide}.
\newblock UNICEF: \url{www.equist.com}.

\bibitem[\protect\citeauthoryear{{UNICEF - Statistics and Monitoring}}{{UNICEF
  - Statistics and Monitoring}}{2012}]{MICS}
{UNICEF - Statistics and Monitoring} (2012).
\newblock {Multiple Indicator Cluster Surveys (MICS)}.
\newblock {\url{http://www.unicef.org/statistics/index_24302.html}}.

\bibitem[\protect\citeauthoryear{{United Nations}}{{United
  Nations}}{2019}]{sdgsWeb}
{United Nations} (2019).
\newblock {\em Sustainable Development Goals}.
\newblock \url{http://sustainabledevelopment.un.org/owg.html}.

\bibitem[\protect\citeauthoryear{USAID}{USAID}{2019}]{dhs}
USAID (2019).
\newblock {\em Demographic and Health Surveys}.
\newblock \url{http://www.dhsprogram.com}: {United States Agency for
  International Development}.

\bibitem[\protect\citeauthoryear{Utazi, Thorley, Alegana, Ferrari, Takahashi,
  Metcalf, Lessler, and Tatem}{Utazi et~al.}{2018}]{utazi:etal:18}
Utazi, C.~E., J.~Thorley, V.~A. Alegana, M.~J. Ferrari, S.~Takahashi, C.~J.~E.
  Metcalf, J.~Lessler, and A.~J. Tatem (2018).
\newblock High resolution age-structured mapping of childhood vaccination
  coverage in low and middle income countries.
\newblock {\em Vaccine\/}~{\em 36}, 1583--1591.

\bibitem[\protect\citeauthoryear{Vandendijck, Faes, Kirby, Lawson, and
  Hens}{Vandendijck et~al.}{2016}]{vandendijck:etal:16}
Vandendijck, Y., C.~Faes, R.~S. Kirby, A.~Lawson, and N.~Hens (2016).
\newblock Model-based inference for small area estimation with sampling
  weights.
\newblock {\em Spatial Statistics\/}~{\em 18}, 455--473.

\bibitem[\protect\citeauthoryear{Wagner, Heft-Neal, Bhutta, Black, Burke, and
  Bendavid}{Wagner et~al.}{2018}]{wagner:etal:18}
Wagner, Z., S.~Heft-Neal, Z.~A. Bhutta, R.~E. Black, M.~Burke, and E.~Bendavid
  (2018).
\newblock Armed conflict and child mortality in {A}frica: a geospatial
  analysis.
\newblock {\em The Lancet\/}~{\em 392}, 857--865.

\bibitem[\protect\citeauthoryear{Wakefield, Fuglstad, Riebler, Godwin, Wilson,
  and Clark}{Wakefield et~al.}{2019}]{wakefield:etal:19}
Wakefield, J., G.-A. Fuglstad, A.~Riebler, J.~Godwin, K.~Wilson, and S.~Clark
  (2019).
\newblock Estimating under five mortality in space and time in a developing
  world context.
\newblock {\em Statistical Methods in Medical Research\/}~{\em 28}, 2614--2634.

\bibitem[\protect\citeauthoryear{Wardrop, Jochem, Bird, Chamberlain, Clarke,
  Kerr, Bengtsson, Juran, Seaman, and Tatem}{Wardrop
  et~al.}{2018}]{wardrop:etal:18}
Wardrop, N., W.~Jochem, T.~Bird, H.~Chamberlain, D.~Clarke, D.~Kerr,
  L.~Bengtsson, S.~Juran, V.~Seaman, and A.~Tatem (2018).
\newblock Spatially disaggregated population estimates in the absence of
  national population and housing census data.
\newblock {\em Proceedings of the National Academy of Sciences\/}~{\em 115},
  3529--3537.

\bibitem[\protect\citeauthoryear{Watjou, Faes, Lawson, Kirby, Aregay, Carroll,
  and Vandendijck}{Watjou et~al.}{2017}]{watjou:etal:17}
Watjou, K., C.~Faes, A.~Lawson, R.~Kirby, M.~Aregay, R.~Carroll, and
  Y.~Vandendijck (2017).
\newblock Spatial small area smoothing models for handling survey data with
  nonresponse.
\newblock {\em Statistics in Medicine\/}~{\em 36}, 3708--3745.

\bibitem[\protect\citeauthoryear{{World Bank}}{{World Bank}}{2019}]{LSMS}
{World Bank} (2019).
\newblock Living standards measurement study (lsms) | surveyunit.
\newblock {http://surveys.worldbank.org/lsms}.

\bibitem[\protect\citeauthoryear{You and Zhou}{You and
  Zhou}{2011}]{you:zhou:11}
You, Y. and Q.~M. Zhou (2011).
\newblock Hierarchical {B}ayes small area estimation under a spatial model with
  application to health survey data.
\newblock {\em Survey Methodology\/}~{\em 37}, 25--37.

\end{thebibliography}


\begin{thebibliography}{}

\bibitem[\protect\citeauthoryear{Deville and Tille}{Deville and
  Tille}{1998}]{deville1998unequal}
Deville, J.-C. and Y.~Tille (1998).
\newblock Unequal probability sampling without replacement through a splitting
  method.
\newblock {\em Biometrika\/}~{\em 85}, 89--101.

\bibitem[\protect\citeauthoryear{Midzuno}{Midzuno}{1951}]{midzuno1951sampling}
Midzuno, H. (1951).
\newblock On the sampling system with probability proportionate to sum of
  sizes.
\newblock {\em Annals of the Institute of Statistical Mathematics\/}~{\em 3},
  99--107.

\end{thebibliography}

\end{document}

\title{Supplementary Material: Design- and Model-Based Approaches to Small-Area Estimation in a
Low and Middle Income Country Context: Comparisons and Recommendations}

\author{John Paige, Geir-Arne Fuglstad, Andrea Riebler, Jon Wakefield\thanks{John Paige was supported by The National Science Foundation Graduate Research Fellowship Program under award DGE-1256082, and Jon Wakefield was supported by the National Institute of Health under award R01CAO95994.}}
\date{}
\maketitle

\section*{Appendix A: Spatial Aggregation}

\subsection*{Appendix A.1: BYM2 Model}

Although spatial aggregation for the BYM2 models without cluster effects is relatively straightforward, it is less obvious how to produce county level estimates for the BYM2 models that include cluster effects. There are census estimates of the proportion of population in each county that is urban versus rural, and the number of EAs that are urban and rural within each county is also known; we will use this information when calculating county-level estimates.

It is possible to account for excess-variation due to cluster effects when producing estimates for each modeled stratum (county level estimates for BYM2\textsubscript{uC} model and county~$\times$~urban/rural for the BYM2\textsubscript{UC} model) by averaging random variation over the known number of EAs:
\begin{equation}
\hat{p}^j_S = \frac{1}{n_S} \sum_{c \ : \ s[c] = S} \hat{p}_c^j, 
\label{eq:stratumEstimate}
\end{equation}
where $\hat{p}^j_S$ is the $j$th drawn predicted probability from the posterior for a given stratum $S$, $s[c]$ is the stratum of EA $c$, $n_S$ is the number of EAs in stratum $S$, and $\hat{p}_c^j$ is the $j$th drawn predicted probability from the posterior for EA $c$. This assumes that each EA is approximately equal-sized, but if the number of EAs within any given stratum is large and their size is iid and independent of the predictive distribution for the EA, then this method will also be a good approximation to the true county level posterior. Since the number of people in each EA is not known in practice, it is unclear how to better aggregate cluster level results to the modeled stratum level. For the BYM2\textsubscript{UC} model, we can denote the two strata within each county as $U$ and $R$ for urban and rural with respective estimates $\hat{p}^j_U$ and $\hat{p}^j_R$. We can this method to generate draws from the county level posterior for the BYM2\textsubscript{uC} model and from the county~$\times$~urban/rural level for the BYM2\textsubscript{UC} model.

For the BYM2\textsubscript{UC} model, we then use, 
$$ \hat{p}^j = q_U \hat{p}_{U}^j + (1 - q_U) \hat{p}_{R}^j, $$
to sample from the  county level posterior distribution for each county, where $q_U$ is the proportion of the target population (i.e., children within the first month of life or women aged 20--29) in the county that is urban. Note that this requires knowledge of $q_U$, which might only be known approximately in practice.

\subsection*{Appendix A.2: SPDE Model}

Recall,
\begin{equation}
p_i = \int_{A_i} p(\bmx) \times q(\bmx) ~ d\bmx \approx \sum_{j=1}^{m_i} p(\bmx_j) \times q(\bmx_j).
\label{eq:popDensityIntegralcopy}
\end{equation}
We would  like to aggregate predictions over the `target' population in a county. The target population might be children within the first month of birth or women aged 20--29. In either case, it is necessary to adjust a population density surface so that it is representative of the target population.

Let $i$ represent the area for which we want to make a prediction, and $q( \cdot )$ be the population density throughout the county as a function of space. We assume the number of enumeration areas within the strata in the area is known, as is the case with the 2014 Kenya DHS at the county level. Let $C_{iU}$ and $C_{iR}$ be the number of enumeration areas in the urban and rural strata within area $i$. From the 2009 Kenya Population and Housing Census, we have an empirical distribution for the amount of our target population within urban and rural EAs in Kenya as a whole. Based on these empirical distributions, let the expected value of the number of our target population per enumeration area be $E_U$ and $E_R$ in urban and rural areas, respectively. Assuming these are constant across space, we then obtain the expected total target population in the urban and rural parts of area $i$ are, respectively, $C_{iU} \times E_U$ and $C_{iR} \times E_R$.

Ideally, we would like our population density surface, $q$, to integrate to approximately $C_{iU} \times E_U$ and $C_{iR} \times E_R$ in the urban and rural parts of area $i$ so that it is more representative of our target population. If $A_i$ is the spatial domain of area $i$, we can partition it into urban and rural parts: $A_i = A_{iU} \cup A_{iR}$. We can therefore adjust the population density surface, creating a new surface:
\begin{equation}
\tilde{q}(\bmx) = \begin{cases}
\left[ \int_{A_{iU}}q(\bmx) ~d\bmx \right]^{-1}
C_{iU} \times E_U \times  q(\bmx) & \bmx \in A_{iU}, \\
\left[ \int_{A_{iR}}q(\bmx) ~d\bmx \right]^{-1}
C_{iR} \times E_R\times q(\bmx), & \bmx \in A_{iR}.
\end{cases}
\label{eq:adjustedPopDensity}
\end{equation}

Plugging this adjusted density surface into our area level estimator, we have:
\begin{equation}
p_i = \int_{A_i} p(\bmx) \times \tilde{q}(\bmx) ~ d\bmx \approx \sum_{j=1}^{m_i} p(\bmx_j) \times \tilde{q}(\bmx_j).
\label{eq:adjustedPopDensityIntegral}
\end{equation}
This is the foundation of the county level estimator that we use for predictions for the SPDE `U' models. At this point, we may or may not marginalize out any cluster effect in order to reduce bias in the predictions.

\section*{Appendix B: Simulation Details}

\subsection*{Appendix B.1: Sampling}

Since the number of enumeration areas selected within each of the 92 strata is fixed based on the empirical number of clusters in that strata for the 2014 DHS, it is sufficient to consider the sampling weights within any given stratum. Let $H_c$ be the number of households in cluster $c$, and let $N_c$ be the number of children that we sampled within that cluster. Since we use probability proportional to size (PPS) sampling or simple random sampling (SRS) to sample the clusters for the Stratified  and Unstratified sampling designs, respectively, the probability of picking any given cluster is chosen to be either equal in the representative sampling case or proportional to the number of households within it in the PPS case. Given that $n$ clusters are sampled from a given strata with $N$ EAs in total, the probability of including cluster $c$ is either:
$$p_c = \frac{n H_c}{\sum_{c}H_c} $$
in the PPS case or
$$ p_c = \frac{n}{N} $$
in the SRS case. In the PPS case, we use Midzuno's method \citep{midzuno1951sampling, deville1998unequal} for sampling without replacement in such a way  as to match these inclusion probabilities, which in our case are all well-defined. The probability of including any given child in the sample given that the cluster was sampled is just the probability of selecting the household he/she is living in. Exactly 25 households are selected in each cluster, and so this probability is,  
$$p_{ck \vert c} = \frac{25}{H_c}, $$
where $k$ is the index of the child. Since a given child can only be selected if its respective cluster is selected, the probability of including any given child in our sample is either 
$$p_{ck} = p_c  \times p_{ck \vert c} =  \frac{25n}{\sum_{c}H_c} $$
under the Stratified design or
$$p_{ck} = p_c  \times p_{ck \vert c} =  \frac{25n}{N H_c} $$
for the Unstratified design, where the sum is taken to be over all clusters within the considered stratum. This yields a sampling weight for each child of $\frac{\sum_{c}H_c}{25n}$ or $\frac{N H_c}{25n}$ in the Stratified or Unstratified sampling cases respectively, and these can be summed over all the children sampled within any given cluster to yield respective cluster sampling weights of $N_c \sum_{c}H_c/25n$ or $N_c N H_c/25n$.

\subsection*{Appendix B.2: Additional Simulation Study Plots}

 \FloatBarrier
\begin{figure}[H]
\centering
\image{width=3in}{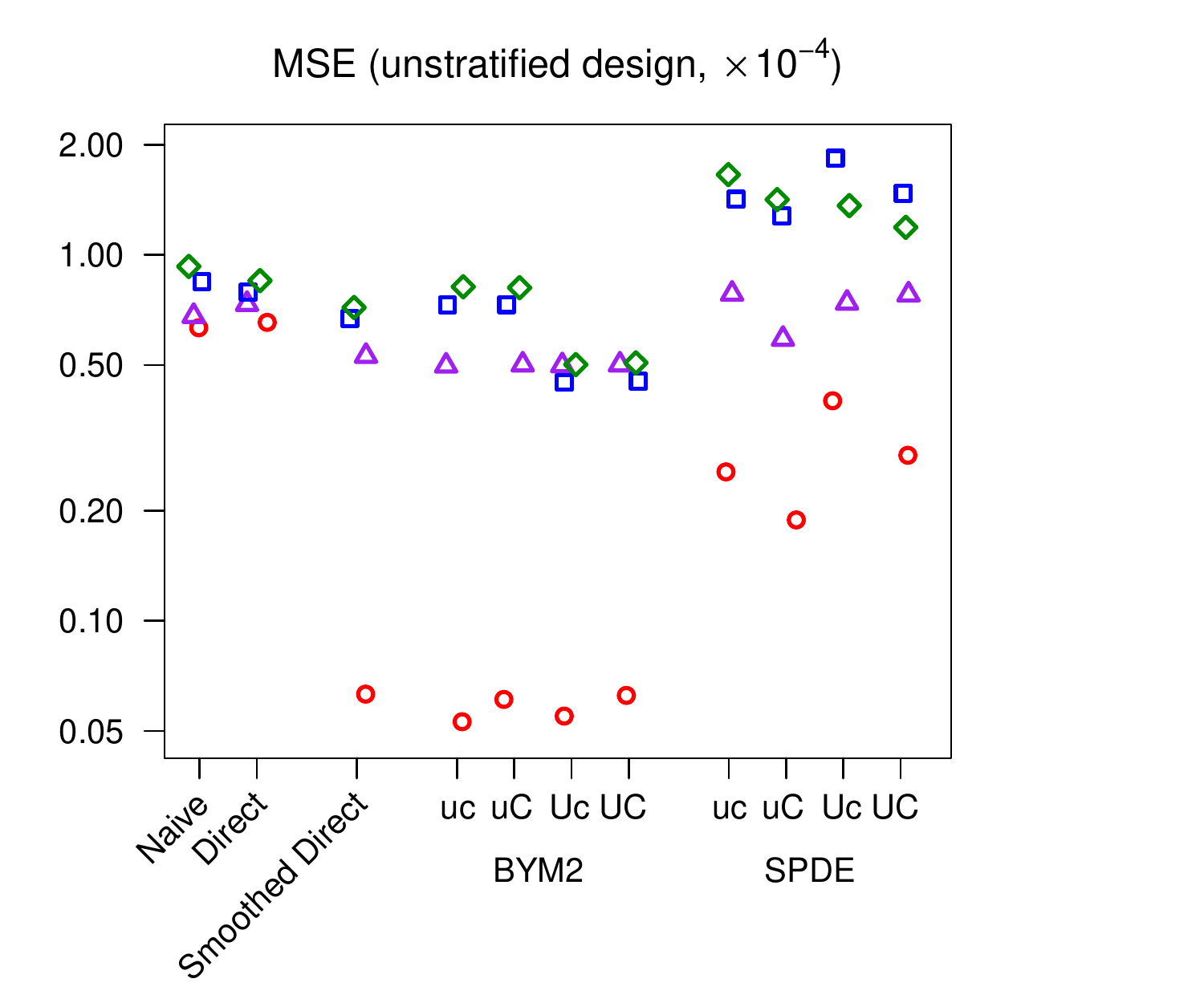} \image{width=3in}{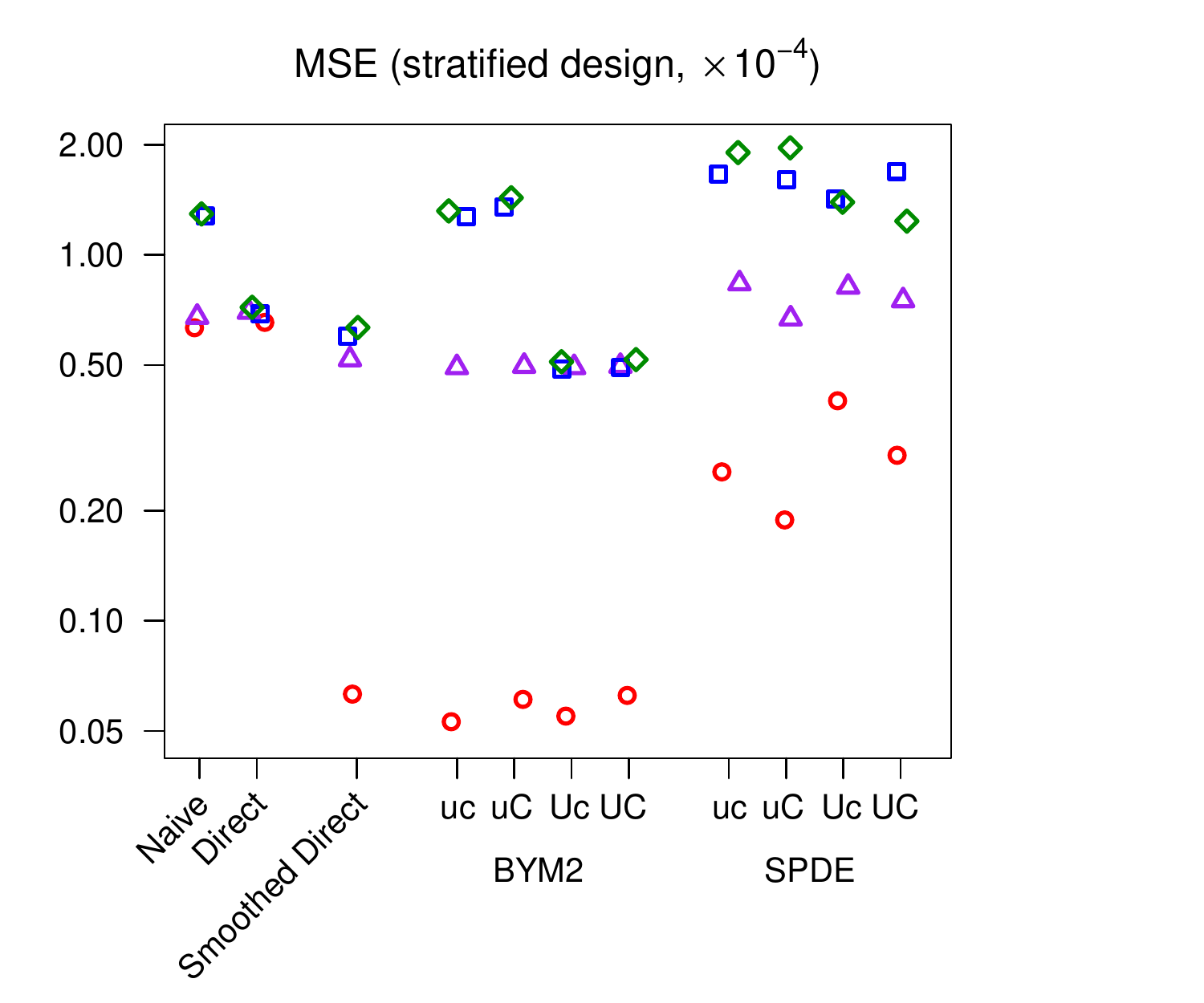} \\
\image{width=3in}{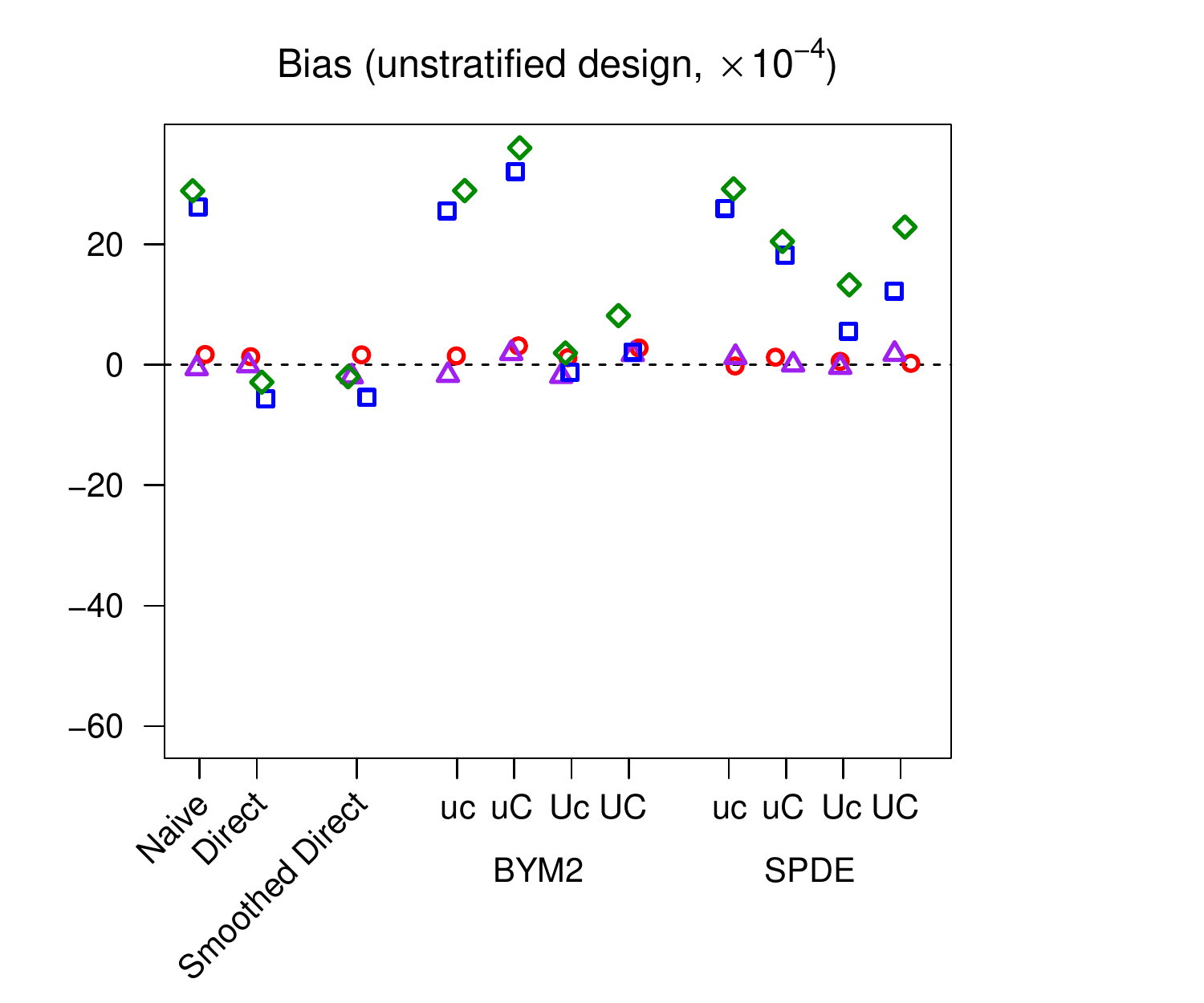} \image{width=3in}{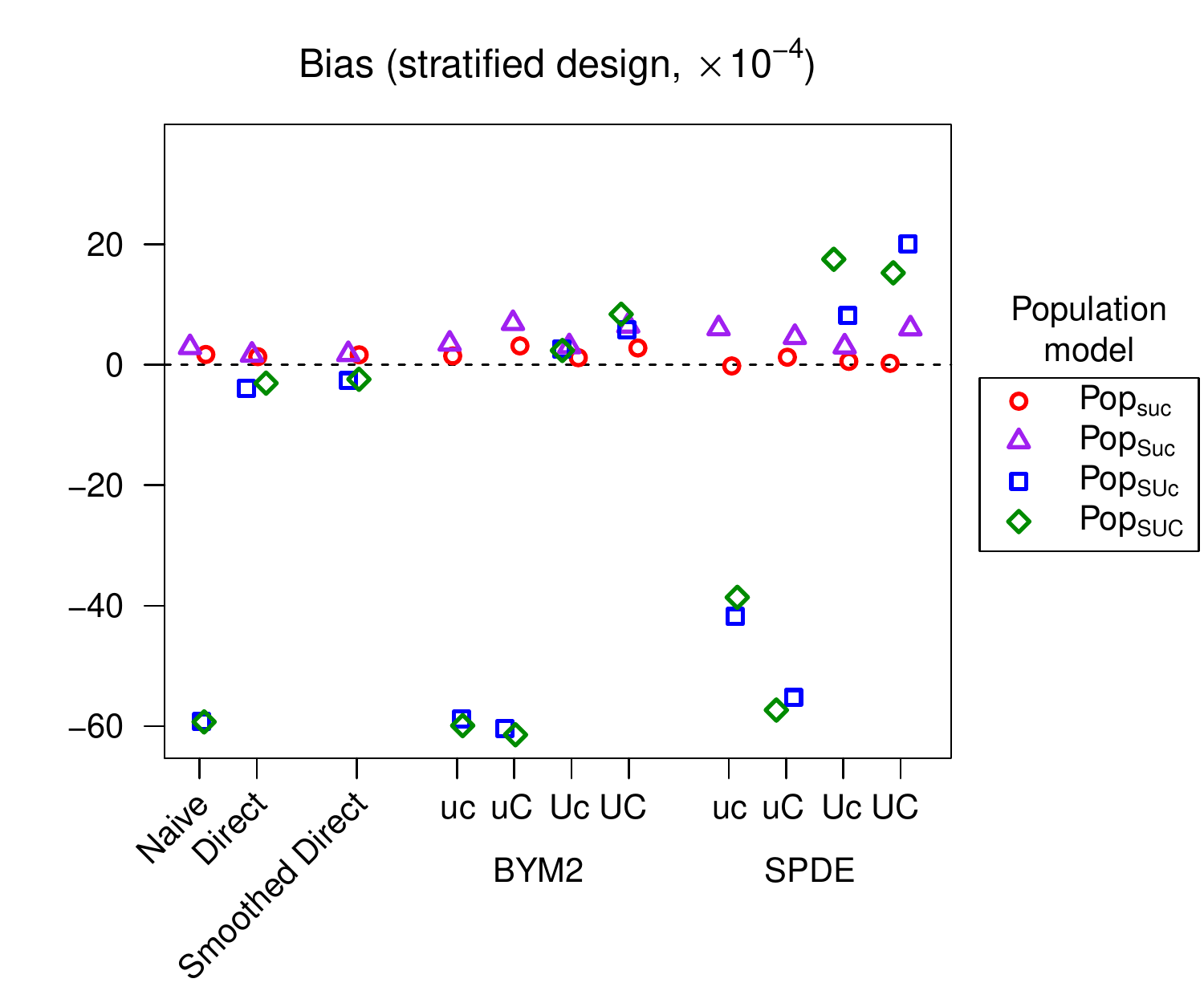}
%\image{width=3in}{figures/varPlotSRS.pdf} \image{width=3in}{figures/varPlotDHS.pdf} \\
\caption{County level scoring rules for each of the simulated populations and the main models considered. Scoring rules for the simulated unstratified and stratified surveys are respectively in the left and right columns. The labels s/S, u/U, and c/C denote whether or not spatial, urban, and cluster effects are included respectively.}
\label{fig:scoringRules1Full}
\end{figure}

\begin{figure}
\centering
\image{width=3in}{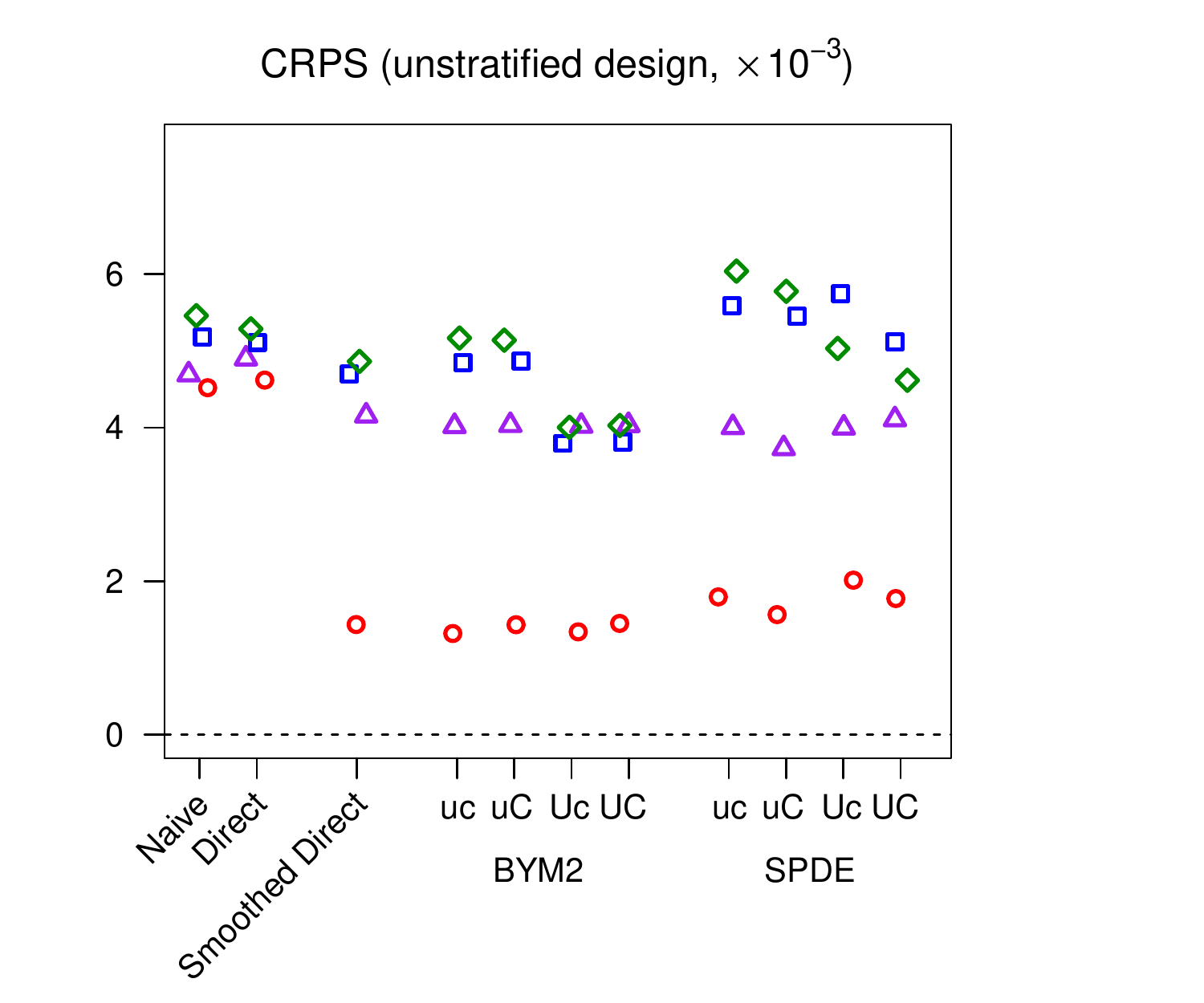} \image{width=3in}{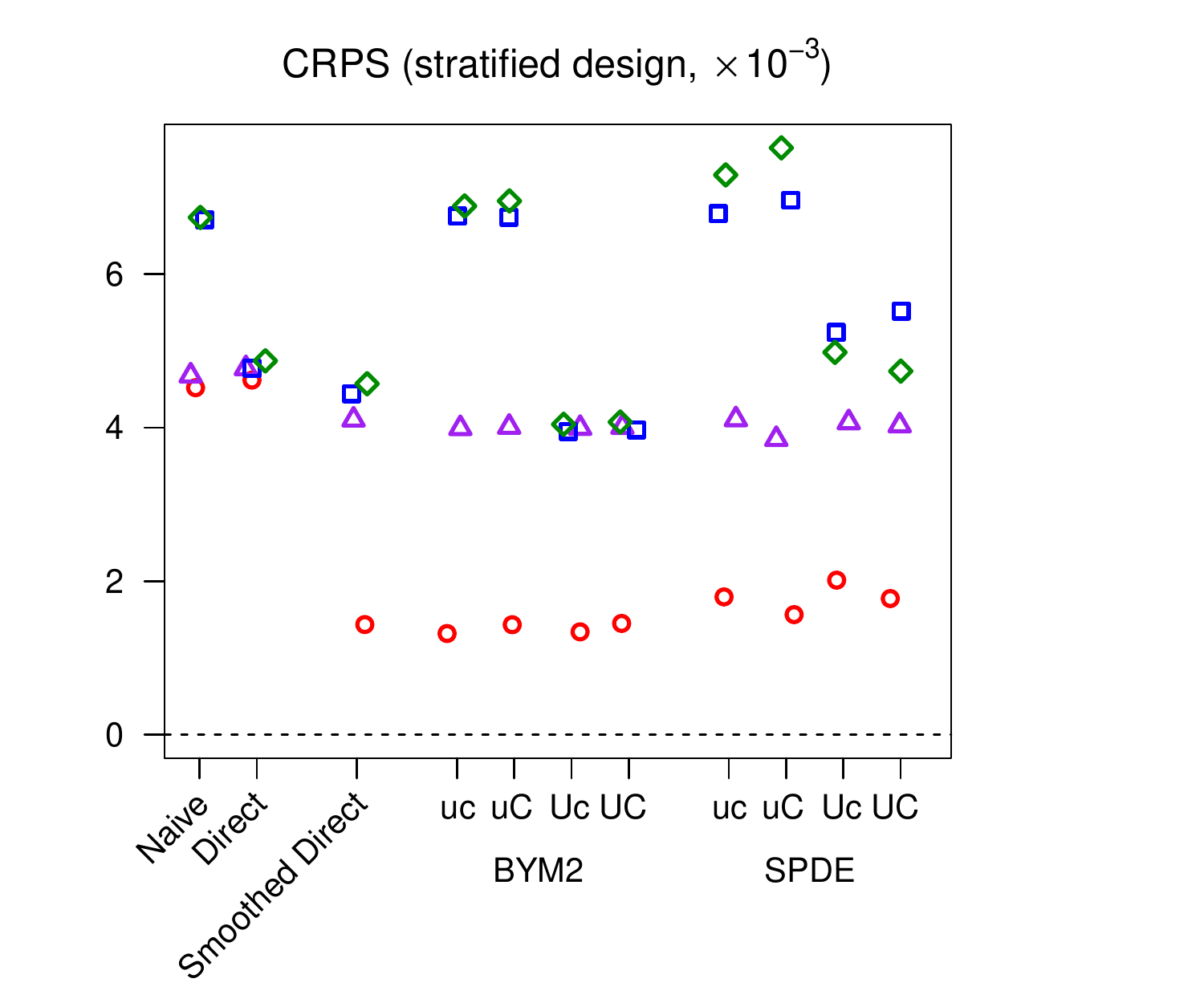} \\
\image{width=3in}{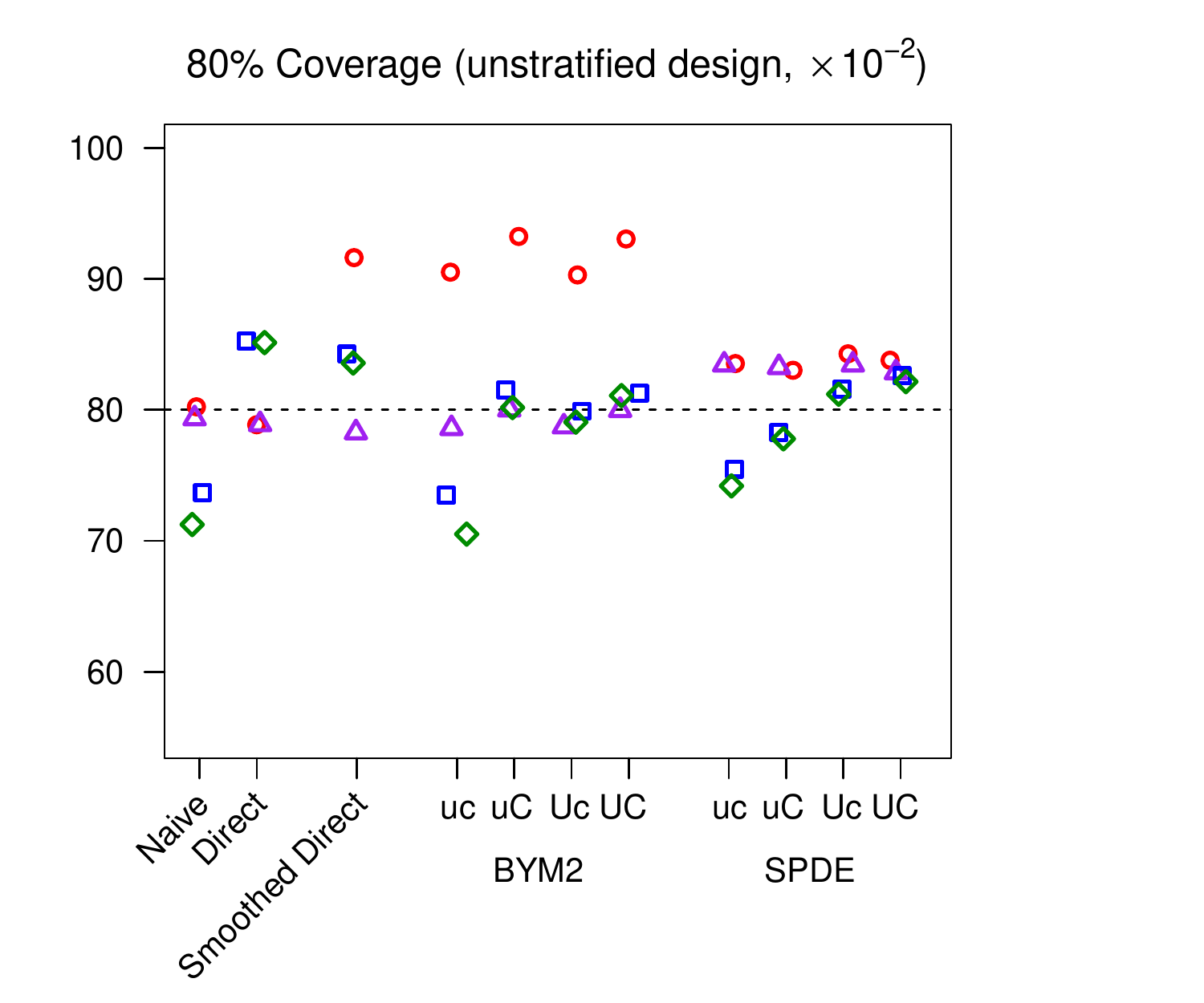} \image{width=3in}{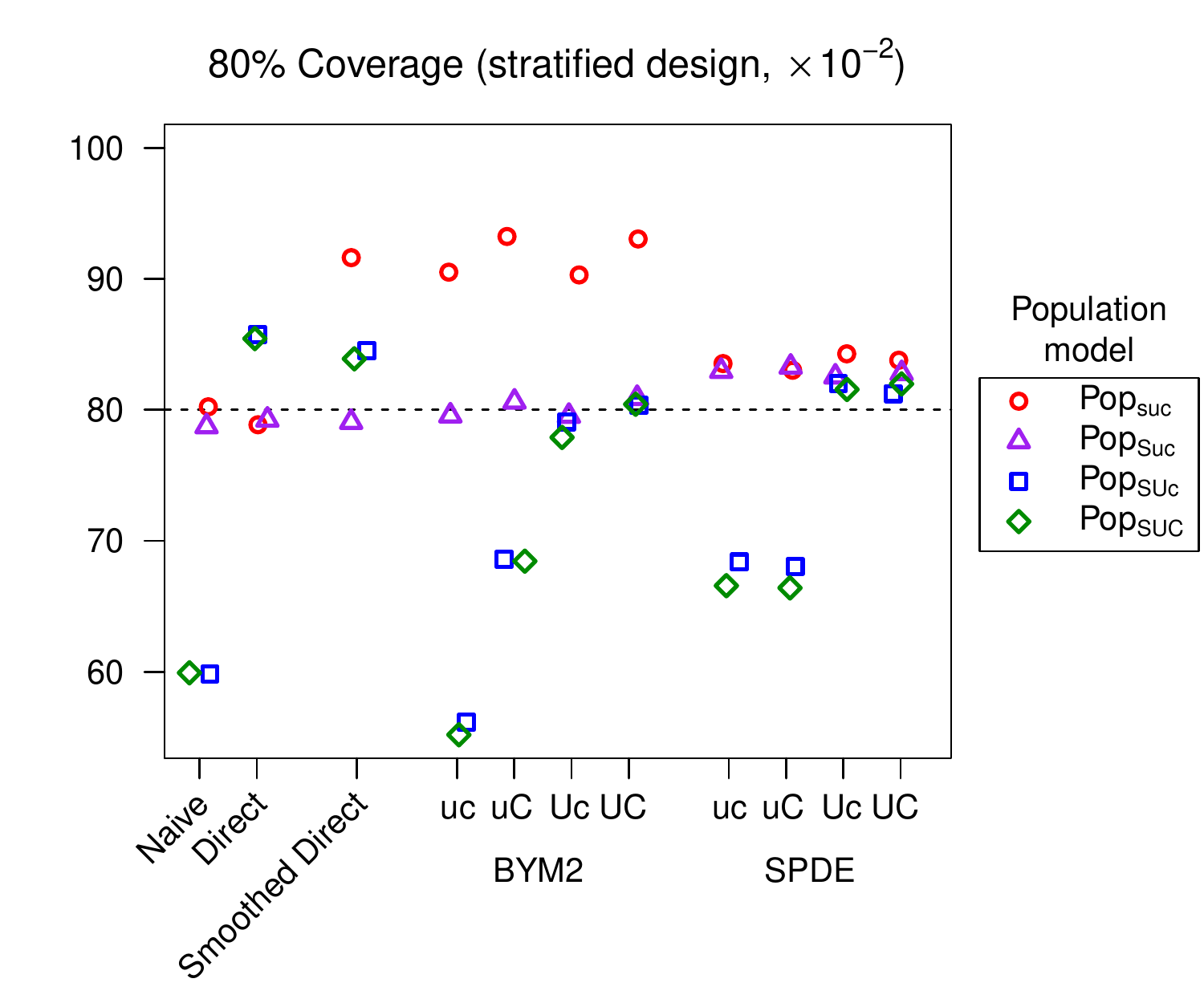} \\
\image{width=3in}{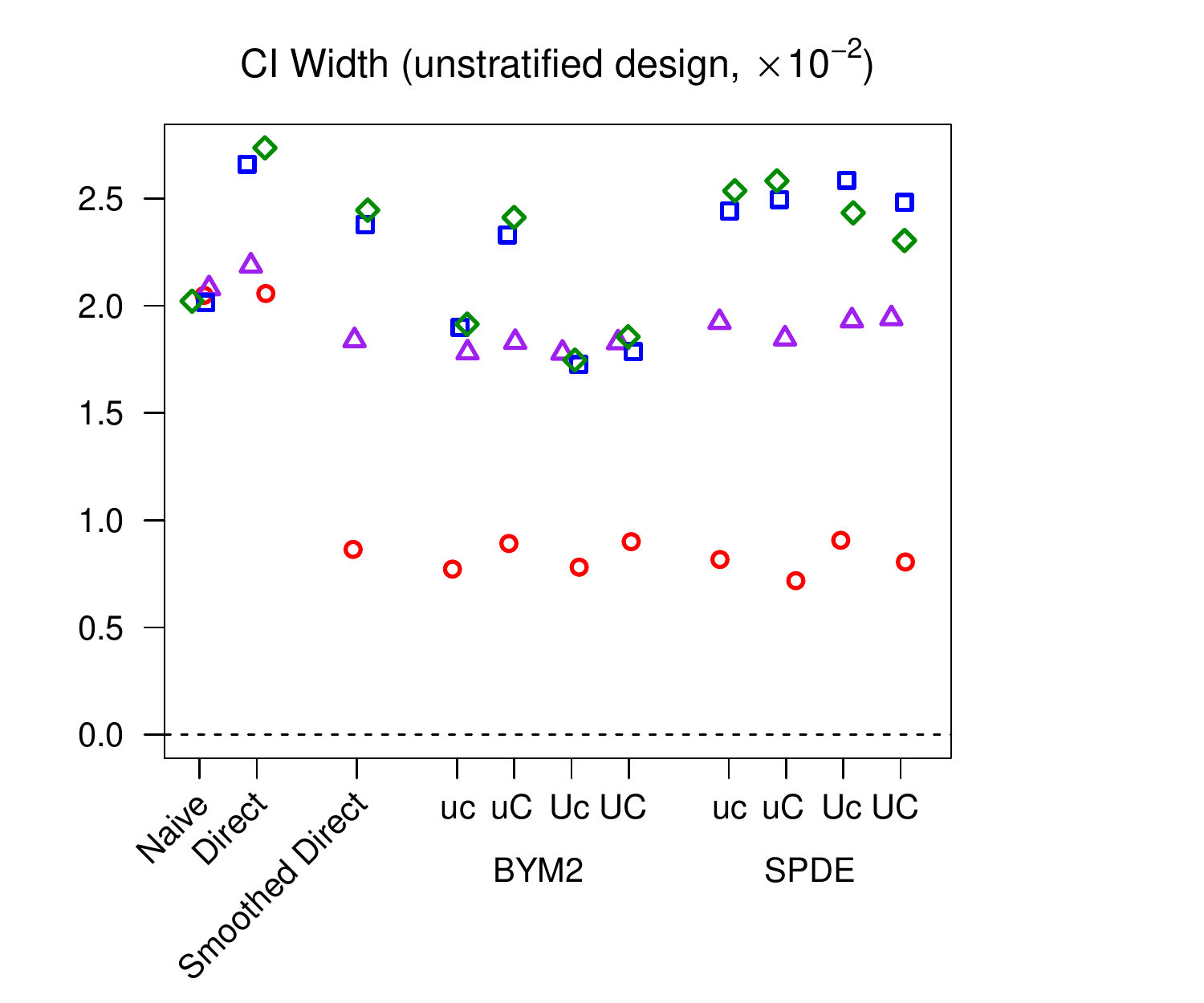} \image{width=3in}{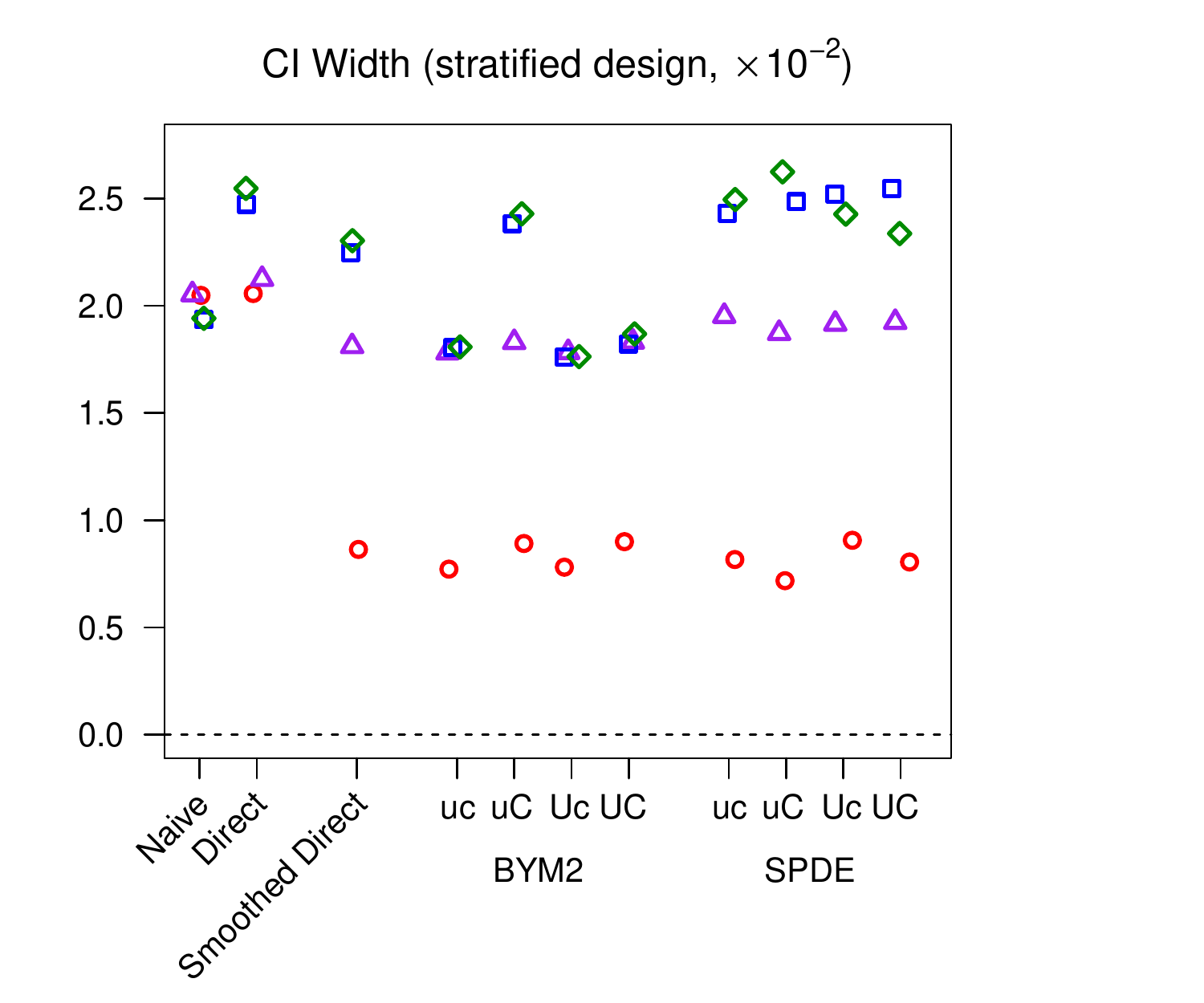} \\
\caption{County level scoring rules plotted for each of the simulated populations and the main models considered. Scoring rules for the simulated unstratified and stratified surveys are respectively plotted in the left and right columns. The labels s/S, u/U, and c/C denote whether or not spatial, urban, and cluster effects are included in the models respectively.}
\label{fig:scoringRules2Full}
\end{figure}

 \FloatBarrier
 
\subsection*{Appendix B.3: Simulation Result Tables}

\subsubsection*{Appendix B.3.1: Unstratified Sampling Results}

\begin{table}[H]
\centering
\begin{tabular}{lllllll}
\toprule
\multicolumn{1}{c}{\textbf{ }} & \multicolumn{1}{c}{\textbf{Bias}} & \multicolumn{1}{c}{\textbf{Var}} & \multicolumn{1}{c}{\textbf{MSE}} & \multicolumn{1}{c}{\textbf{CRPS}} & \multicolumn{1}{c}{\textbf{80\% Cvg}} & \multicolumn{1}{c}{\textbf{CI Width}} \\
  & ($\times 10^{-4}$) & ($\times 10^{-5}$) & ($\times 10^{-4}$) & ($\times 10^{-3}$) & ($\times 10^{-2}$) & ($\times 10^{-2}$)\\
\midrule
\addlinespace[0.3em]
\multicolumn{7}{l}{\textit{\textbf{Naive}}}\\
\hspace{1em} & \textbf{-0.6} & 6.2 & 0.64 & 4.5 & \textbf{81} & 2.08\\
\addlinespace[0.3em]
\multicolumn{7}{l}{\textit{\textbf{Direct}}}\\
\hspace{1em} & -0.7 & \em{6.7} & \em{0.69} & \em{4.7} & \textbf{79} & \em{2.11}\\
\addlinespace[0.3em]
\multicolumn{7}{l}{\textit{\textbf{Smoothed Direct}}}\\
\hspace{1em} & -2.4 & 0.48 & 0.06 & 1.4 & 90 & 0.85\\
\addlinespace[0.3em]
\multicolumn{7}{l}{\textit{\textbf{BYM2}}}\\
\hspace{1em}uc & -2.6 & \textbf{0.35} & \textbf{0.05} & \textbf{1.3} & 90 & 0.74\\
\hspace{1em}uC & -1.0 & 0.40 & \textbf{0.05} & 1.4 & \em{93} & 0.86\\
\hspace{1em}Uca & -2.5 & 0.38 & \textbf{0.05} & \textbf{1.3} & 90 & 0.76\\
\hspace{1em}UCa & -0.8 & 0.44 & 0.06 & 1.4 & \em{93} & 0.87\\
\hspace{1em}UcA & -2.5 & 0.38 & \textbf{0.05} & \textbf{1.3} & 90 & 0.76\\
\hspace{1em}UCA & -0.9 & 0.44 & 0.06 & 1.4 & \em{93} & 0.87\\
\addlinespace[0.3em]
\multicolumn{7}{l}{\textit{\textbf{SPDE}}}\\
\hspace{1em}uc & -2.6 & 3.3 & 0.35 & 1.9 & 84 & 0.82\\
\hspace{1em}uC & \em{-3.5} & 1.2 & 0.14 & 1.5 & 84 & \textbf{0.72}\\
\hspace{1em}Uc & -2.6 & 3.7 & 0.39 & 2.0 & 84 & 0.88\\
\hspace{1em}UC & \em{-3.5} & 4.0 & 0.43 & 2.0 & 83 & 0.86\\
\bottomrule
\end{tabular}
\caption{Scoring rules for all models when predictions are aggregated to the county level. Based on 250 simulated unstratified surveys for the naive and direct models and 100 for the others for the simulated population with constant risk (Pop\textsubscript{suc}). ``Cvg'' stands for coverage.}
\end{table}

\begin{longtable}[t]{lllrrrrr}
\caption{\label{tab:simulationParametersSRS_suc} Parameter summary statistics for all spatial smoothing models when predictions are aggregated to the county level. Averaged over 100 simulated unstratified surveys for the simulated population with constant risk (Pop\textsubscript{suc}).}\\
\toprule
 &   &   & Est & SD & Q10 & Q50 & Q90\\
\midrule
\endfirsthead
\caption[]{Parameter summary statistics for all spatial smoothing models when predictions are aggregated to the county level. Averaged over 100 simulated unstratified surveys for the simulated population with constant risk (Pop\textsubscript{suc}). \textit{(continued)}}\\
\toprule
 &   &   & Est & SD & Q10 & Q50 & Q90\\
\midrule
\endhead
\
\endfoot
\bottomrule
\endlastfoot
\addlinespace[0.3em]
\multicolumn{8}{l}{\textit{\textbf{Smoothed Direct}}}\\
\hspace{1em} &  & Intercept & -1.755 & 0.01 & -1.767 & -1.754 & -1.742\\

\hspace{1em} &  & Phi & 0.278 & 0.238 & 0.032 & 0.207 & 0.653\\

\hspace{1em} &  & Tot. Var & 0.00094 & 0.000744 & 0.00024 & 0.00074 & 0.00187\\

\hspace{1em} &  & Spatial Var & 0.000237 & 0.000309 & 0.000017 & 0.000126 & 0.000599\\

\hspace{1em} &  & iid Var & 0.0007 & 0.000639 & 0.00014 & 0.00052 & 0.00149\\
\cmidrule{1-8}
\addlinespace[0.3em]
\multicolumn{8}{l}{\textit{\textbf{BYM2}}}\\
\hspace{1em} & uc & Intercept & -1.755 & 0.01 & -1.767 & -1.767 & -1.742\\
\end{longtable}

\begin{table}[H]
\centering
\begin{tabular}{lllllll}
\toprule
\multicolumn{1}{c}{\textbf{ }} & \multicolumn{1}{c}{\textbf{Bias}} & \multicolumn{1}{c}{\textbf{Var}} & \multicolumn{1}{c}{\textbf{MSE}} & \multicolumn{1}{c}{\textbf{CRPS}} & \multicolumn{1}{c}{\textbf{80\% Cvg}} & \multicolumn{1}{c}{\textbf{CI Width}} \\
  & ($\times 10^{-4}$) & ($\times 10^{-5}$) & ($\times 10^{-4}$) & ($\times 10^{-3}$) & ($\times 10^{-2}$) & ($\times 10^{-2}$)\\
\midrule
\addlinespace[0.3em]
\multicolumn{7}{l}{\textit{\textbf{Naive}}}\\
\hspace{1em} & -0.6 & 6.7 & 0.68 & 4.7 & 79 & 2.08\\
\addlinespace[0.3em]
\multicolumn{7}{l}{\textit{\textbf{Direct}}}\\
\hspace{1em} & -0.1 & 7.2 & 0.73 & \em{4.9} & 79 & \em{2.19}\\
\addlinespace[0.3em]
\multicolumn{7}{l}{\textit{\textbf{Smoothed Direct}}}\\
\hspace{1em} & \em{-2.0} & 5.1 & 0.53 & 4.2 & 78 & 1.84\\
\addlinespace[0.3em]
\multicolumn{7}{l}{\textit{\textbf{BYM2}}}\\
\hspace{1em}uc & -1.7 & \textbf{4.8} & \textbf{0.50} & 4.0 & 79 & \textbf{1.78}\\
\hspace{1em}uC & 1.9 & 4.9 & \textbf{0.50} & 4.0 & \textbf{80} & 1.83\\
\hspace{1em}Uca & -1.9 & \textbf{4.8} & \textbf{0.50} & 4.0 & 79 & \textbf{1.78}\\
\hspace{1em}UCa & 1.7 & 4.9 & \textbf{0.50} & 4.0 & \textbf{80} & 1.83\\
\hspace{1em}UcA & -1.9 & \textbf{4.8} & \textbf{0.50} & 4.0 & 79 & \textbf{1.78}\\
\hspace{1em}UCA & 1.8 & 4.9 & \textbf{0.50} & 4.0 & \textbf{80} & 1.83\\
\addlinespace[0.3em]
\multicolumn{7}{l}{\textit{\textbf{SPDE}}}\\
\hspace{1em}uc & 1.3 & \em{7.6} & \em{0.78} & 4.0 & \em{83} & 1.92\\
\hspace{1em}uC & \textbf{0.0} & 5.7 & 0.59 & \textbf{3.7} & \em{83} & 1.85\\
\hspace{1em}Uc & -0.4 & 7.1 & 0.74 & 4.0 & \em{83} & 1.93\\
\hspace{1em}UC & 1.8 & 7.5 & \em{0.78} & 4.1 & \em{83} & 1.94\\
\bottomrule
\end{tabular}
\caption{Scoring rules for all models when predictions are aggregated to the county level. Based on 250 simulated unstratified surveys for the naive and direct models and 100 for the others for the simulated population without urban or cluster effects (Pop\textsubscript{Suc}). ``Cvg'' stands for coverage.}
\end{table}

\begin{longtable}[t]{lllrrrrr}
\caption{\label{tab:simulationParametersSRS_Suc} Parameter summary statistics for all spatial smoothing models when predictions are aggregated to the county level. Averaged over 100 simulated unstratified surveys for the simulated population without urban or cluster effects (Pop\textsubscript{Suc}).}\\
\toprule
 &   &   & Est & SD & Q10 & Q50 & Q90\\
\midrule
\endfirsthead
\caption[]{Parameter summary statistics for all spatial smoothing models when predictions are aggregated to the county level. Averaged over 100 simulated unstratified surveys for the simulated population without urban or cluster effects (Pop\textsubscript{Suc}). \textit{(continued)}}\\
\toprule
 &   &   & Est & SD & Q10 & Q50 & Q90\\
\midrule
\endhead
\
\endfoot
\bottomrule
\endlastfoot
\addlinespace[0.3em]
\multicolumn{8}{l}{\textit{\textbf{Smoothed Direct}}}\\
\hspace{1em} &  & Intercept & -1.748 & 0.015 & -1.766 & -1.748 & -1.729\\

\hspace{1em} &  & Phi & 0.497 & 0.249 & 0.161 & 0.494 & 0.839\\

\hspace{1em} &  & Tot. Var & 0.0134 & 0.00398 & 0.0088 & 0.0128 & 0.0186\\

\hspace{1em} &  & Spatial Var & 0.0068 & 0.00437 & 0.0018 & 0.0061 & 0.0128\\

\hspace{1em} &  & iid Var & 0.0065 & 0.00369 & 0.0021 & 0.0061 & 0.0114\\
\cmidrule{1-8}
\addlinespace[0.3em]
\multicolumn{8}{l}{\textit{\textbf{BYM2}}}\\
\hspace{1em} & uc & Intercept & -1.748 & 0.015 & -1.766 & -1.766 & -1.73\\
\end{longtable}

\begin{table}[H]
\centering
\begin{tabular}{lllllll}
\toprule
\multicolumn{1}{c}{\textbf{ }} & \multicolumn{1}{c}{\textbf{Bias}} & \multicolumn{1}{c}{\textbf{Var}} & \multicolumn{1}{c}{\textbf{MSE}} & \multicolumn{1}{c}{\textbf{CRPS}} & \multicolumn{1}{c}{\textbf{80\% Cvg}} & \multicolumn{1}{c}{\textbf{CI Width}} \\
  & ($\times 10^{-4}$) & ($\times 10^{-5}$) & ($\times 10^{-4}$) & ($\times 10^{-3}$) & ($\times 10^{-2}$) & ($\times 10^{-2}$)\\
\midrule
\addlinespace[0.3em]
\multicolumn{7}{l}{\textit{\textbf{Naive}}}\\
\hspace{1em} & 26.2 & 7.6 & 0.85 & 5.2 & \em{74} & 2.01\\
\addlinespace[0.3em]
\multicolumn{7}{l}{\textit{\textbf{Direct}}}\\
\hspace{1em} & -5.7 & 7.7 & 0.79 & 5.1 & 85 & \em{2.66}\\
\addlinespace[0.3em]
\multicolumn{7}{l}{\textit{\textbf{Smoothed Direct}}}\\
\hspace{1em} & -5.4 & 6.5 & 0.67 & 4.7 & 84 & 2.38\\
\addlinespace[0.3em]
\multicolumn{7}{l}{\textit{\textbf{BYM2}}}\\
\hspace{1em}uc & 25.6 & 6.5 & 0.73 & 4.8 & \em{74} & 1.90\\
\hspace{1em}uC & \em{32.1} & 6.1 & 0.73 & 4.9 & 82 & 2.33\\
\hspace{1em}Uca & \textbf{-1.2} & \textbf{4.3} & \textbf{0.45} & \textbf{3.8} & \textbf{80} & \textbf{1.73}\\
\hspace{1em}UCa & 2.1 & 4.4 & \textbf{0.45} & \textbf{3.8} & 81 & 1.79\\
\hspace{1em}UcA & -1.3 & \textbf{4.3} & \textbf{0.45} & \textbf{3.8} & \textbf{80} & \textbf{1.73}\\
\hspace{1em}UCA & 2.1 & 4.4 & \textbf{0.45} & \textbf{3.8} & 81 & 1.79\\
\addlinespace[0.3em]
\multicolumn{7}{l}{\textit{\textbf{SPDE}}}\\
\hspace{1em}uc & 26.0 & 13.2 & 1.42 & 5.6 & 76 & 2.44\\
\hspace{1em}uC & 18.2 & 12.1 & 1.28 & 5.5 & 78 & 2.50\\
\hspace{1em}Uc & 5.6 & \em{16.6} & \em{1.84} & \em{5.7} & 82 & 2.58\\
\hspace{1em}UC & 12.3 & 13.8 & 1.47 & 5.1 & 83 & 2.48\\
\bottomrule
\end{tabular}
\caption{Scoring rules for all models when predictions are aggregated to the county level. Based on 250 simulated unstratified surveys for the naive and direct models and 100 for the others for the simulated population with spatial and urban effects but without cluster effect (Pop\textsubscript{SUc}). ``Cvg'' stands for coverage.}
\end{table}

\begin{longtable}[t]{lllrrrrr}
\caption{\label{tab:simulationParametersSRS_SUc} Parameter summary statistics for all spatial smoothing models when predictions are aggregated to the county level. Averaged over 100 simulated unstratified surveys for the simulated population with spatial and urban effects but without cluster effect (Pop\textsubscript{SUc}).}\\
\toprule
 &   &   & Est & SD & Q10 & Q50 & Q90\\
\midrule
\endfirsthead
\caption[]{Parameter summary statistics for all spatial smoothing models when predictions are aggregated to the county level. Averaged over 100 simulated unstratified surveys for the simulated population with spatial and urban effects but without cluster effect (Pop\textsubscript{SUc}). \textit{(continued)}}\\
\toprule
 &   &   & Est & SD & Q10 & Q50 & Q90\\
\midrule
\endhead
\
\endfoot
\bottomrule
\endlastfoot
\addlinespace[0.3em]
\multicolumn{8}{l}{\textit{\textbf{Smoothed Direct}}}\\
\hspace{1em} &  & Intercept & -1.866 & 0.019 & -1.89 & -1.866 & -1.843\\

\hspace{1em} &  & Phi & 0.744 & 0.201 & 0.444 & 0.794 & 0.961\\

\hspace{1em} &  & Tot. Var & 0.0477 & 0.0127 & 0.033 & 0.0459 & 0.0645\\

\hspace{1em} &  & Spatial Var & 0.0361 & 0.0152 & 0.0174 & 0.0349 & 0.0561\\

\hspace{1em} &  & iid Var & 0.0116 & 0.0092 & 0.0019 & 0.0093 & 0.0246\\
\cmidrule{1-8}
\addlinespace[0.3em]
\multicolumn{8}{l}{\textit{\textbf{BYM2}}}\\
\hspace{1em} & uc & Intercept & -1.841 & 0.017 & -1.862 & -1.862 & -1.821\\
\end{longtable}

\begin{table}[H]
\centering
\begin{tabular}{lllllll}
\toprule
\multicolumn{1}{c}{\textbf{ }} & \multicolumn{1}{c}{\textbf{Bias}} & \multicolumn{1}{c}{\textbf{Var}} & \multicolumn{1}{c}{\textbf{MSE}} & \multicolumn{1}{c}{\textbf{CRPS}} & \multicolumn{1}{c}{\textbf{80\% Cvg}} & \multicolumn{1}{c}{\textbf{CI Width}} \\
  & ($\times 10^{-4}$) & ($\times 10^{-5}$) & ($\times 10^{-4}$) & ($\times 10^{-3}$) & ($\times 10^{-2}$) & ($\times 10^{-2}$)\\
\midrule
\addlinespace[0.3em]
\multicolumn{7}{l}{\textit{\textbf{Naive}}}\\
\hspace{1em} & 28.9 & 8.3 & 0.93 & 5.5 & 71 & 2.02\\
\addlinespace[0.3em]
\multicolumn{7}{l}{\textit{\textbf{Direct}}}\\
\hspace{1em} & -2.9 & 8.3 & 0.85 & 5.3 & 85 & \em{2.74}\\
\addlinespace[0.3em]
\multicolumn{7}{l}{\textit{\textbf{Smoothed Direct}}}\\
\hspace{1em} & -2.0 & 7.0 & 0.72 & 4.9 & 84 & 2.45\\
\addlinespace[0.3em]
\multicolumn{7}{l}{\textit{\textbf{BYM2}}}\\
\hspace{1em}uc & 28.9 & 7.2 & 0.82 & 5.2 & \em{70} & 1.91\\
\hspace{1em}uC & \em{36.0} & 6.7 & 0.81 & 5.1 & \textbf{80} & 2.41\\
\hspace{1em}Uca & 2.0 & \textbf{4.8} & \textbf{0.50} & \textbf{4.0} & 79 & \textbf{1.75}\\
\hspace{1em}UCa & 8.2 & \textbf{4.8} & 0.51 & \textbf{4.0} & 81 & 1.86\\
\hspace{1em}UcA & \textbf{1.9} & \textbf{4.8} & \textbf{0.50} & \textbf{4.0} & 79 & \textbf{1.75}\\
\hspace{1em}UCA & 8.2 & \textbf{4.8} & 0.51 & \textbf{4.0} & 81 & 1.85\\
\addlinespace[0.3em]
\multicolumn{7}{l}{\textit{\textbf{SPDE}}}\\
\hspace{1em}uc & 29.2 & \em{15.5} & \em{1.66} & \em{6.0} & 74 & 2.54\\
\hspace{1em}uC & 20.5 & 13.3 & 1.42 & 5.8 & 78 & 2.58\\
\hspace{1em}Uc & 13.3 & 11.9 & 1.36 & 5.0 & 81 & 2.43\\
\hspace{1em}UC & 22.9 & 10.9 & 1.19 & 4.6 & 82 & 2.30\\
\bottomrule
\end{tabular}
\caption{Scoring rules for all models when predictions are aggregated to the county level. Based on 250 simulated unstratified surveys for the naive and direct models and 100 for the others for the simulated population with all effects present in the risk model (Pop\textsubscript{SUC}). ``Cvg'' stands for coverage.}
\end{table}

\begin{longtable}[t]{lllrrrrr}
\caption{\label{tab:simulationParametersSRS_SUC} Parameter summary statistics for all spatial smoothing models when predictions are aggregated to the county level. Averaged over 100 simulated unstratified surveys for the simulated population all effects present in the risk model.}\\
\toprule
 &   &   & Est & SD & Q10 & Q50 & Q90\\
\midrule
\endfirsthead
\caption[]{Parameter summary statistics for all spatial smoothing models when predictions are aggregated to the county level. Averaged over 100 simulated unstratified surveys for the simulated population all effects present in the risk model. \textit{(continued)}}\\
\toprule
 &   &   & Est & SD & Q10 & Q50 & Q90\\
\midrule
\endhead
\
\endfoot
\bottomrule
\endlastfoot
\addlinespace[0.3em]
\multicolumn{8}{l}{\textit{\textbf{Smoothed Direct}}}\\
\hspace{1em} &  & Intercept & -1.855 & 0.02 & -1.879 & -1.855 & -1.831\\

\hspace{1em} &  & Phi & 0.737 & 0.202 & 0.435 & 0.785 & 0.958\\

\hspace{1em} &  & Tot. Var & 0.0489 & 0.0131 & 0.0339 & 0.0471 & 0.0662\\

\hspace{1em} &  & Spatial Var & 0.0366 & 0.0156 & 0.0176 & 0.0354 & 0.0571\\

\hspace{1em} &  & iid Var & 0.0123 & 0.00958 & 0.0021 & 0.01 & 0.0257\\
\cmidrule{1-8}
\addlinespace[0.3em]
\multicolumn{8}{l}{\textit{\textbf{BYM2}}}\\
\hspace{1em} & uc & Intercept & -1.83 & 0.018 & -1.852 & -1.852 & -1.809\\
\end{longtable}

 \FloatBarrier
\subsection*{Appendix B.3.2: Stratified Sampling Results}
 \FloatBarrier

\begin{table}[H]
\centering
\begin{tabular}{lllllll}
\toprule
\multicolumn{1}{c}{\textbf{ }} & \multicolumn{1}{c}{\textbf{Bias}} & \multicolumn{1}{c}{\textbf{Var}} & \multicolumn{1}{c}{\textbf{MSE}} & \multicolumn{1}{c}{\textbf{CRPS}} & \multicolumn{1}{c}{\textbf{80\% Cvg}} & \multicolumn{1}{c}{\textbf{CI Width}} \\
  & ($\times 10^{-4}$) & ($\times 10^{-5}$) & ($\times 10^{-4}$) & ($\times 10^{-3}$) & ($\times 10^{-2}$) & ($\times 10^{-2}$)\\
\midrule
\addlinespace[0.3em]
\multicolumn{7}{l}{\textit{\textbf{Naive}}}\\
\hspace{1em} & 1.7 & 6.2 & 0.63 & 4.5 & \textbf{80} & 2.05\\
\addlinespace[0.3em]
\multicolumn{7}{l}{\textit{\textbf{Direct}}}\\
\hspace{1em} & 1.3 & \em{6.4} & \em{0.65} & \em{4.6} & 79 & \em{2.06}\\
\addlinespace[0.3em]
\multicolumn{7}{l}{\textit{\textbf{Smoothed Direct}}}\\
\hspace{1em} & 1.6 & 0.48 & 0.06 & 1.4 & 92 & 0.86\\
\addlinespace[0.3em]
\multicolumn{7}{l}{\textit{\textbf{BYM2}}}\\
\hspace{1em}uc & 1.5 & \textbf{0.39} & \textbf{0.05} & \textbf{1.3} & 90 & 0.77\\
\hspace{1em}uC & \em{3.1} & 0.45 & 0.06 & 1.4 & \em{93} & 0.89\\
\hspace{1em}Uca & 1.2 & 0.41 & \textbf{0.05} & \textbf{1.3} & 90 & 0.78\\
\hspace{1em}UCa & 2.8 & 0.48 & 0.06 & 1.4 & \em{93} & 0.90\\
\hspace{1em}UcA & 1.2 & 0.41 & \textbf{0.05} & \textbf{1.3} & 90 & 0.78\\
\hspace{1em}UCA & 2.8 & 0.48 & 0.06 & 1.4 & \em{93} & 0.90\\
\addlinespace[0.3em]
\multicolumn{7}{l}{\textit{\textbf{SPDE}}}\\
\hspace{1em}uc & \textbf{-0.2} & 2.3 & 0.26 & 1.8 & 84 & 0.82\\
\hspace{1em}uC & 1.2 & 1.7 & 0.19 & 1.6 & 83 & \textbf{0.72}\\
\hspace{1em}Uc & 0.6 & 3.8 & 0.40 & 2.0 & 84 & 0.91\\
\hspace{1em}UC & \textbf{0.2} & 2.6 & 0.28 & 1.8 & 84 & 0.81\\
\bottomrule
\end{tabular}
\caption{Scoring rules for all models when predictions are aggregated to the county level. Based on 250 simulated stratified surveys for the naive and direct models and 100 for the others for the simulated population with constant risk (Pop\textsubscript{suc}). ``Cvg'' stands for coverage.}
\end{table}

\begin{longtable}[t]{lllrrrrr}
\caption{\label{tab:simulationParametersDHS_suc} Parameter summary statistics for all spatial smoothing models when predictions are aggregated to the county level. Averaged over 100 simulated stratified surveys for the simulated population with constant risk (Pop\textsubscript{suc}).}\\
\toprule
 &   &   & Est & SD & Q10 & Q50 & Q90\\
\midrule
\endfirsthead
\caption[]{Parameter summary statistics for all spatial smoothing models when predictions are aggregated to the county level. Averaged over 100 simulated stratified surveys for the simulated population with constant risk (Pop\textsubscript{suc}). \textit{(continued)}}\\
\toprule
 &   &   & Est & SD & Q10 & Q50 & Q90\\
\midrule
\endhead
\
\endfoot
\bottomrule
\endlastfoot
\addlinespace[0.3em]
\multicolumn{8}{l}{\textit{\textbf{Smoothed Direct}}}\\
\hspace{1em} &  & Intercept & -1.751 & 0.01 & -1.764 & -1.751 & -1.739\\

\hspace{1em} &  & Phi & 0.274 & 0.235 & 0.032 & 0.204 & 0.642\\

\hspace{1em} &  & Tot. Var & 0.00099 & 0.00076 & 0.00026 & 0.00079 & 0.00195\\

\hspace{1em} &  & Spatial Var & 0.000244 & 0.000317 & 0.000017 & 0.00013 & 0.000614\\

\hspace{1em} &  & iid Var & 0.00074 & 0.000655 & 0.00015 & 0.00055 & 0.00157\\
\cmidrule{1-8}
\addlinespace[0.3em]
\multicolumn{8}{l}{\textit{\textbf{BYM2}}}\\
\hspace{1em} & uc & Intercept & -1.752 & 0.01 & -1.764 & -1.764 & -1.739\\
\end{longtable}

\begin{table}[H]
\centering
\begin{tabular}{lllllll}
\toprule
\multicolumn{1}{c}{\textbf{ }} & \multicolumn{1}{c}{\textbf{Bias}} & \multicolumn{1}{c}{\textbf{Var}} & \multicolumn{1}{c}{\textbf{MSE}} & \multicolumn{1}{c}{\textbf{CRPS}} & \multicolumn{1}{c}{\textbf{80\% Cvg}} & \multicolumn{1}{c}{\textbf{CI Width}} \\
  & ($\times 10^{-4}$) & ($\times 10^{-5}$) & ($\times 10^{-4}$) & ($\times 10^{-3}$) & ($\times 10^{-2}$) & ($\times 10^{-2}$)\\
\midrule
\addlinespace[0.3em]
\multicolumn{7}{l}{\textit{\textbf{Naive}}}\\
\hspace{1em} & 2.9 & 6.6 & 0.68 & 4.7 & 79 & 2.05\\
\addlinespace[0.3em]
\multicolumn{7}{l}{\textit{\textbf{Direct}}}\\
\hspace{1em} & \textbf{1.6} & 6.8 & 0.70 & \em{4.8} & 79 & \em{2.12}\\
\addlinespace[0.3em]
\multicolumn{7}{l}{\textit{\textbf{Smoothed Direct}}}\\
\hspace{1em} & 1.8 & 5.0 & 0.52 & 4.1 & 79 & 1.81\\
\addlinespace[0.3em]
\multicolumn{7}{l}{\textit{\textbf{BYM2}}}\\
\hspace{1em}uc & 3.5 & \textbf{4.8} & \textbf{0.49} & 4.0 & \textbf{80} & \textbf{1.78}\\
\hspace{1em}uC & \em{6.9} & \textbf{4.8} & 0.50 & 4.0 & 81 & 1.83\\
\hspace{1em}Uca & 3.0 & \textbf{4.8} & \textbf{0.49} & 4.0 & \textbf{80} & \textbf{1.78}\\
\hspace{1em}UCa & 6.5 & \textbf{4.8} & 0.50 & 4.0 & 81 & 1.83\\
\hspace{1em}UcA & 3.0 & \textbf{4.8} & \textbf{0.49} & 4.0 & \textbf{80} & \textbf{1.78}\\
\hspace{1em}UCA & 6.5 & \textbf{4.8} & 0.50 & 4.0 & 81 & 1.83\\
\addlinespace[0.3em]
\multicolumn{7}{l}{\textit{\textbf{SPDE}}}\\
\hspace{1em}uc & 6.0 & \em{8.1} & \em{0.84} & 4.1 & \em{83} & 1.95\\
\hspace{1em}uC & 4.6 & 6.4 & 0.67 & \textbf{3.8} & \em{83} & 1.87\\
\hspace{1em}Uc & 3.0 & 7.8 & 0.82 & 4.1 & 82 & 1.92\\
\hspace{1em}UC & 6.0 & 7.3 & 0.75 & 4.0 & \em{83} & 1.92\\
\bottomrule
\end{tabular}
\caption{Scoring rules for all models when predictions are aggregated to the county level. Based on 250 simulated stratified surveys for the naive and direct models and 100 for the others for the simulated population without urban or cluster effects (Pop\textsubscript{Suc}). ``Cvg'' stands for coverage.}
\end{table}

\begin{longtable}[t]{lllrrrrr}
\caption{\label{tab:simulationParametersDHS_Suc} Parameter summary statistics for all spatial smoothing models when predictions are aggregated to the county level. Averaged over 100 simulated stratified surveys for the simulated population without urban or cluster effects (Pop\textsubscript{Suc}).}\\
\toprule
 &   &   & Est & SD & Q10 & Q50 & Q90\\
\midrule
\endfirsthead
\caption[]{Parameter summary statistics for all spatial smoothing models when predictions are aggregated to the county level. Averaged over 100 simulated stratified surveys for the simulated population without urban or cluster effects (Pop\textsubscript{Suc}). \textit{(continued)}}\\
\toprule
 &   &   & Est & SD & Q10 & Q50 & Q90\\
\midrule
\endhead
\
\endfoot
\bottomrule
\endlastfoot
\addlinespace[0.3em]
\multicolumn{8}{l}{\textit{\textbf{Smoothed Direct}}}\\
\hspace{1em} &  & Intercept & -1.745 & 0.015 & -1.763 & -1.745 & -1.726\\

\hspace{1em} &  & Phi & 0.499 & 0.249 & 0.163 & 0.496 & 0.841\\

\hspace{1em} &  & Tot. Var & 0.0137 & 0.00403 & 0.0091 & 0.0132 & 0.0191\\

\hspace{1em} &  & Spatial Var & 0.007 & 0.00446 & 0.0019 & 0.0063 & 0.0131\\

\hspace{1em} &  & iid Var & 0.0067 & 0.00376 & 0.0021 & 0.0063 & 0.0117\\
\cmidrule{1-8}
\addlinespace[0.3em]
\multicolumn{8}{l}{\textit{\textbf{BYM2}}}\\
\hspace{1em} & uc & Intercept & -1.745 & 0.015 & -1.763 & -1.763 & -1.726\\
\end{longtable}

\begin{table}[H]
\centering
\begin{tabular}{lllllll}
\toprule
\multicolumn{1}{c}{\textbf{ }} & \multicolumn{1}{c}{\textbf{Bias}} & \multicolumn{1}{c}{\textbf{Var}} & \multicolumn{1}{c}{\textbf{MSE}} & \multicolumn{1}{c}{\textbf{CRPS}} & \multicolumn{1}{c}{\textbf{80\% Cvg}} & \multicolumn{1}{c}{\textbf{CI Width}} \\
  & ($\times 10^{-4}$) & ($\times 10^{-5}$) & ($\times 10^{-4}$) & ($\times 10^{-3}$) & ($\times 10^{-2}$) & ($\times 10^{-2}$)\\
\midrule
\addlinespace[0.3em]
\multicolumn{7}{l}{\textit{\textbf{Naive}}}\\
\hspace{1em} & -59.3 & 9.2 & 1.28 & 6.7 & 60 & 1.93\\
\addlinespace[0.3em]
\multicolumn{7}{l}{\textit{\textbf{Direct}}}\\
\hspace{1em} & -4.0 & 6.7 & 0.69 & 4.8 & 86 & 2.47\\
\addlinespace[0.3em]
\multicolumn{7}{l}{\textit{\textbf{Smoothed Direct}}}\\
\hspace{1em} & \textbf{-2.6} & 5.9 & 0.6 & 4.4 & 84 & 2.25\\
\addlinespace[0.3em]
\multicolumn{7}{l}{\textit{\textbf{BYM2}}}\\
\hspace{1em}uc & -58.8 & 9.1 & 1.27 & 6.8 & \em{56} & 1.80\\
\hspace{1em}uC & \em{-60.4} & 9.8 & 1.35 & 6.7 & 69 & 2.38\\
\hspace{1em}Uca & \textbf{2.6} & \textbf{4.7} & \textbf{0.49} & \textbf{4.0} & 79 & \textbf{1.76}\\
\hspace{1em}UCa & 5.8 & 4.8 & \textbf{0.49} & \textbf{4.0} & \textbf{80} & 1.82\\
\hspace{1em}UcA & \textbf{2.6} & \textbf{4.7} & \textbf{0.49} & \textbf{4.0} & 79 & \textbf{1.76}\\
\hspace{1em}UCA & 5.8 & 4.8 & \textbf{0.49} & \textbf{4.0} & \textbf{80} & 1.82\\
\addlinespace[0.3em]
\multicolumn{7}{l}{\textit{\textbf{SPDE}}}\\
\hspace{1em}uc & -41.8 & 14.6 & 1.66 & 6.8 & 68 & 2.43\\
\hspace{1em}uC & -55.2 & 12.8 & 1.61 & \em{7.0} & 68 & 2.49\\
\hspace{1em}Uc & 8.2 & 12.4 & 1.42 & 5.2 & 82 & 2.52\\
\hspace{1em}UC & 20.1 & \em{15.1} & \em{1.69} & 5.5 & 81 & \em{2.55}\\
\bottomrule
\end{tabular}
\caption{Scoring rules for all models when predictions are aggregated to the county level. Based on 250 simulated stratified surveys for the naive and direct models and 100 for the others for the simulated population with spatial and urban effects but without cluster effect (Pop\textsubscript{SUc}). ``Cvg'' stands for coverage.}
\end{table}

\begin{longtable}[t]{lllrrrrr}
\caption{\label{tab:simulationParametersDHS_SUc} Parameter summary statistics for all spatial smoothing models when predictions are aggregated to the county level. Averaged over 100 simulated stratified surveys for the simulated population with spatial and urban effects but without cluster effect (Pop\textsubscript{SUc}).}\\
\toprule
 &   &   & Est & SD & Q10 & Q50 & Q90\\
\midrule
\endfirsthead
\caption[]{Parameter summary statistics for all spatial smoothing models when predictions are aggregated to the county level. Averaged over 100 simulated stratified surveys for the simulated population with spatial and urban effects but without cluster effect (Pop\textsubscript{SUc}). \textit{(continued)}}\\
\toprule
 &   &   & Est & SD & Q10 & Q50 & Q90\\
\midrule
\endhead
\
\endfoot
\bottomrule
\endlastfoot
\addlinespace[0.3em]
\multicolumn{8}{l}{\textit{\textbf{Smoothed Direct}}}\\
\hspace{1em} &  & Intercept & -1.864 & 0.019 & -1.887 & -1.864 & -1.842\\

\hspace{1em} &  & Phi & 0.745 & 0.201 & 0.444 & 0.796 & 0.962\\

\hspace{1em} &  & Tot. Var & 0.0494 & 0.0129 & 0.0345 & 0.0476 & 0.0665\\

\hspace{1em} &  & Spatial Var & 0.0375 & 0.0157 & 0.018 & 0.0364 & 0.0579\\

\hspace{1em} &  & iid Var & 0.0119 & 0.00942 & 0.002 & 0.0096 & 0.0252\\
\cmidrule{1-8}
\addlinespace[0.3em]
\multicolumn{8}{l}{\textit{\textbf{BYM2}}}\\
\hspace{1em} & uc & Intercept & -1.912 & 0.016 & -1.931 & -1.931 & -1.893\\
\end{longtable}

\begin{table}[H]
\centering
\begin{tabular}{lllllll}
\toprule
\multicolumn{1}{c}{\textbf{ }} & \multicolumn{1}{c}{\textbf{Bias}} & \multicolumn{1}{c}{\textbf{Var}} & \multicolumn{1}{c}{\textbf{MSE}} & \multicolumn{1}{c}{\textbf{CRPS}} & \multicolumn{1}{c}{\textbf{80\% Cvg}} & \multicolumn{1}{c}{\textbf{CI Width}} \\
  & ($\times 10^{-4}$) & ($\times 10^{-5}$) & ($\times 10^{-4}$) & ($\times 10^{-3}$) & ($\times 10^{-2}$) & ($\times 10^{-2}$)\\
\midrule
\addlinespace[0.3em]
\multicolumn{7}{l}{\textit{\textbf{Naive}}}\\
\hspace{1em} & -59.3 & 9.3 & 1.29 & 6.7 & 60 & 1.94\\
\addlinespace[0.3em]
\multicolumn{7}{l}{\textit{\textbf{Direct}}}\\
\hspace{1em} & -3.0 & 7.0 & 0.72 & 4.9 & 85 & 2.55\\
\addlinespace[0.3em]
\multicolumn{7}{l}{\textit{\textbf{Smoothed Direct}}}\\
\hspace{1em} & \textbf{-2.4} & 6.2 & 0.63 & 4.6 & 84 & 2.30\\
\addlinespace[0.3em]
\multicolumn{7}{l}{\textit{\textbf{BYM2}}}\\
\hspace{1em}uc & -59.9 & 9.5 & 1.32 & 6.9 & \em{55} & 1.81\\
\hspace{1em}uC & \em{-61.4} & 10.4 & 1.43 & 7.0 & 68 & 2.43\\
\hspace{1em}Uca & \textbf{2.4} & \textbf{4.9} & \textbf{0.51} & \textbf{4.0} & 78 & \textbf{1.76}\\
\hspace{1em}UCa & 8.4 & \textbf{4.9} & 0.52 & 4.1 & \textbf{80} & 1.87\\
\hspace{1em}UcA & \textbf{2.4} & \textbf{4.9} & \textbf{0.51} & \textbf{4.0} & 78 & \textbf{1.76}\\
\hspace{1em}UCA & 8.4 & \textbf{4.9} & 0.52 & 4.1 & \textbf{80} & 1.87\\
\addlinespace[0.3em]
\multicolumn{7}{l}{\textit{\textbf{SPDE}}}\\
\hspace{1em}uc & -38.6 & \em{17.2} & 1.90 & 7.3 & 67 & 2.50\\
\hspace{1em}uC & -57.3 & 16.0 & \em{1.96} & \em{7.6} & 66 & \em{2.62}\\
\hspace{1em}Uc & 17.5 & 12.9 & 1.39 & 5.0 & 82 & 2.43\\
\hspace{1em}UC & 15.3 & 10.9 & 1.24 & 4.7 & 82 & 2.34\\
\bottomrule
\end{tabular}
\caption{Scoring rules for all models when predictions are aggregated to the county level. Based on 250 simulated stratified surveys for the naive and direct models and 100 for the others for the simulated population with all effects present in the risk model (Pop\textsubscript{SUC}). ``Cvg'' stands for coverage.}
\label{tab:scoresStratifiedSUC}
\end{table}

\begin{longtable}[t]{lllrrrrr}
\caption{\label{tab:simulationParametersDHS_SUC} Parameter summary statistics for all spatial smoothing models when predictions are aggregated to the county level. Averaged over 100 simulated stratified surveys for the simulated population with all effects present in the risk model (Pop\textsubscript{SUC}).}\\
\toprule
 &   &   & Est & SD & Q10 & Q50 & Q90\\
\midrule
\endfirsthead
\caption[]{Parameter summary statistics for all spatial smoothing models when predictions are aggregated to the county level. Averaged over 100 simulated stratified surveys for the simulated population with all effects present in the risk model (Pop\textsubscript{SUC}). \textit{(continued)}}\\
\toprule
 &   &   & Est & SD & Q10 & Q50 & Q90\\
\midrule
\endhead
\
\endfoot
\bottomrule
\endlastfoot
\addlinespace[0.3em]
\multicolumn{8}{l}{\textit{\textbf{Smoothed Direct}}}\\
\hspace{1em} &  & Intercept & -1.855 & 0.019 & -1.878 & -1.855 & -1.832\\

\hspace{1em} &  & Phi & 0.745 & 0.2 & 0.445 & 0.795 & 0.962\\

\hspace{1em} &  & Tot. Var & 0.0481 & 0.0127 & 0.0335 & 0.0463 & 0.0649\\

\hspace{1em} &  & Spatial Var & 0.0364 & 0.0153 & 0.0175 & 0.0352 & 0.0565\\

\hspace{1em} &  & iid Var & 0.0117 & 0.00922 & 0.002 & 0.0095 & 0.0247\\
\cmidrule{1-8}
\addlinespace[0.3em]
\multicolumn{8}{l}{\textit{\textbf{BYM2}}}\\
\hspace{1em} & uc & Intercept & -1.903 & 0.016 & -1.922 & -1.922 & -1.884\\
\end{longtable}

\clearpage
\section*{Appendix C: Additional Results from Application to Secondary Education in Kenya}

\begin{sidewaysfigure}
\centering
\includegraphics[width=1.05\textwidth]{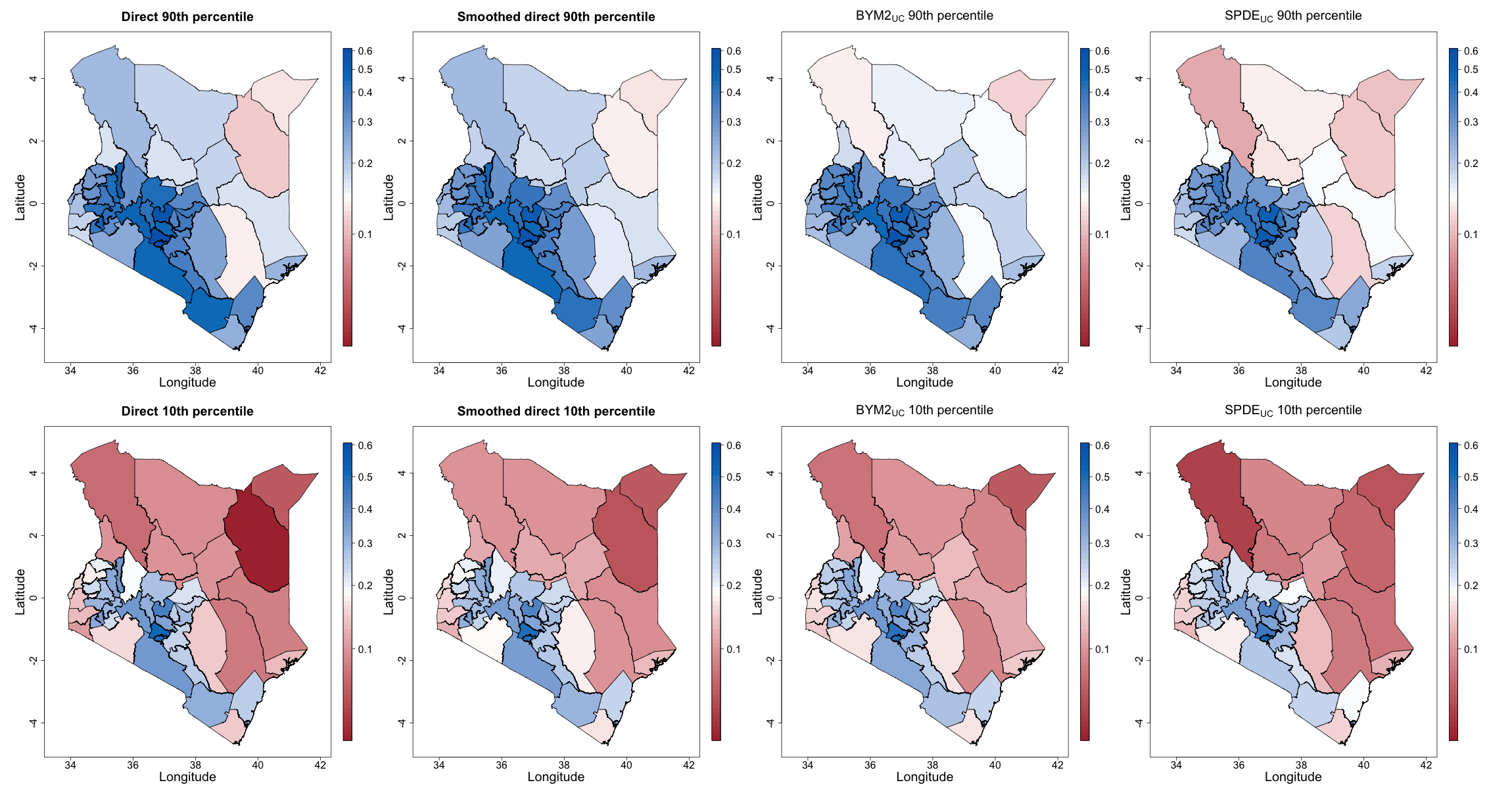}
\caption{Kenya county level 2014 secondary education 90th (top) and 10th (bottom) quantiles for women aged 20--29. The BYM2 and SPDE model predictions include both urban and cluster effects.}\label{fig:Kenya-comparison-quantiles-admin1-ed}
\end{sidewaysfigure}

\begin{longtable}[t]{lrrrrr}
\caption{\label{tab:parameterPredictionsSEP}Parameter and hyperparameter estimate summary statistics from the BYM2 and SPDE models including both urban and cluster effects, fit to the 2014 secondary education prevalence data for young women from the 2014 Kenya DHS.}\\
\toprule
 & Est & SD & Q10 & Q50 & Q90\\
\midrule
\endfirsthead
\caption[]{Parameter and hyperparameter estimate summary statistics from the BYM2 and SPDE models including both urban and cluster effects, fit to the 2014 secondary education prevalence data for young women from the 2014 Kenya DHS. \textit{(continued)}}\\
\toprule
 & Est & SD & Q10 & Q50 & Q90\\
\midrule
\endhead
\endfoot
\bottomrule
\endlastfoot
\addlinespace[0.3em]
\multicolumn{6}{l}{\textit{\textbf{Smoothed Direct}}}\\
\hspace{1em}Intercept & -1.044 & 0.045 & -1.099 & -1.044 & -0.989\\
\hspace{1em}BYM2 Phi & 0.803 & 0.159 & 0.571 & 0.844 & 0.969\\
\hspace{1em}BYM2 Tot. Var & 0.315 & 0.078 & 0.224 & 0.305 & 0.418\\
\hspace{1em}BYM2 Spatial Var & 0.255 & 0.089 & 0.145 & 0.247 & 0.367\\
\hspace{1em}BYM2 iid Var & 0.059 & 0.048 & 0.010 & 0.046 & 0.132\\
\hspace{1em}BYM2 Tot. SD & 0.557 & 0.068 & 0.474 & 0.552 & 0.647\\
\hspace{1em}BYM2 Spatial SD & 0.498 & 0.089 & 0.380 & 0.497 & 0.606\\
\bottomrule
\end{longtable}

\begin{longtable}[t]{lrrrrrrrrr}
\caption{\label{tab:countyPredictionsSEP}Predicted secondary education prevalences and 80\% credible intervals for each of the 47 counties in Kenya for women aged 20--29 in 2014.}\\
\toprule
\multicolumn{1}{c}{\em{\textbf{ }}} & \multicolumn{3}{c}{\em{\textbf{Smoothed Direct}}} & \multicolumn{3}{c}{\em{\textbf{BYM2\textsubscript{UC}}}} & \multicolumn{3}{c}{\em{\textbf{SPDE\textsubscript{UC}}}} \\
\cmidrule(l{3pt}r{3pt}){2-4} \cmidrule(l{3pt}r{3pt}){5-7} \cmidrule(l{3pt}r{3pt}){8-10}
County & Est & Q10 & Q90 & Est & Q10 & Q90 & Est & Q10 & Q90\\
\midrule
\endfirsthead
\caption[]{Predicted secondary education prevalences and 80\% credible intervals for each of the 47 counties in Kenya for women aged 20--29 in 2014. \textit{(continued)}}\\
\toprule
\multicolumn{1}{c}{\em{\textbf{ }}} & \multicolumn{3}{c}{\em{\textbf{Smoothed Direct}}} & \multicolumn{3}{c}{\em{\textbf{BYM2\textsubscript{UC}}}} & \multicolumn{3}{c}{\em{\textbf{SPDE\textsubscript{UC}}}} \\
\cmidrule(l{3pt}r{3pt}){2-4} \cmidrule(l{3pt}r{3pt}){5-7} \cmidrule(l{3pt}r{3pt}){8-10}
County & Est & Q10 & Q90 & Est & Q10 & Q90 & Est & Q10 & Q90\\
\midrule
\endhead
\endfoot
\bottomrule
\endlastfoot
Baringo & 0.250 & 0.202 & 0.304 & 0.240 & 0.203 & 0.282 & 0.250 & 0.218 & 0.283\\
Bomet & 0.276 & 0.240 & 0.316 & 0.284 & 0.245 & 0.326 & 0.267 & 0.236 & 0.300\\
Bungoma & 0.231 & 0.186 & 0.283 & 0.287 & 0.249 & 0.330 & 0.259 & 0.232 & 0.286\\
Busia & 0.193 & 0.156 & 0.237 & 0.182 & 0.150 & 0.218 & 0.176 & 0.150 & 0.203\\
Elgeyo Marakwet & 0.403 & 0.348 & 0.460 & 0.357 & 0.313 & 0.406 & 0.282 & 0.251 & 0.313\\
\addlinespace
Embu & 0.333 & 0.292 & 0.376 & 0.346 & 0.298 & 0.395 & 0.295 & 0.262 & 0.330\\
Garissa & 0.125 & 0.093 & 0.166 & 0.146 & 0.114 & 0.183 & 0.110 & 0.077 & 0.147\\
Homa Bay & 0.167 & 0.142 & 0.196 & 0.184 & 0.152 & 0.217 & 0.168 & 0.146 & 0.192\\
\end{longtable}

\clearpage
\section*{Appendix D: Estimating NMRs in Kenya}

\begin{figure}
\centering
\image{width=3in}{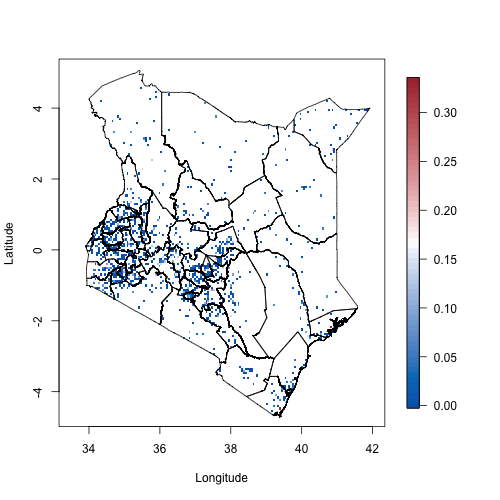} \image{width=3in}{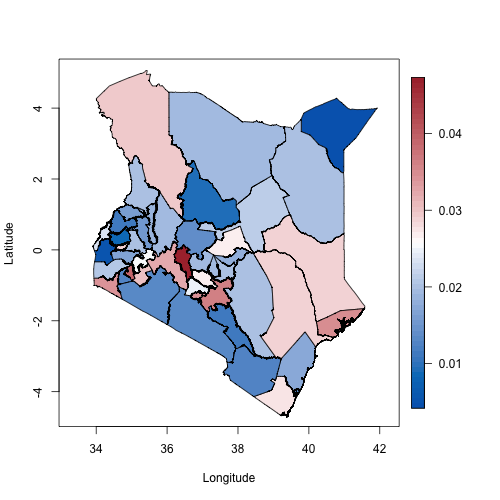}
\caption{Empirical average of neonatal mortality rates in Kenya from 2010-2014 based on data from the 2014 Kenya DHS. Values are shown at both the cluster (left) and county levels (right).}
\label{fig:NMRs}
\end{figure}

\begin{sidewaysfigure}
\centering
\includegraphics[width=.8\textwidth]{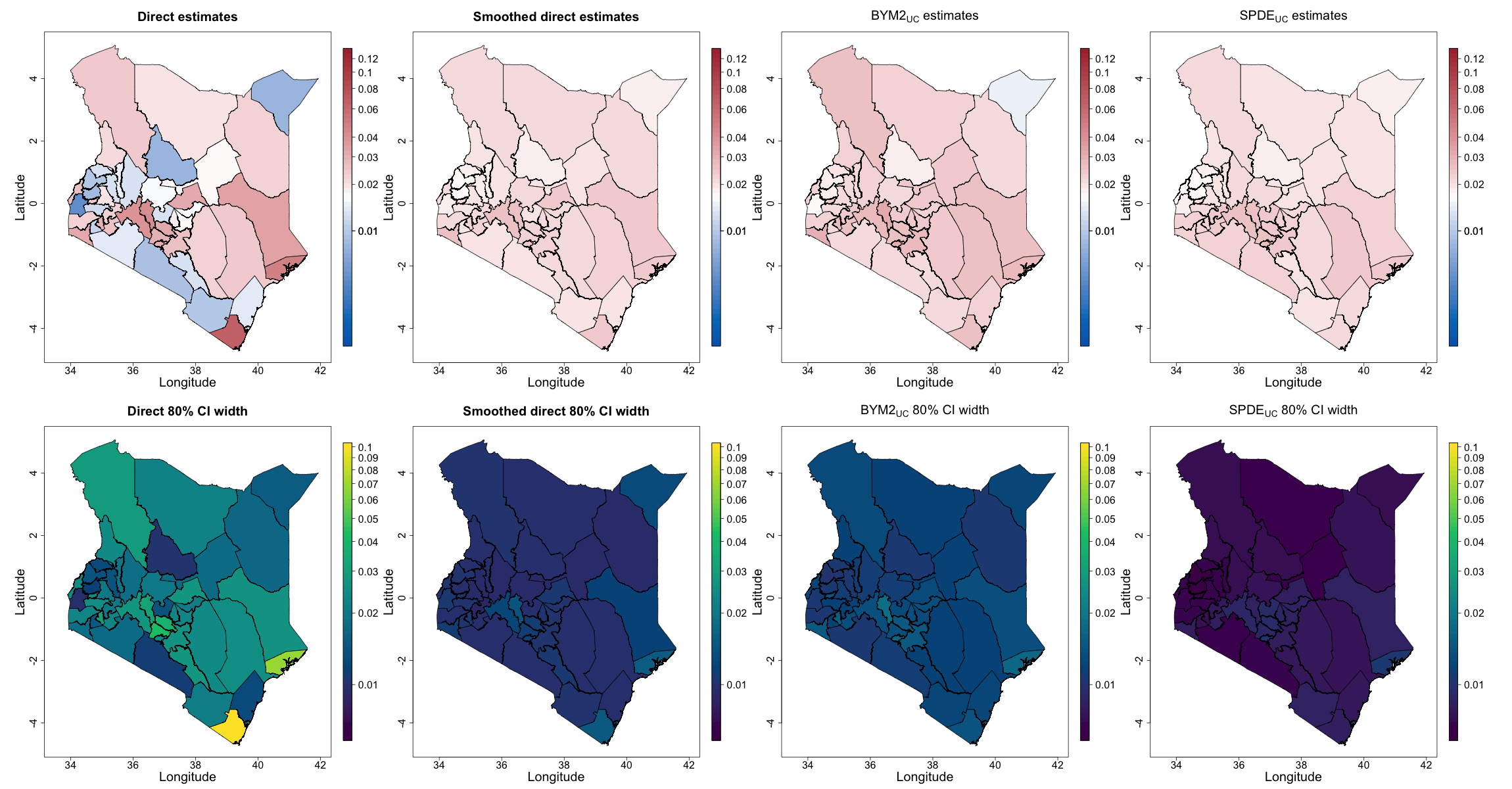}\\
\includegraphics[width=.8\textwidth]{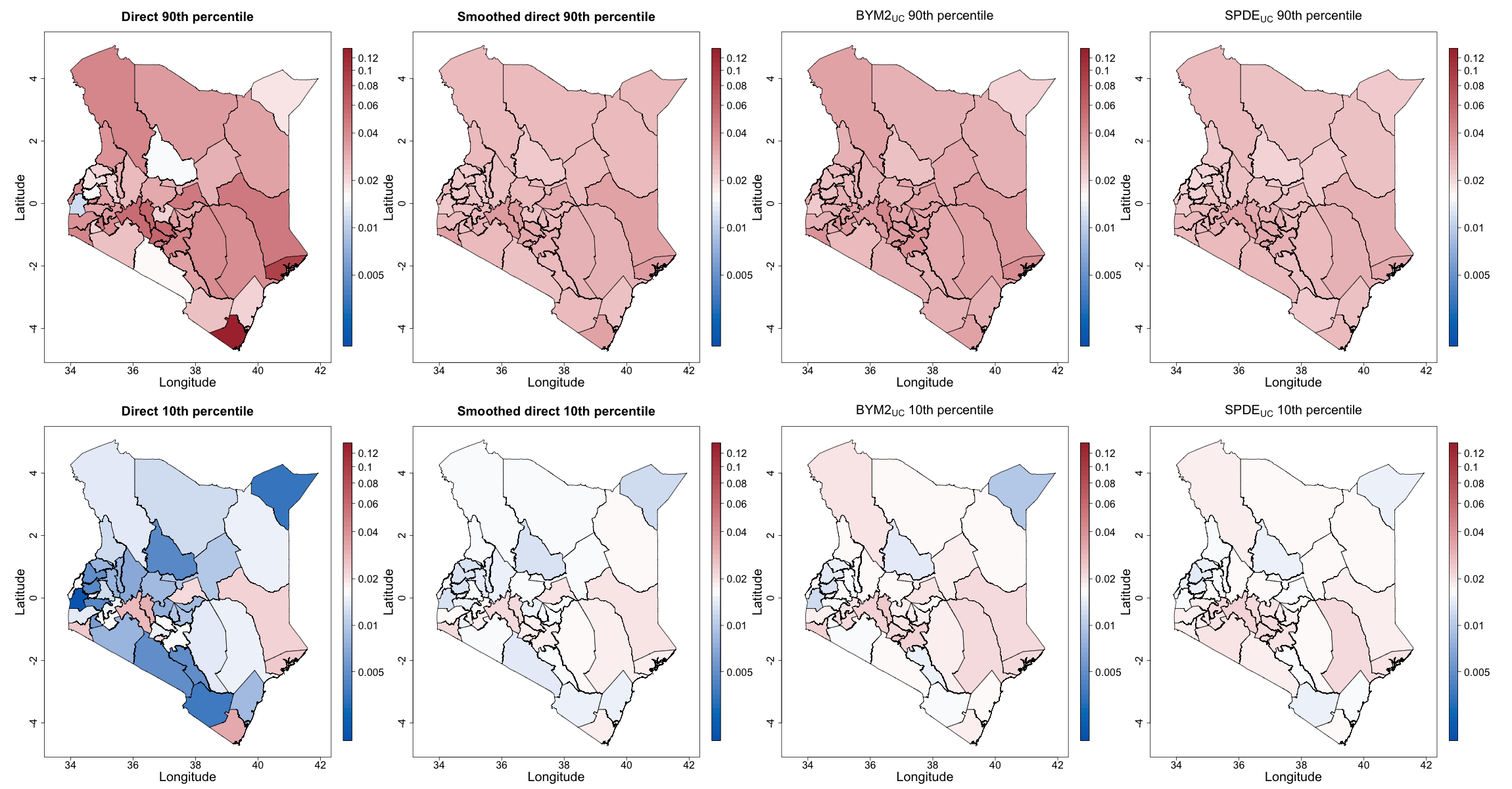}
\caption{Kenya county level neonatal mortality rate predictive mean (top row) and 80\% uncertainty interval widths (second row), 90th percentile (third row), and 10th percentile (bottom row) from 2010-2014. The BYM2 and SPDE model predictions include both urban and cluster effects.}\label{fig:Kenya-comparison-admin1-mort}
\end{sidewaysfigure}

In the section, we estimate neonatal mortality rates (NMRs) for children in Kenya from 2010--2014 based on data from the 2014 Kenya DHS plotted in Figure \ref{fig:NMRs}. The figure shows that the vast majority of clusters have empirical NMRs very close to zero, though there are some clusters that have much higher NMRs with some even above 30\%.  Central NMR estimates, 80\% uncertainty interval widths, and 80\% uncertainty intervals are plotted in Figure \ref{fig:Kenya-comparison-admin1-mort}, and individual county level predictions are given Table \ref{tab:countyPredictionsNMR}. The largest NMRs were estimated to be in several counties just northwest of Nairobi as well as in central eastern Kenya, and the lowest NMRs were estimated to be in the counties near the central western and southwestern borders. Although we expected to find a significant urban effect, we found little evidence suggesting any difference in NMRs between rural and urban areas.

Compared to our analysis of the secondary education prevalence data, the smoothed direct, BYM2\textsubscript{UC}, and SPDE\textsubscript{UC} models estimated from the data had smaller spatial effect variances relative to the predictive uncertainties. For the SPDE\textsubscript{UC} model, for instance, this is evidenced by the fact that the median 80\% credible interval width for the NMR data is 0.0074, whereas the point estimates have a range of 0.0098.  The equivalent values in the secondary education application, which are respectively 0.062 and 0.476, indicate much greater variability across space. The estimated variances of the county level random effects are also relatively small, and are estimated to be 0.059 and 0.060 for the smoothed direct and BYM2\textsubscript{UC} models, respectively. The variance of the spatial component of the SPDE\textsubscript{UC} model was estimated to be 0.069. The cluster effect variance, however,  was estimated to be comparatively larger, with BYM2 and SPDE\textsubscript{UC} estimates of 0.183 and 0.225 respectively, further indicating the large amount of noise relative to any spatial signal in the data. In spite of the lack of spatial signal, the spatial smoothing models helped to reduce the predictive uncertainties relative to the naive and direct models as evidenced from the plotted credible interval widths of the predictions in Figure \ref{fig:Kenya-comparison-admin1-mort}.

Considering the difficulty of including cluster level variation when aggregating SPDE\textsubscript{UC} predictions to the county level, combined with the fact that a substantial majority of the variation in the data occurs at the cluster level as opposed to spatial level variation, we believe the smoothed direct and BYM2 models might be better suited for this particular application. Not including variation due to cluster effects in the spatial aggregation aside from integrating out cluster level variation is likely what leads the SPDE\textsubscript{UC} model predicted NMRs having such narrow credible intervals relative to the other models.

We validate the spatial smoothing models that produce predictions at the cluster level by leaving out data from each county, one county at a time, and making predictions at the cluster level. The results are given in Table \ref{tab:leaveOutCountyNMR}. This table shows that the SPDE\textsubscript{UC} model performs just as well as any of the other spatial smoothing models when making cluster level predictions. In fact, all of the models perform very similarly, again suggesting that variation in the NMR data is primarily at the cluster level and at spatial scales too fine to easily identify.

\clearpage

\begin{longtable}[t]{lrrrrr}
\caption{\label{tab:parameterPredictionsNMR}Parameter and hyperparameter estimate summary statistics from the BYM2 and SPDE models including both urban and cluster effects, fit to the 2010-2014 neonatal mortality rate data from the 2014 Kenya DHS.}\\
\toprule
 & Est & SD & Q10 & Q50 & Q90\\
\midrule
\endfirsthead
\caption[]{Parameter and hyperparameter estimate summary statistics from the BYM2 and SPDE models including both urban and cluster effects, fit to the 2010-2014 neonatal mortality rate data from the 2014 Kenya DHS. \textit{(continued)}}\\\\
\toprule
 & Est & SD & Q10 & Q50 & Q90\\
\midrule
\endhead
\
\endfoot
\bottomrule
\endlastfoot
\addlinespace[0.3em]
\multicolumn{6}{l}{\textit{\textbf{Smoothed Direct}}}\\
\hspace{1em}Intercept & -3.849 & 0.070 & -3.939 & -3.848 & -3.76\\
\hspace{1em}BYM2 Phi & 0.302 & 0.250 & 0.036 & 0.23 & 0.696\\
\hspace{1em}BYM2 Tot. Var & 0.059 & 0.041 & 0.017 & 0.049 & 0.112\\
\hspace{1em}BYM2 Spatial Var & 0.018 & 0.022 & 0.001 & 0.01 & 0.046\\
\hspace{1em}BYM2 iid Var & 0.041 & 0.034 & 0.009 & 0.032 & 0.084\\
\hspace{1em}BYM2 Tot. SD & 0.229 & 0.08 & 0.131 & 0.222 & 0.334\\
\hspace{1em}BYM2 Spatial SD & 0.114 & 0.070 & 0.038 & 0.099 & 0.215\\
\hspace{1em}BYM2 iid SD & 0.187 & 0.078 & 0.092 & 0.180 & 0.290\\
\addlinespace[0.3em]
\multicolumn{6}{l}{\textit{\textbf{BYM2\textsubscript{UC}}}}\\
\hspace{1em}Intercept & -3.989 & 0.090 & -4.106 & -3.989 & -3.872\\
\hspace{1em}Urban & 0.077 & 0.112 & -0.066 & 0.078 & 0.220\\
\hspace{1em}Cluster Var & 0.183 & 0.095 & 0.073 & 0.168 & 0.310\\
\hspace{1em}BYM2 Phi & 0.415 & 0.309 & 0.044 & 0.359 & 0.884\\
\hspace{1em}BYM2 Tot. Var & 0.060 & 0.038 & 0.021 & 0.052 & 0.109\\
\hspace{1em}BYM2 Spatial Var & 0.024 & 0.024 & 0.002 & 0.016 & 0.054\\
\hspace{1em}BYM2 iid Var & 0.037 & 0.034 & 0.004 & 0.028 & 0.081\\
\hspace{1em}Cluster SD & 0.413 & 0.111 & 0.269 & 0.410 & 0.557\\
\hspace{1em}BYM2 Tot. SD & 0.235 & 0.073 & 0.146 & 0.229 & 0.331\\
\hspace{1em}BYM2 Spatial SD & 0.135 & 0.073 & 0.046 & 0.125 & 0.233\\
\hspace{1em}BYM2 iid SD & 0.172 & 0.084 & 0.066 & 0.166 & 0.285\\
\addlinespace[0.3em]
\multicolumn{6}{l}{\textit{\textbf{SPDE\textsubscript{UC}}}}\\
\hspace{1em}Intercept & -4.000 & 0.089 & -4.115 & -4.000 & -3.887\\
\hspace{1em}Urban & 0.079 & 0.111 & -0.064 & 0.079 & 0.221\\
\hspace{1em}Range & 241 & 195 & 79 & 178 & 451\\
\hspace{1em}Spatial Var & 0.069 & 0.062 & 0.017 & 0.051 & 0.140\\
\hspace{1em}Spatial SD & 0.243 & 0.101 & 0.132 & 0.225 & 0.375\\
\hspace{1em}Nugget Var & 0.225 & 0.103 & 0.110 & 0.211 & 0.367\\
\hspace{1em}Nugget SD & 0.463 & 0.105 & 0.331 & 0.459 & 0.606\\*
\end{longtable}

\begin{table}[H]
\centering
\begin{tabular}{lrrrrrrrr}
\toprule
\multicolumn{1}{c}{\em{\textbf{ }}} & \multicolumn{4}{c}{\em{\textbf{BYM2}}} & \multicolumn{4}{c}{\em{\textbf{SPDE}}} \\
\cmidrule(l{3pt}r{3pt}){2-5} \cmidrule(l{3pt}r{3pt}){6-9}
  & uc & uC & Uc & UC & uc & uC & Uc & UC\\
\midrule
\addlinespace[0.3em]
\multicolumn{9}{l}{\textbf{MSE ($\times 10^{-4}$)}}\\
\hspace{1em}Avg & 24.5 & 24.6 & 24.6 & 24.6 & 24.6 & 24.6 & 24.6 & 24.6\\
\hspace{1em}Urban & 29.6 & 29.6 & 29.6 & 29.7 & 29.6 & 29.6 & 29.7 & 29.7\\
\hspace{1em}Rural & 21.3 & 21.4 & 21.3 & 21.4 & 21.4 & 21.4 & 21.4 & 21.4\\
\addlinespace[0.3em]
\multicolumn{9}{l}{\textbf{Var ($\times 10^{-4}$)}}\\
\hspace{1em}Avg & 24.5 & 24.5 & 24.6 & 24.6 & 24.6 & 24.6 & 24.6 & 24.6\\
\hspace{1em}Urban & 29.6 & 29.6 & 29.6 & 29.6 & 29.6 & 29.6 & 29.6 & 29.7\\
\hspace{1em}Rural & 21.3 & 21.3 & 21.3 & 21.3 & 21.4 & 21.4 & 21.4 & 21.4\\
\addlinespace[0.3em]
\multicolumn{9}{l}{\textbf{Bias ($\times 10^{-4}$)}}\\
\hspace{1em}Avg & 8.2 & 26.0 & 8.8 & 27.1 & 4.1 & 3.3 & 5.1 & 7.5\\
\hspace{1em}Urban & -0.1 & 17.7 & 11.0 & 29.0 & -3.0 & -4.0 & 8.0 & 10.9\\
\hspace{1em}Rural & 13.4 & 31.3 & 7.4 & 25.9 & 8.6 & 7.9 & 3.2 & 5.3\\
\addlinespace[0.3em]
\multicolumn{9}{l}{\textbf{CPO}}\\
\hspace{1em}Avg & 0.657 & 0.649 & 0.657 & 0.649 & 0.659 & 0.662 & 0.659 & 0.660\\
\addlinespace[0.3em]
\multicolumn{9}{l}{\textbf{CRPS}}\\
\hspace{1em}Avg & 0.024 & 0.025 & 0.024 & 0.025 & 0.024 & 0.025 & 0.024 & 0.025\\
\bottomrule
\end{tabular}
\caption{Validation results calculated at the cluster level when leaving out one county at a time for the 2014 Kenya DHS NMR data from 2010-2014.}
\label{tab:leaveOutCountyNMR}
\end{table}

\begin{longtable}[t]{lrrrrrrrrr}
\caption{\label{tab:countyPredictionsNMR}Predicted neonatal mortality rates (NMRs) and 80\% credible intervals for each of the 47 counties in Kenya.}\\
\toprule
\multicolumn{1}{c}{\em{\textbf{ }}} & \multicolumn{3}{c}{\em{\textbf{Smoothed Direct}}} & \multicolumn{3}{c}{\em{\textbf{BYM2\textsubscript{UC}}}} & \multicolumn{3}{c}{\em{\textbf{SPDE\textsubscript{UC}}}} \\
\cmidrule(l{3pt}r{3pt}){2-4} \cmidrule(l{3pt}r{3pt}){5-7} \cmidrule(l{3pt}r{3pt}){8-10}
County & Est & Q10 & Q90 & Est & Q10 & Q90 & Est & Q10 & Q90\\
\midrule
\endfirsthead
\caption{
%\label{tab:countyPredictionsNMR}
Predicted neonatal mortality rates (NMRs) and 80\% credible intervals for each of the 47 counties in Kenya. (\textit{continued})}\\
\toprule
\multicolumn{1}{c}{\em{\textbf{ }}} & \multicolumn{3}{c}{\em{\textbf{Smoothed Direct}}} & \multicolumn{3}{c}{\em{\textbf{BYM2\textsubscript{UC}}}} & \multicolumn{3}{c}{\em{\textbf{SPDE\textsubscript{UC}}}} \\
\cmidrule(l{3pt}r{3pt}){2-4} \cmidrule(l{3pt}r{3pt}){5-7} \cmidrule(l{3pt}r{3pt}){8-10}
County & Est & Q10 & Q90 & Est & Q10 & Q90 & Est & Q10 & Q90\\
\midrule
\endhead
\endfoot
\bottomrule
\caption[]{Predicted neonatal mortality rates (NMRs) and 80\% credible intervals for each of the 47 counties in Kenya. \textit{(continued)}}\\
\endlastfoot
Baringo & 0.0194 & 0.0149 & 0.0242 & 0.0219 & 0.0172 & 0.0282 & 0.0206 & 0.0171 & 0.0241\\
Bomet & 0.0224 & 0.0181 & 0.0281 & 0.0262 & 0.0204 & 0.0339 & 0.0234 & 0.0195 & 0.0275\\
Bungoma & 0.0177 & 0.0128 & 0.0230 & 0.0179 & 0.0135 & 0.0235 & 0.0169 & 0.0140 & 0.0199\\
Busia & 0.0207 & 0.0160 & 0.0262 & 0.0223 & 0.0168 & 0.0290 & 0.0176 & 0.0144 & 0.0211\\
Elgeyo Marakwet & 0.0189 & 0.0142 & 0.0241 & 0.0210 & 0.0160 & 0.0272 & 0.0193 & 0.0159 & 0.0228\\
\addlinespace
Embu & 0.0209 & 0.0163 & 0.0264 & 0.0237 & 0.0181 & 0.0309 & 0.0221 & 0.0181 & 0.0264\\
Garissa & 0.0243 & 0.0195 & 0.0313 & 0.0265 & 0.0207 & 0.0342 & 0.0227 & 0.0188 & 0.0273\\
Homa Bay & 0.0212 & 0.0168 & 0.0264 & 0.0228 & 0.0180 & 0.0288 & 0.0221 & 0.0187 & 0.0256\\
Isiolo & 0.0203 & 0.0159 & 0.0251 & 0.0236 & 0.0186 & 0.0300 & 0.0200 & 0.0168 & 0.0234\\
Kajiado & 0.0191 & 0.0143 & 0.0241 & 0.0217 & 0.0165 & 0.0275 & 0.0219 & 0.0186 & 0.0254\\
\addlinespace
Kakamega & 0.0176 & 0.0128 & 0.0228 & 0.0178 & 0.0131 & 0.0236 & 0.0176 & 0.0147 & 0.0208\\
Kericho & 0.0211 & 0.0170 & 0.0259 & 0.0244 & 0.0192 & 0.0309 & 0.0245 & 0.0207 & 0.0284\\
Kiambu & 0.0227 & 0.0179 & 0.0292 & 0.0262 & 0.0203 & 0.0339 & 0.0257 & 0.0214 & 0.0303\\
Kilifi & 0.0192 & 0.0146 & 0.0242 & 0.0218 & 0.0166 & 0.0286 & 0.0203 & 0.0164 & 0.0243\\
Kirinyaga & 0.0210 & 0.0164 & 0.0265 & 0.0238 & 0.0180 & 0.0311 & 0.0230 & 0.0189 & 0.0277\\
\addlinespace
Kisii & 0.0196 & 0.0148 & 0.0248 & 0.0209 & 0.0159 & 0.0276 & 0.0223 & 0.0188 & 0.0260\\
Kisumu & 0.0194 & 0.0145 & 0.0248 & 0.0213 & 0.0161 & 0.0276 & 0.0217 & 0.0183 & 0.0253\\
Kitui & 0.0217 & 0.0173 & 0.0271 & 0.0235 & 0.0182 & 0.0301 & 0.0210 & 0.0173 & 0.0252\\
Kwale & 0.0239 & 0.0182 & 0.0326 & 0.0250 & 0.0189 & 0.0328 & 0.0209 & 0.0170 & 0.0252\\
Laikipia & 0.0204 & 0.0160 & 0.0254 & 0.0217 & 0.0167 & 0.0280 & 0.0207 & 0.0172 & 0.0245\\
\addlinespace
Lamu & 0.0246 & 0.0187 & 0.0339 & 0.0286 & 0.0217 & 0.0385 & 0.0244 & 0.0194 & 0.0301\\
Machakos & 0.0223 & 0.0179 & 0.0280 & 0.0284 & 0.0221 & 0.0368 & 0.0235 & 0.0195 & 0.0277\\
Makueni & 0.0206 & 0.0156 & 0.0265 & 0.0206 & 0.0151 & 0.0270 & 0.0203 & 0.0165 & 0.0244\\
Mandera & 0.0183 & 0.0124 & 0.0252 & 0.0148 & 0.0097 & 0.0218 & 0.0184 & 0.0149 & 0.0221\\
Marsabit & 0.0205 & 0.0160 & 0.0258 & 0.0222 & 0.0167 & 0.0286 & 0.0203 & 0.0172 & 0.0235\\
\addlinespace
Meru & 0.0234 & 0.0191 & 0.0296 & 0.0248 & 0.0192 & 0.0321 & 0.0199 & 0.0159 & 0.0240\\
Migori & 0.0252 & 0.0205 & 0.0318 & 0.0292 & 0.0228 & 0.0375 & 0.0227 & 0.0187 & 0.0267\\
Mombasa & 0.0199 & 0.0146 & 0.0262 & 0.0219 & 0.0158 & 0.0300 & 0.0217 & 0.0169 & 0.0270\\
Murang'a & 0.0230 & 0.0183 & 0.0295 & 0.0257 & 0.0196 & 0.0340 & 0.0249 & 0.0206 & 0.0300\\
Nairobi & 0.0240 & 0.0194 & 0.0307 & 0.0305 & 0.0234 & 0.0397 & 0.0258 & 0.0210 & 0.0310\\
\addlinespace
Nakuru & 0.0258 & 0.0208 & 0.0333 & 0.0277 & 0.0219 & 0.0352 & 0.0267 & 0.0224 & 0.0315\\
Nandi & 0.0199 & 0.0156 & 0.0249 & 0.0215 & 0.0166 & 0.0280 & 0.0201 & 0.0168 & 0.0236\\
Narok & 0.0202 & 0.0156 & 0.0253 & 0.0209 & 0.0161 & 0.0265 & 0.0223 & 0.0191 & 0.0257\\
Nyamira & 0.0224 & 0.0179 & 0.0284 & 0.0284 & 0.0215 & 0.0373 & 0.0231 & 0.0197 & 0.0268\\
Nyandarua & 0.0265 & 0.0209 & 0.0349 & 0.0314 & 0.0238 & 0.0423 & 0.0257 & 0.0215 & 0.0301\\
\addlinespace
Nyeri & 0.0197 & 0.0151 & 0.0248 & 0.0236 & 0.0183 & 0.0310 & 0.0237 & 0.0192 & 0.0283\\
Samburu & 0.0179 & 0.0131 & 0.0230 & 0.0188 & 0.0141 & 0.0250 & 0.0179 & 0.0146 & 0.0217\\
Siaya & 0.0178 & 0.0128 & 0.0233 & 0.0173 & 0.0124 & 0.0230 & 0.0188 & 0.0155 & 0.0223\\
Taita Taveta & 0.0202 & 0.0153 & 0.0257 & 0.0214 & 0.0158 & 0.0282 & 0.0191 & 0.0149 & 0.0234\\
Tana River & 0.0221 & 0.0178 & 0.0278 & 0.0262 & 0.0210 & 0.0329 & 0.0240 & 0.0203 & 0.0278\\
\addlinespace
Tharaka-Nithi & 0.0205 & 0.0158 & 0.0261 & 0.0231 & 0.0174 & 0.0311 & 0.0212 & 0.0170 & 0.0257\\
Trans-Nzoia & 0.0179 & 0.0134 & 0.0229 & 0.0200 & 0.0152 & 0.0257 & 0.0179 & 0.0146 & 0.0214\\
Turkana & 0.0210 & 0.0164 & 0.0266 & 0.0254 & 0.0199 & 0.0325 & 0.0215 & 0.0181 & 0.0254\\
Uasin Gishu & 0.0183 & 0.0139 & 0.0231 & 0.0203 & 0.0154 & 0.0260 & 0.0196 & 0.0164 & 0.0230\\
Vihiga & 0.0207 & 0.0161 & 0.0262 & 0.0239 & 0.0180 & 0.0319 & 0.0194 & 0.0161 & 0.0231\\
\addlinespace
Wajir & 0.0212 & 0.0171 & 0.0263 & 0.0215 & 0.0167 & 0.0274 & 0.0202 & 0.0167 & 0.0241\\
West Pokot & 0.0201 & 0.0155 & 0.0254 & 0.0219 & 0.0168 & 0.0287 & 0.0190 & 0.0156 & 0.0227\\*
\end{longtable}

\bibliographystyle{chicago}
\bibliography{spatepi.bib}